\documentstyle[twoside,epsf]{article}
\catcode`\@=11
\long\def\@makefntext#1{
\protect\noindent \hbox to 3.2pt {\hskip-.9pt
$^{{\eightrm\@thefnmark}}$\hfil}#1\hfill}		%CAN BE USED

\def\@makefnmark{\hbox to 0pt{$^{\@thefnmark}$\hss}}	%ORIGINAL

\def\ps@myheadings{\let\@mkboth\@gobbletwo
\def\@oddhead{\hbox{}
\rightmark\hfil\eightrm\thepage}
\def\@oddfoot{}\def\@evenhead{\eightrm\thepage\hfil
\leftmark\hbox{}}\def\@evenfoot{}
\def\sectionmark##1{}\def\subsectionmark##1{}}

\oddsidemargin=\evensidemargin
\newcounter{sectionc}\newcounter{subsectionc}\newcounter{subsubsectionc}
\renewcommand{\section}[1] {\vspace{12pt}\addtocounter{sectionc}{1}
\setcounter{subsectionc}{0}\setcounter{subsubsectionc}{0}\noindent
	{\tenbf\thesectionc. #1}\par\vspace{5pt}}
\renewcommand{\subsection}[1] {\vspace{12pt}\addtocounter{subsectionc}{1}
	\setcounter{subsubsectionc}{0}\noindent
	{\bf\thesectionc.\thesubsectionc. {\kern1pt \bfit #1}}\par\vspace{5pt}}
\renewcommand{\subsubsection}[1] {\vspace{12pt}\addtocounter{subsubsectionc}{1}
	\noindent{\tenrm\thesectionc.\thesubsectionc.\thesubsubsectionc.
	{\kern1pt \tenit #1}}\par\vspace{5pt}}

\newcounter{appendixc}
\newcounter{subappendixc}[appendixc]
\newcounter{subsubappendixc}[subappendixc]
\renewcommand{\thesubappendixc}{\Alph{appendixc}.\arabic{subappendixc}}
\renewcommand{\thesubsubappendixc}
	{\Alph{appendixc}.\arabic{subappendixc}.\arabic{subsubappendixc}}

\renewcommand{\appendix}[1] {\vspace{12pt}
        \refstepcounter{appendixc}
        \setcounter{figure}{0}
        \setcounter{table}{0}
        \setcounter{lemma}{0}
        \setcounter{theorem}{0}
        \setcounter{corollary}{0}
        \setcounter{definition}{0}
        \setcounter{equation}{0}
        \renewcommand{\thefigure}{\Alph{appendixc}.\arabic{figure}}
        \renewcommand{\thetable}{\Alph{appendixc}.\arabic{table}}
        \renewcommand{\theappendixc}{\Alph{appendixc}}
        \renewcommand{\thelemma}{\Alph{appendixc}.\arabic{lemma}}
        \renewcommand{\thetheorem}{\Alph{appendixc}.\arabic{theorem}}
        \renewcommand{\thedefinition}{\Alph{appendixc}.\arabic{definition}}
        \renewcommand{\thecorollary}{\Alph{appendixc}.\arabic{corollary}}
        \renewcommand{\theequation}{\Alph{appendixc}.\arabic{equation}}
        \noindent{\tenbf Appendix \theappendixc #1}\par\vspace{5pt}}
\newcommand{\subappendix}[1] {\vspace{12pt}
        \refstepcounter{subappendixc}
        \noindent{\bf Appendix \thesubappendixc. {\kern1pt \bfit #1}}
	\par\vspace{5pt}}
\newcommand{\subsubappendix}[1] {\vspace{12pt}
        \refstepcounter{subsubappendixc}
        \noindent{\rm Appendix \thesubsubappendixc. {\kern1pt \tenit #1}}
	\par\vspace{5pt}}

\newcommand{\textlineskip}{\baselineskip=13pt}
\newcommand{\smalllineskip}{\baselineskip=10pt}

\def\eightcirc{
\begin{picture}(0,0)
\put(4.4,1.8){\circle{6.5}}
\end{picture}}
\def\eightcopyright{\eightcirc\kern2.7pt\hbox{\eightrm c}}

\newcommand{\copyrightheading}[1]
	{\vspace*{-2.5cm}\smalllineskip{\flushleft
	{\footnotesize International Journal of Modern Physics B, #1}\\
	{\footnotesize $\eightcopyright$\, World Scientific Publishing
	 Company}\\
	 }}

\newcommand{\publisher}[2]{{\begin{center}\footnotesize\smalllineskip
	Received #1\\
	Revised #2
	\end{center}
	}}

\def\abstracts#1#2#3{{
	\centering{\begin{minipage}{4.5in}\baselineskip=10pt\footnotesize
	\parindent=0pt #1\par
	\parindent=15pt #2\par
	\parindent=15pt #3
	\end{minipage}}\par}}

\def\keywords#1{{
	\centering{\begin{minipage}{4.5in}\baselineskip=10pt\footnotesize
	{\footnotesize\it Keywords}\/: #1
	\end{minipage}}\par}}

\newcommand{\bibit}{\nineit}
\newcommand{\bibbf}{\ninebf}
\renewenvironment{thebibliography}[1]			%ALL CHANGES DD 13/3/92
	{\frenchspacing
	 \ninerm\baselineskip=11pt
	 \begin{list}{\arabic{enumi}.}
	{\usecounter{enumi}\setlength{\parsep}{0pt}
	 \setlength{\leftmargin 12.7pt}{\rightmargin 0pt} %FOR 1--9 ITEMS
	 \setlength{\itemsep}{0pt} \settowidth
	{\labelwidth}{#1.}\sloppy}}{\end{list}}

\newcounter{itemlistc}
\newcounter{romanlistc}
\newcounter{alphlistc}
\newcounter{arabiclistc}

\newcommand{\fcaption}[1]{
        \refstepcounter{figure}
        \setbox\@tempboxa = \hbox{\footnotesize Fig.~\thefigure. #1}
        \ifdim \wd\@tempboxa > 5in
           {\begin{center}
        \parbox{5in}{\footnotesize\smalllineskip Fig.~\thefigure. #1}
            \end{center}}
        \else
             {\begin{center}
             {\footnotesize Fig.~\thefigure. #1}
              \end{center}}
        \fi}

\newcommand{\tcaption}[1]{
        \refstepcounter{table}
        \setbox\@tempboxa = \hbox{\footnotesize Table~\thetable. #1}
        \ifdim \wd\@tempboxa > 5in
           {\begin{center}
        \parbox{5in}{\footnotesize\smalllineskip Table~\thetable. #1}
            \end{center}}
        \else
             {\begin{center}
             {\footnotesize Table~\thetable. #1}
              \end{center}}
        \fi}

\def\@citex[#1]#2{\if@filesw\immediate\write\@auxout
	{\string\citation{#2}}\fi
\def\@citea{}\@cite{\@for\@citeb:=#2\do
	{\@citea\def\@citea{,}\@ifundefined
	{b@\@citeb}{{\bf ?}\@warning
	{Citation `\@citeb' on page \thepage \space undefined}}
	{\csname b@\@citeb\endcsname}}}{#1}}

\newif\if@cghi
\def\cite{\@cghitrue\@ifnextchar [{\@tempswatrue
	\@citex}{\@tempswafalse\@citex[]}}
\def\citelow{\@cghifalse\@ifnextchar [{\@tempswatrue
	\@citex}{\@tempswafalse\@citex[]}}
\def\@cite#1#2{{$\null^{#1}$\if@tempswa\typeout
	{IJCGA warning: optional citation argument
	ignored: `#2'} \fi}}

\def\pmb#1{\setbox0=\hbox{#1}
	\kern-.025em\copy0\kern-\wd0
	\kern.05em\copy0\kern-\wd0
	\kern-.025em\raise.0433em\box0}

\def\fnt#1#2{\footnotetext{\kern-.3em
	{$^{\mbox{\scriptsize #1}}$}{#2}}}

\def\fpage#1{\begingroup
\voffset=.3in
\thispagestyle{empty}\begin{table}[b]\centerline{\footnotesize #1}
	\end{table}\endgroup}

\def\runninghead#1#2{\pagestyle{myheadings}
\markboth{{\protect\footnotesize\it{\quad #1}}\hfill}
{\hfill{\protect\footnotesize\it{#2\quad}}}}
\font\tenrm=cmr10
\font\tenit=cmti10
\font\tenbf=cmbx10
\font\bfit=cmbxti10 at 10pt
\font\ninerm=cmr9
\font\nineit=cmti9
\font\ninebf=cmbx9
\font\eightrm=cmr8

\catcode`\@=12

\def\qed{\hbox{${\vcenter{\vbox{			%HOLLOW SQUARE
   \hrule height 0.4pt\hbox{\vrule width 0.4pt height 6pt
   \kern5pt\vrule width 0.4pt}\hrule height 0.4pt}}}$}}

\renewcommand{\section}[1] {\vspace{12pt}\addtocounter{sectionc}{1}
\setcounter{equation}{0}
\setcounter{subsectionc}{0}\setcounter{subsubsectionc}{0}\noindent
	{\tenbf\thesectionc. #1}\par\vspace{5pt}}
\renewcommand\theequation{\thesectionc.\arabic{equation}}

\begin{document}

\runninghead{A.O. Anokhin  \/ \& \/ M.I. Katsnelson}
{On the Superconductivity of Disordered Alloys}

\normalsize\textlineskip
\thispagestyle{empty}
\setcounter{page}{1}

\copyrightheading{}			%{Vol. 0, No. 0 (1993) 000---000}

\vspace*{0.88truein}

\fpage{1}
\centerline{\bf ON THE PHONON--INDUCED SUPERCONDUCTIVITY}
\vspace*{0.035truein}
\centerline{\bf OF DISORDERED ALLOYS}
\vspace*{0.37truein}
\centerline{\footnotesize A.O. ANOKHIN
            \normalsize and \footnotesize M.I. KATSNELSON}
\vspace*{0.015truein}
\centerline{\footnotesize\it Institute of Metal Physics}
\baselineskip=10pt
\centerline{\footnotesize\it S.Kovalevskaya Str.~18,
           620219 Ekaterinburg, Russia}
\vspace*{0.225truein}
\publisher{(received date)}{(revised date)}

\vspace*{0.21truein}
\abstracts{
	A model of alloy is considered which includes both quenched
disorder in the electron subsystem (``alloy'' subsystem) and
electron-phonon interaction.
	For given approximate solution for the
alloy part of the problem, which is assumed  to  be  conserving  in
Baym's sense, we construct the generating functional and derive the
Eliashberg-type equations which are valid to the  lowest  order  in
the adiabatic parameter.
	The  renormalization  of  bare  electron-phonon
interaction vertices  by  disorder  is  taken  into  account
consistently with the approximation for the alloy self-energy.
For the case of exact configurational averaging the same set of
equations is established within the  usual  $T$-matrix  approach.
	We demonstrate that for any conserving approximation for the alloy
part of the self-energy the Anderson's theorem holds in the case of
isotropic singlet pairing provided disorder renormalizations of the
electron-phonon interaction vertices are neglected.
	Taking  account
of the disorder renormalization of the electron-phonon  interaction
we  analyze  general  equations  qualitatively  and   present   the
expressions for $T_{c}$  for the case of weak and intermediate
electron-phonon  coupling.
	Disorder  renormalizations  of  the  logarithmic
corrections  to  the  effective  coupling,  which  arise  when  the
effective interaction kernel for the Cooper channel has the  second
energy  scale,  as  well  as  the  renormalization  of  the  dilute
paramagnetic impurity suppression are discussed.}{}{}

\vspace*{10pt}

%\textlineskip			%) USE THIS MEASUREMENT WHEN THERE IS
%\vspace*{12pt}			%) NO SECTION HEADING

\vspace*{1pt}\textlineskip	%) USE THIS MEASUREMENT WHEN THERE IS

\section{Introduction}
\noindent
	A  problem  of  how  disorder   influences   a   superconductive
transition temperature $T_{c}$, first studied in pioneering  papers  by
Anderson \cite{c1,c2} and by Abrikosov and Gor'\-kov,~\cite{c3}
draws a great deal  of
physicists attention.\cite{c4,c5,c6,c7,c8,c9,c10}
However,  even
if  a  usual  mechanism   of   the   superconductivity   owing   to
electron-phonon interaction (EPI) and isotropic $s$-type pairing  are
concerned --- it is this subject  present  paper  deals  with  ---  one
cannot regard the problem as being solved completely. The situation
where  disorder  scattering  is  weak  or  the   concentration   of
impurities (both magnetic and nonmagnetic)  is  small  was  studied
rigorously.\cite{c3,c4,c11}

As for strong disordering, in early works\cite{c6,c7,c10}
this problem was considered within
the framework of semiphenomenological approaches which were called for
explaining qualitatively empirical rules (e.g., Miedema
rules\cite{c12,c13,c14})
determining superconductive transition temperature in alloys.
In a pioneering paper by Anderson et al.\cite{c15} a mechanism
of universal $T_c$--degradation was proposed which was based
on the enhancement of screened Coulomb pseudopotential
due to disordering effects in compounds with A-15 structure.
Further a great number of works appeared touching interrelations
between the phonon-induced superconductivity and  Anderson's localization,
and other effects of strong disorder, see papers,\cite{c16,c17,c18,c19,c20}
and also a recent review by Belitz and Kirkpatrick\cite{c21} where
mapping on the non-linear matrix $\sigma$-model was used when treating
the Anderson's localization and related problems.

Recently a new impetus has appeared for studying the influence of disorder
on $T_c$. This impetus comes mainly from discussing mechanisms of
superconductivity in novel Cu-oxide based compounds (see, e.g.,
papers by Abrikosov,\cite{c22} where the influence of
non-magnetic impurities on $T_c$ is discussed in the case of strongly
quasimomentum-dependent singlet order parameter). These substances
are supposed to be strongly correlated systems  which
may be described in the simplest way by an ``alloy  analogy''
within some approaches, as for instance Hubbard-III
approximation \cite{c23} or static approximation within functional integral
approach\cite{c24,c25,c26,c27} where this analogy
emerges in a rather natural way.
	In particular, the renormalization of  the  phonon-induced
$T_{c}$   was discussed  in  Ref.\cite{c28} where the use was made of
Hubbard-III approximation when treating the effects of strong
electron on-site repulsion, however only a part of contributions
was taken into account. Although the approximations of Hubbard-III type
are considered now as not being quite adequate, from the formal point of
view they have many similarities with modern approaches to highly
correlated systems, which are based on the $d\to\infty$
limit:\cite{c29,c30,c31,c32,c33}
the electron self-energy possesses strong energy dependence in the normal
phase, while the quasimomentum dependence is absent.

Thus, long time after creating BCS theory the classical problem of
the disorder influence on $T_c$ attracts attention of researchers.

At the same time, existence of large number of works and approximations
has lead to that it is difficult to compare results  of various approaches
between each other. Even the order parameter is understand in different ways.
Therefore it is instructive to study the problem in a general form by which
we mean the following.
Given an approximate self-energy expression for the
averaged electron Green's function in the normal phase and without
electron-phonon interaction, we search for the equations
determining the order parameter in the superconductive phase
for the electron-phonon pairing mechanism,
which are consistent (in some still undetermined sense)
with the approximation.

	In the present paper we study systematically the question of how
strong quenched substitutional disorder (components of an alloy are
assumed  to  be  nonmagnetic)  influences  $T_{c}$, leaving aside an
interesting in its own right and fascinating  problem  of  an
interplay between superconductivity and localization  phenomenon
(the latter is known may occure in disordered systems\cite{c34}).
We shall demonstrate
that for a given approximation describing normal state of
an alloy in the absence of EPI and satisfying  natural  requirement
of being thermodynamically consistent in Baym-Kadanoff sense\cite{c35,c36}
(later on rather the terms ``$\phi$-derivable'' or ``conserving'' will be
used interchangeably when referring to such approximations), and with
suitable meaning of the term ``consistent'', the problem can be solved
in an unique way. We propose a procedure which enables constructing
the  Eliashberg-type equations, which are valid to the lowest  order
in the adiabatic parameter and properly take into account the disorder
renormalization of EPI vertices. The structure of these equations
is the same for all $\phi$-derivable approximations and such approximations,
as Virtual Crystal, Diffusional, Averaged $T$--Matrix and Coherent Potential
approximations, etc., can be discussed within an unified scheme.
	However, the linearized version of the  equations
is so complicated that further simplifications are required and  we
establish also what kind of simplifications leads to the results
previously obtained within some (in fact, within the most of already
mentioned) approximations widely accepted in treating the ``alloy''
problems.

We should note that the problem in a similar form has been considered by
Belitz\cite{c16,c17,c18} with the use of exact eigenfunction representation.
However, in this approach the very definition of the order parameter differs
from the traditional one, which exploits quasimomentum or coordinate
representation, and establishing relations between these approaches is,
in general, a very difficult problem. The approach being
proposed in this paper seems to be more  directly related with observable
characteristics, example of which is the quasimomentum dependence of the
superconductive gap. In this connection, the traditional and
often not specially discussed supposition, that the order parameter
is constant over the Fermi surface in dirty metals and, therefore,
the type of averaging procedure is not crucially important,
does not, generally speaking, hold (see Ref.\cite{c22}).

	The paper is organized as follows. In Section~2
we discuss the model.
	The generating functional is constructed and  the
equations of the  strong  coupling  theory  are  then  derived.
	An alternative approach with the use of standard $T$-matrix
formulation for the configurational averaging is given in Section~3
where we discuss also the relationships between the two approaches.
	In Section~4 we demonstrate that for the case of the
isotropic singlet
pairing owing to electron-phonon interaction the Anderson's theorem
may be valid when formulated analogously to that for the BCS model,
and we explicitly specify the conditions for the  theorem  to  hold
(all earlier treatments appear  to  be  incomplete).
	In  Section~5
reduced isotropic equations are  derived  in  which  the  anomalous
self-energy   contribution   coming   from   disorder    scattering
(``disorder'' contribution to the anomalous part of the self-energy)
is eliminated exactly.
	Section~6 deals with qualitative analysis of
the  reduced  equations  within  certain  approximations  for   the
isotropic contributions of  electron-phonon  vertices  and  general
expressions for $T_{c}$ are presented there for the  case  of  weak
and intermediate  electron-phonon  coupling.
	We  establish   relations
between effective electron-phonon coupling and generating  function
for    arbitrary    thermodynamically    consistent     single-site
approximation for the alloy self-energy  and  write  down  explicit
expressions for  renormalized  electron-phonon  coupling  within  a
number of widely known single-site
approximations in Section~7.
	In Appendix~A we give a derivation of Ward-type identities.
	Appendix~B deals with renormalizations of the $T_{c}$--suppression
by dilute paramagnetic impurities and in  Appendix~C some useful relations
between generating functions for ``disorder'' self-energy  and  for
vertex corrections are  established  within  arbitrary  single-site
approximation.

\section{The Model, Generating Functional and Eliashberg-type Equations}
\noindent
  To begin with, we consider the model of alloy with
substitutional disorder described by the Fr\"olich-type model
Hamiltonian
\begin{equation}
        H=H_{e}+H_{e-ph}+H_{ph},
\label{eq2.1}
\end{equation}
where $H_{e}$ is Hamiltonian of an  electron  subsystem,
which  in  the Wannier representation has the form
\begin{eqnarray}
& \displaystyle
        H_{e}=\sum_{ij,\sigma} h_{ij} c^{\dag}_{i\sigma}c_{j\sigma}, &
\label{eq2.2} \\
& \displaystyle
        h_{ij}=t_{ij}-\mu\delta_{ij}+\xi_{ij}+\varepsilon_{i}\delta_{ij}, &
\label{eq2.3}
\end{eqnarray}
and $c^{\dag}_{i\sigma}$ ($c_{i\sigma}$) is electron creation
(annihilation) operator  in  the Wannier state with spin
$\sigma$ on  site  $\bf i$;
$t_{ij}$,  non-random  part  of transfer integral;
$\xi_{ij}$,  random  part  of  transfer  integral  and
$<\xi_{ij}>=0$ ($<\ldots>$  means  configurational  average
with   certain probability distribution);
$\varepsilon_{i}$, on-site electron energy; $\delta_{ij}$, usual
Kronecker delta symbol and $\mu$,  the chemical potential.

  In the case of substitutional disorder of a  general  type  both
$\xi_{ij}$  and $\varepsilon_{i}$ are random variables
which  depend on kinds  of  atoms
placed at the corresponding lattice sites and particular  cases  of
alloy  disorder  may  be  obtained  by  appropriate  choosing   the
quantities $\xi_{ij}$ and $\varepsilon_{i}$ to be random or not.
The so  called  diagonal
disorder emerges when one sets all the $\xi_{ij}$ to be identically zero
and quantities $\varepsilon_{i}$ are random and
depend on the  sort of atoms at the corresponding lattice sites.
If, on the  contrary,  $\xi_{ij}$ are  random,
the Hamiltonian  (\ref{eq2.2})  describes  the  case  of
off-diagonal alloy disorder.\cite{c37}

        To complete the description of  a  disorder  type  one
should fix distributions for the  random  quantities  entering  the
Hamiltonian (\ref{eq2.2}).
  For the case of random binary alloy model with diagonal disorder
Eq.~(\ref{eq2.3}) takes the form
\begin{eqnarray}
&  h_{ij}=t_{ij}-\mu\delta_{ij}+\varepsilon_{A}\delta_{ij} +
                V \eta_{i}\delta_{ij},
\label{eq2.4} \\
&  {\rm Prob}(\eta_{i} = 1) = c,
   \quad
   {\rm Prob}(\eta_{i} = 0) = 1 - c,
\label{eq2.5}
\end{eqnarray}
with $c$ being the concentration of B-type component of the alloy  in
the host which is  assumed  to  be  of  A-type,  and
$V=\varepsilon_{B} -\varepsilon_{A}$ is the
scattering potential of B-type atoms measured relatively to the host potential.
We will use this specific model of alloy in Sections~6 and~7
and in Appendices~B and~C. In the remaining part of the paper a discussion
is rather general and does not rely upon any specific alloy model.

  The Hamiltonian of the phonon subsystem is of the following form
\begin{equation}
H_{ph}=\sum_{{\bf q},s}\omega_{{\bf q},s}
                 b^{\dag}_{{\bf q},s}b_{{\bf q},s},
\label{eq2.6}
\end{equation}
with $b^{\dag}_{{\bf q},s}$  ($b_{{\bf q},s}$)
being phonon creation  (annihilation)
operator  with quasimomentum  ${\bf q}$ and  the  branch  index  $s$,
$\omega_{{\bf q},s}$ being the phonon
frequencies and we assume that  $\omega_{{\bf q},s}$
are  non-random functions despite they are renormalized
by all the interactions in our model.

  Finally, the  Hamiltonian  of  the  electron-phonon  interaction
reads
\begin{equation}
   H_{e-ph}=\sum_{ij,\sigma} \sum_{l,s} M^{(s)}_{l;ij}
            c^{\dag}_{i\sigma} c_{j\sigma}
            \frac{1}{\sqrt{N}}\sum_{\bf q}
            \exp(-{\rm i}{\bf q}{\bf R}_{l})
            \left( b^{\dag}_{{\bf q},s} + b_{-{\bf q},s}\right),
\label{eq2.7}
\end{equation}
where $M^{(s)}_{l;ij}$ is matrix element of  the  electron-phonon
interaction (EPI) and supposed to be independent of disorder.\cite{c1}

Deriving explicite expressions for $M^{(s)}_{l;ij}$ requires some care
since because the scattering by impurities
is ellastic only in the coordinate frame moving with
crystall lattice as was pointed out by Blount.\cite{c38}
We will not write down the expressions for $M^{(s)}_{l;ij}$ here;
corresponding detalies and the results of calculations are given
in Refs.\cite{c39,c40,c41,c42}.

  As is known\cite{c43} one should treat $\omega_{{\bf q},s}$ as
completely renormalized by all the
relevant interactions when considering  the
phenomenon of superconductivity in such model, and effects  of  EPI
on electron spectrum should be taken into account in the lowest (in
fact, second) order in $M^{(s)}_{l;ij}$ for each disorder
configuration.
This actually corresponds to the lowest order of the pertubation  theory
in addiabatic parameter as was demonstrated by Migdal.\cite{c44}
Of course the two simplifying approximations about non-random character of
$M^{(s)}_{l;ij}$ and $\omega_{{\bf q},s}$ cannot be proven regorously
if the real situation  is kept in mind,
and therefore they have a model character within the framework of
the Fr\"olich-type model.
However, as we will see in what follows,
even with these approximations the model remains complicated and
interesting enough for studying.

  In the Gorkov-Nambu representation\cite{c4,c45}
(\ref{eq2.2}) and~(\ref{eq2.7}) take the form
\begin{equation}
   H_e = \sum_{ij} h_{ij} {\bf C}^{\dag}_{i} \tau_{3} {\bf C}_{j},
\label{eq2.8}
\end{equation}
\begin{equation}
   H_{e-ph}=\sum_{ij} \sum_{l,s} M^{(s)}_{l;ij}
        {\bf C}^{\dag}_{i} \tau_{3} {\bf C}_{j}
        \frac{1}{\sqrt{N}}\sum_{\bf q}
        \exp(-{\rm i}{\bf q}{\bf R}_{l})
        \left( b^{\dag}_{{\bf q},s} + b_{-{\bf q},s}\right),
\label{eq2.9}
\end{equation}
where we have introduced Nambu row- and column- spinors
\begin{equation}
{\bf C}^{\dag}_{i} = \left( c^{\dag}_{i\uparrow},\, c_{i\downarrow} \right),
\qquad
{\bf C}_{i} = \left(
                   \begin{array}{l}
                       c_{i\uparrow} \\
                       c^{\dag}_{i\downarrow}
                   \end{array}
              \right)
\label{eq2.10}
\end{equation}
and
\begin{equation}
\tau_{0} = \left(\begin{array}{cc}
                  1  &  0 \\
                  0  &  1
                 \end{array}\right),
\quad
\tau_{1} = \left(\begin{array}{cc}
                  0  &  1 \\
                  1  &  0
                 \end{array}\right),
\quad
\tau_{2} = \left(\begin{array}{cr}
                  0        & -{\rm i}  \\
                  {\rm i}  &  0
                 \end{array}\right),
\quad
\tau_{3} = \left(\begin{array}{cr}
                  1  &  0 \\
                  0  &  -1
                 \end{array}\right)
\label{eq2.11}
\end{equation}
are usual Pauli matrices.

  To derive equations of  strong  coupling  theory,  or  so-called
Eliashberg-type e\-qua\-tions,\cite{Eli} we use the following trick. We
construct a functional $W[G,D]$ of  Luttinger-Ward type.\cite{c46}
As is known, the functional derivative of such a functional with respect
to one-particle Green's function $G$
give the self-energy $\Sigma [G,D]$ (or mass operator),
as functionals of $G$ and  $D$,  and,
along with Dyson's equation for Green's function  $G$,  this
procedure leads to a set of self-consistent equations on  $\Sigma[G,D]$
and $G$.\cite{c36} Within the Fr\"olich-type model being used here,
``bare'' phonon Green's
function $D$ is fully  determined  by  $H_{e-ph}$ Eq.~(\ref{eq2.6}), and
coinsides with the renormalized one.
We suppose also that the functional $W^{e}[G]$ determining
certain approximate self-energy $\Sigma^{e}[G]$ for the  alloy  part  of  the
problem, Eqs.~(\ref{eq2.2},\ref{eq2.3},\ref{eq2.8}) (i.e., without EPI),
is known and what remains is to find the contribution of EPI to $W[G,D]$
up to the lowest order in EPI coupling with account of those renormalizations
of EPI vertices by disorder which are consistent  with  given  approximate
$\Sigma^{e}[G]$ (or, which is the same, with the approximation to $W^{e}[G]$).

  More formally, we seek for the set of equations of the following
general form
\begin{eqnarray}
&  W[G,D] = W^{e}[G] + W^{e-ph}[G,D] , &
\label{eq2.12} \\
&  \Sigma[G,D] = \Sigma^{e}[G] + \Sigma^{e-ph}[G,D] , &
\label{eq2.13} \\
& \displaystyle
   \Sigma^{e}[G] = \frac{\delta W^{e}[G]}{\delta G^T} , &
\label{eq2.14} \\
& \displaystyle
   \Sigma^{e-ph}[G,D] = \frac{\delta W^{e-ph}[G,D]}{\delta G^T} , &
\label{eq2.15} \\
&  G = G_{0} + G_{0} \Sigma[G,D] G , &
\label{eq2.16}
\end{eqnarray}
where $W^{e}[G]$ and $W^{e-ph}[G,D]$ are the contributions to  the
functional
$W[G,D]$ which are due to disorder scattering and EPI respectively;
$\Sigma^{e}[G]$  and  $\Sigma^{e-ph}[G,D]$, the  corresponding
contributions to the
self-energy $\Sigma[G,D]$, and in (\ref{eq2.16}) $G_{0}$ is ``bare''
one-particle  Green's
function determined by non-random part of $H_{e}$ Eq.~(\ref{eq2.2}).

To construct the functional $W^{e-ph}[G,D]$ we note that to lowest order
in EPI
\begin{eqnarray}
\Pi[G,D] & = & \displaystyle
          2\frac{\delta W[G,D]}{\delta D^T}
\nonumber \\
         & = & \displaystyle
          2\frac{\delta W^{e-ph}[G,D]}{\delta D^T} ,
\label{eq2.17}
\end{eqnarray}
where $\Pi[G,D]$ is the polarization operator, and
\begin{equation}
\frac{\delta\Pi[G,D]}{\delta D^T} = 0,
\label{eq2.18}
\end{equation}
so that $\Pi[G,D]$ is a functional of $G$ only, $\Pi=\Pi[G]$.
The functional $W^{e-ph}[G,D]$ may then be writen as
\begin{equation}
W^{e-ph}[G,D]=\frac{1}{2} {\rm Sp} D\Pi[G]
\label{eq2.19}
\end{equation}
in some formal notation.

Now we may give the definition of the renormalizations of EPI which are
consistent with the approximtion for $\Sigma^{e}[G]$. We will call
the contributions to $\Sigma^{e-ph}[G,D]$ consistent with the approximation
to $\Sigma^{e}[G]$ if this contributions are derived
from the functional (\ref{eq2.19})
where the polarization operator $\Pi[G]$ is consistent with
the approximation for $\Sigma^{e}[G]$ in Boym-Kadanoff sence.\cite{c35,c36}

To obtain explicite expression we devide $W^{e-ph}[G,D]$ into two
parts
\begin{equation}
        W^{e-ph}[D,G]=W^{H}[D,G] + W^{F}[D,G] ,
\label{1}
\end{equation}
where $W^{H}[G,D]$ and $W^{F}[G,D]$ are Hartree- and Fock-type contributions
to the generating functional respectively.

Consider Fock-type contribution first. Contributions to the
polarization operator
\begin{equation}
\Pi^{F}[G] = 2\frac{\delta W^{F}[D,G]}{\delta D^T}
\label{eq2.21}
\end{equation}
are shown in Fig.~1.
%%%%%%%%%%%%%%%%%%%%%%%%%%%%%%%%%%%%%%%%%%%%%%%%%%%%%%%%%%%%%%%%%%%%%%%%%%%%
\begin{figure}[htbp]
\vspace*{13pt}
\epsfysize=0.5truein
\centerline{\epsfbox{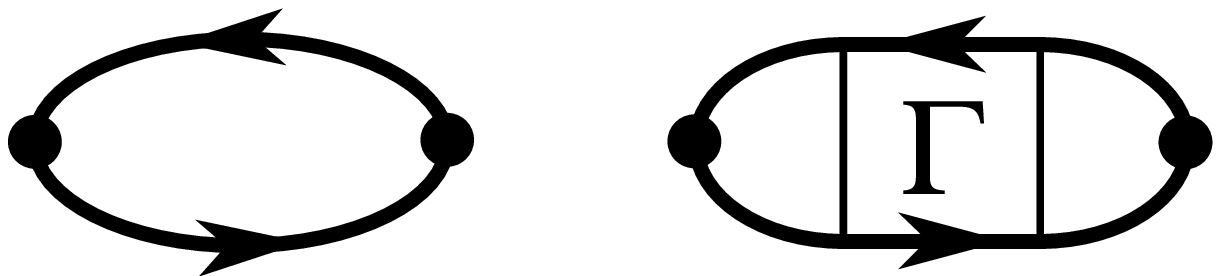}}
\vspace*{13pt}
\fcaption{Fock-type contributions to the
polarization operator. Black circle denotes ``bare'' electron-phonon
three-leg vertex; bold line, the renormalized electron Green's function;
the square labeled with \protect{${\bf \Gamma}$}, full disorder four-leg
vertex. Arrows points the direction of index arrangement.}
\end{figure}
%%%%%%%%%%%%%%%%%%%%%%%%%%%%%%%%%%%%%%%%%%%%%%%%%%%%%%%%%%%%%%%%%%%%%%%%%%%%
Restoreing the generating functional $W^{F}[G,D]$
with the use of Eq.(\ref{eq2.19}) yields
\begin{eqnarray}
& \displaystyle
   W^{F}[G,D]=
   -\frac{T}{2} \sum_{{\rm i}p_s,{\rm i}p_m}
   D_{\gamma\gamma'}({\rm i}p_s - {\rm i}p_m)
   M^{\gamma}_{\alpha\alpha'} \Bigl\{
   G_{\alpha\beta}({\rm i}p_m)G_{\beta'\alpha'}({\rm i}p_s) +
\nonumber \\
&  G_{\alpha\mu}({\rm i}p_m)G_{\mu'\alpha'}({\rm i}p_s)
   \Gamma_{\mu\nu;\nu'\mu'}({\rm i}p_m,{\rm i}p_s)
   G_{\beta'\nu'}({\rm i}p_s)G_{\nu\beta}({\rm i}p_m)
   \Bigr\}
   M^{\gamma'}_{\beta'\beta}
\label{eq2.22}
\end{eqnarray}
and we depicted the functional $W^{F}[G,D]$ in Fig.~2.
%%%%%%%%%%%%%%%%%%%%%%%%%%%%%%%%%%%%%%%%%%%%%%%%%%%%%%%%%%%%%%%%%%%%%%%%%%%%
\begin{figure}[htbp]
\vspace*{13pt}
\epsfysize=1.0truein
\centerline{\epsfbox{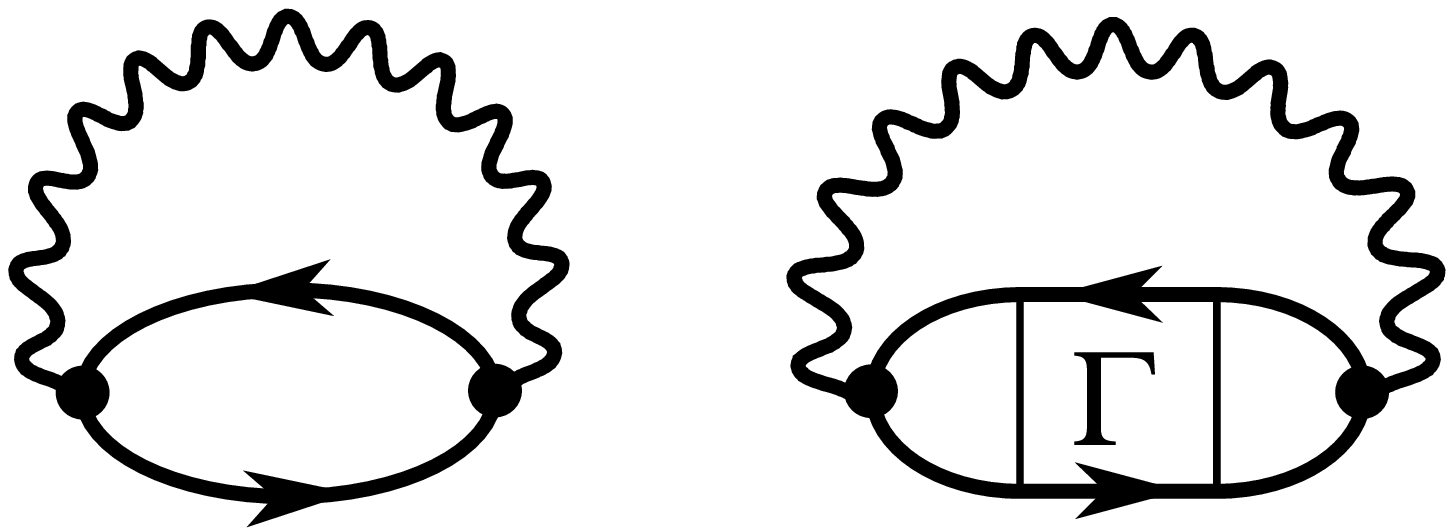}}
\vspace*{13pt}
\fcaption{Fock-type contributions to the generating
functional. Bold wavy line denotes ``bare'' phonon Green's function and
meaning of the other graphical elements is the same as in Fig.~1.}
\end{figure}
%%%%%%%%%%%%%%%%%%%%%%%%%%%%%%%%%%%%%%%%%%%%%%%%%%%%%%%%%%%%%%%%%%%%%%%%%%%%
Here and in what follows implicit summation  over  doubly  repeated
Greek indices is implied,
\begin{equation}
   M^{\gamma}_{\alpha\alpha'}=
        M^{s_{\gamma}}_{l_{\gamma};i_{\alpha}j_{\alpha'}}
        \left[ \tau_{3} \right]_{\sigma_{\alpha}\sigma_{\alpha'}}
\label{eq2.23}
\end{equation}
is bare EPI vertex, and we have introduced multy-indices
encapsulating  site  {\it and\ } spinor  indices
(e.g.  $\alpha$,  $\beta$, etc. in (\ref{eq2.22}))
or site {\it and\ }branch indices ($\gamma$, $\gamma'$ in~(\ref{eq2.22})).

The quantity $\Gamma_{\alpha\beta;\beta'\alpha'}({\rm i}p_m,{\rm i}p_s)$
is full disorder four-leg vertex which obeys the Bethe-Salpeter equation
\begin{eqnarray}
& \Gamma_{\alpha\beta;\beta'\alpha'}({\rm i}p_s,{\rm i}p_m) =
  U_{\alpha\beta;\beta'\alpha'}({\rm i}p_s,{\rm i}p_m) + &
\nonumber \\
& + U_{\alpha\mu;\mu'\alpha'}({\rm i}p_s,{\rm i}p_m)
  G_{\mu\nu}({\rm i}p_s)G_{\nu'\mu'}({\rm i}p_m)
  \Gamma_{\nu\beta;\beta'\nu'}({\rm i}p_s,{\rm i}p_m), &
\label{eq2.24}
\end{eqnarray}
and $U_{\mu\nu;\nu'\mu'}({\rm i}p_s,{\rm i}p_m)$ is irreducible
disorder four-leg vertex generated by the approximation used
for $\Sigma^{e}[G]$ (in fact this vertex
is the second variational derivative
of the functional $W^{e}[G]$ with respect to $G$).

  Apart from the standard Fock-type contribution, there is still
a contribution of Hartree-type. This contribution is usually
neglected following Muttalib and Anderson.\cite{c15}
Howevere as was demonstrated in a series of papers by Belitz\cite{c16,c17,c18}
this contribution is important and do play role in final expression
for $T_c$.

To take Hartree-type contribution into account we introduce,
in an analogy with Fock case, a polarization operator $\Pi^{H}[G]$
as
\begin{equation}
\Pi^{H}[G] = 2\frac{\delta W^{H}[D,G]}{\delta D^T}
\label{eq2.25}
\end{equation}
(see Fig.3 for corresponding contributions).
%%%%%%%%%%%%%%%%%%%%%%%%%%%%%%%%%%%%%%%%%%%%%%%%%%%%%%%%%%%%%%%%%%%%%%%%%%%%
\begin{figure}[htbp]
\vspace*{13pt}
\epsfysize=0.5truein
\centerline{\epsfbox{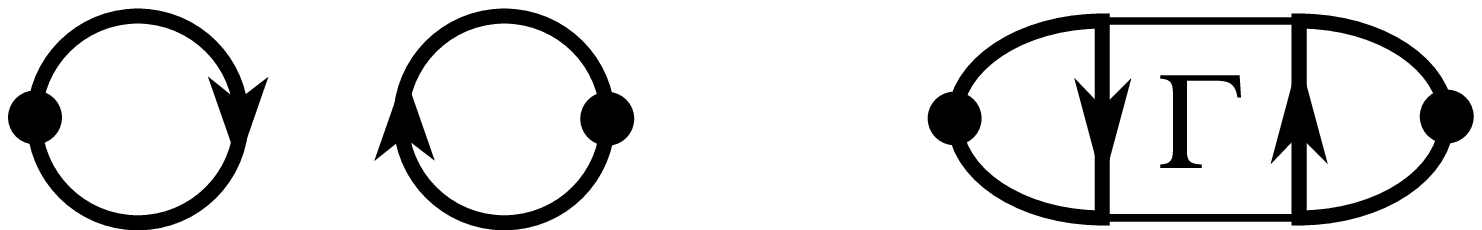}}
\vspace*{13pt}
\fcaption{Hartree-type contributions to the polarization operator.}
\end{figure}
%%%%%%%%%%%%%%%%%%%%%%%%%%%%%%%%%%%%%%%%%%%%%%%%%%%%%%%%%%%%%%%%%%%%%%%%%%%%
We then have
\begin{eqnarray}
&  \displaystyle
   W^{H}[G,D]=
   \frac{T}{2} \sum_{{\rm i}p_s,{\rm i}p_m}
   D_{\gamma\gamma'}(0)
   M^{\gamma}_{\alpha\alpha'} \Bigl\{
   G_{\alpha'\alpha}({\rm i}p_m)G_{\beta\beta'}({\rm i}p_s) +
&
\nonumber \\
&  G_{\alpha'\mu}({\rm i}p_m)G_{\nu\alpha}({\rm i}p_m)
   \Gamma_{\mu\nu;\nu'\mu'}({\rm i}p_m,{\rm i}p_s)
   G_{\beta\mu'}({\rm i}p_s)G_{\nu'\beta'}({\rm i}p_s)
   \Bigr\}
   M^{\gamma'}_{\beta'\beta} , &
\label{eq2.26}
\end{eqnarray}
and the functional $W^{H}[G,D]$ is shown in Fig.~4.
%%%%%%%%%%%%%%%%%%%%%%%%%%%%%%%%%%%%%%%%%%%%%%%%%%%%%%%%%%%%%%%%%%%%%%%%%%%%
\begin{figure}[htbp]
\vspace*{13pt}
\epsfysize=1.0truein
\centerline{\epsfbox{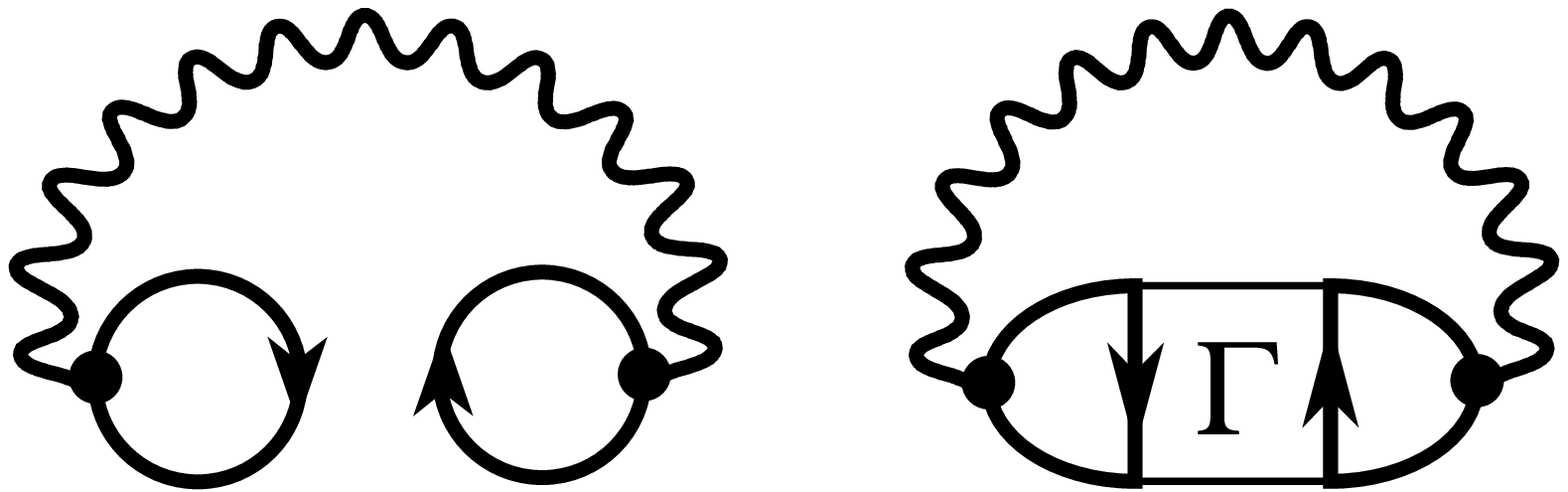}}
\vspace*{13pt}
\fcaption{Hartree-type contributions to the generating functional.}
\end{figure}
%%%%%%%%%%%%%%%%%%%%%%%%%%%%%%%%%%%%%%%%%%%%%%%%%%%%%%%%%%%%%%%%%%%%%%%%%%%%

Putting everything together one obtains for
$W^{e-ph}[G,D]$ the expression
\begin{eqnarray}
& \displaystyle
W^{e-ph}[G,D] =
   \frac{T}{2} \sum_{{\rm i}p_s,{\rm i}p_m}
   D_{\gamma\gamma'}(0)
   M^{\gamma}_{\alpha\alpha'} \Bigl\{
   G_{\alpha'\alpha}({\rm i}p_m)G_{\beta\beta'}({\rm i}p_s) +
&
\nonumber \\
&  G_{\alpha'\mu}({\rm i}p_m)G_{\nu\alpha}({\rm i}p_m)
   \Gamma_{\mu\nu;\nu'\mu'}({\rm i}p_m,{\rm i}p_s)
   G_{\beta\mu'}({\rm i}p_s)G_{\nu'\beta'}({\rm i}p_s)
   \Bigr\}
   M^{\gamma'}_{\beta'\beta}
\nonumber \\
& \displaystyle
   -\frac{T}{2} \sum_{{\rm i}p_s,{\rm i}p_m}
    D_{\gamma\gamma'}({\rm i}p_s - {\rm i}p_m)
    M^{\gamma}_{\alpha\alpha'}G_{\alpha'\beta'}({\rm i}p_m)
    G_{\beta\alpha}({\rm i}p_s)
    M^{\gamma'}_{\beta'\beta}({\rm i}p_m,{\rm i}p_s) &
\label{eq2.27}
\end{eqnarray}
and we have introduced renormalized by the disorder EPI vertex:
\begin{eqnarray}
& M^{\gamma}_{\alpha\alpha'}({\rm i}p_s,{\rm i}p_m) =
  M^{\gamma}_{\alpha\alpha'} + &
\nonumber \\
& +\Gamma_{\alpha\beta;\beta'\alpha'}({\rm i}p_s,{\rm i}p_m)
  G_{\beta\mu}({\rm i}p_s)G_{\mu'\beta'}({\rm i}p_m)
  M^{\gamma}_{\mu\mu'} , &
\label{eq2.28} \\
& M^{\gamma}_{\alpha\alpha'}({\rm i}p_s,{\rm i}p_m) =
  M^{\gamma}_{\alpha\alpha'} + &
\nonumber \\
& U_{\alpha\beta;\beta'\alpha'}({\rm i}p_s,{\rm i}p_m)
  G_{\beta\mu}({\rm i}p_s)G_{\mu'\beta'}({\rm i}p_m)
  M^{\gamma}_{\mu\mu'}({\rm i}p_s,{\rm i}p_m) . &
\label{eq2.29}
\end{eqnarray}
  The use of (\ref{eq2.12}--\ref{eq2.16}) then yields the
following set of coupled equations:
\begin{eqnarray}
&  G_{\alpha\beta}({\rm i}p_s) = G^{0}_{\alpha\beta}({\rm i}p_s) +
   G^{0}_{\alpha\mu}({\rm i}p_s)\Sigma_{\mu\nu}({\rm i}p_s)
   G_{\nu\beta}({\rm i}p_s) , &
\label{eq2.30} \\
&  \left\{G^{-1}_{0}({\rm i}p_s)\right\}_{ij} =
   {\rm i}p_s\tau_{0}\delta_{ij} -
   \left( t_{ij}\tau_{0}-\mu\delta_{ij}
          +\varepsilon_{A}\delta_{ij}\right)\tau_{3} , &
\label{eq2.31} \\
&  \Sigma_{\mu\nu}({\rm i}p_s) =
   \Sigma^{e}_{\mu\nu}({\rm i}p_s) + \Sigma^{e-ph}_{\mu\nu}({\rm i}p_s) , &
\label{eq2.32}
\end{eqnarray}
\begin{equation}
   \Sigma^{e}_{\mu\nu}({\rm i}p_s) =
   \frac{\delta W^{e}[G]}{\delta G_{\nu\mu}({\rm i}p_s)} ,
\label{eq2.33}
\end{equation}
\begin{equation}
  \Sigma^{e-ph}_{\mu\nu}({\rm i}p_s) =
  \Sigma^{H}_{\mu\nu}({\rm i}p_s) +
  \Sigma^{F}_{\mu\nu}({\rm i}p_s) ,
\label{eq2.34}
\end{equation}
\begin{eqnarray}
& \displaystyle
   \Sigma^{F}_{\mu\nu}({\rm i}p_s) =
   -T \sum_{{\rm i}p_m}
   D_{\gamma\gamma'}({\rm i}p_s - {\rm i}p_m) \Bigl\{
   M^{\gamma}_{\mu\alpha}G_{\alpha\beta}({\rm i}p_m)
   M^{\gamma'}_{\beta\nu} +
&
\nonumber \\
& \displaystyle
  +\Gamma_{\mu\mu';\beta\beta'}({\rm i}p_s,{\rm i}p_m)
   G_{\mu'\alpha}({\rm i}p_s) M^{\gamma}_{\alpha\alpha'}
   G_{\alpha\beta}({\rm i}p_m)
   G_{\beta'\nu'}({\rm i}p_m)M^{\gamma'}_{\nu'\nu} +
\nonumber \\
& \displaystyle
  +M^{\gamma}_{\mu\mu'}G_{\mu'\alpha}({\rm i}p_m)
   G_{\alpha'\beta}({\rm i}p_m) M^{\gamma'}_{\beta\beta'}
   G_{\beta'\nu'}({\rm i}p_s)
   \Gamma_{\alpha\alpha';\nu'\nu}({\rm i}p_m,{\rm i}p_s) +
&
\nonumber \\
& \displaystyle
  +\partial
   \Gamma_{\mu\mu';\beta\beta';\nu'\nu}({\rm i}p_s,{\rm i}p_m,{\rm i}p_s)
   \times
&
\nonumber \\
& \displaystyle
   \times
   G_{\mu'\alpha}({\rm i}p_s)
   M^{\gamma}_{\alpha\alpha'}
   G_{\alpha'\beta}({\rm i}p_m)
   G_{\beta'\delta}({\rm i}p_m)
   M^{\gamma'}_{\delta\delta'}
   G_{\delta'\nu'}({\rm i}p_s)
   \Bigr\} ,
&
\label{eq2.35}
\end{eqnarray}
\begin{eqnarray}
& \displaystyle
   \Sigma^{H}_{\mu\nu}({\rm i}p_s) =
   T \sum_{{\rm i}p_m}
   D_{\gamma\gamma'}(0) \Bigl\{
   M^{\gamma}_{\beta\alpha}
   G_{\alpha\beta}({\rm i}p_m)
   M^{\gamma'}_{\mu\nu} +
&
\nonumber \\
& \displaystyle
  +\Gamma_{\mu\mu';\beta'\beta}({\rm i}p_s,{\rm i}p_m)
   G_{\beta\alpha}({\rm i}p_m) M^{\gamma}_{\alpha\alpha'}
   G_{\alpha'\beta'}({\rm i}p_m)
   G_{\mu'\nu'}({\rm i}p_s)M^{\gamma'}_{\nu'\nu} +
&
\nonumber \\
& \displaystyle
  +M^{\gamma}_{\mu\mu'}G_{\mu'\nu'}({\rm i}p_s)
   G_{\alpha\beta}({\rm i}p_m) M^{\gamma'}_{\alpha\alpha'}
   G_{\beta'\alpha'}({\rm i}p_m)
   \Gamma_{\nu'\nu;\beta\beta'}({\rm i}p_s,{\rm i}p_m) +
&
\nonumber \\
& \displaystyle
   +\partial
    \Gamma_{\mu\mu';\beta\beta';\nu'\nu}({\rm i}p_s,{\rm i}p_m,{\rm i}p_s)
   \times
&
\nonumber \\
& \displaystyle
   \times
   G_{\mu'\alpha}({\rm i}p_s)
   M^{\gamma}_{\alpha\alpha'}
   G_{\alpha'\nu'}({\rm i}p_s)
   G_{\beta'\delta}({\rm i}p_m)
   M^{\gamma'}_{\delta\delta'}
   G_{\delta'\beta}({\rm i}p_m)
   \Bigr\} ,
\label{eq2.36}
\end{eqnarray}
where $\partial\Gamma$  denotes the functional derivative
of  four-leg  vertex
\begin{equation}
\partial\Gamma_{\mu\mu';\alpha\beta;\nu'\nu}\\
({\rm i}p_s,{\rm i}p_m,{\rm i}p_s) =
	\frac{
	\delta\Gamma_{\nu'\mu';\alpha\beta}({\rm i}p_s,{\rm i}p_m)
	}{
        \delta G_{\nu\mu}({\rm i}p_s)
	}
\label{eq2.37}
\end{equation}
Note, that in this expression and in similar expressions in what follows
the derivative is rather a partial variational
derivative. A possibility to use partial variational derivatives stems from
the fact that both fermionic loops and ``crossing'' propagation lines are
absent in disorder vertices for the case of quenched disorder. Therefore,
in the right hand side of the expression~(\ref{eq2.37}) the arguments
${\rm i}p_s$ and ${\rm i}p_m$ label two independent functional variables and
should be formally considered as being distinct. Latter on we will exploit
this fact without mentioning explicitly.

  The use of the Bethe-Salpeter equation (\ref{eq2.24}) and definition
(\ref{eq2.37}) gives the equation on $\partial\Gamma$:
\begin{eqnarray}
&  \partial\Gamma_{\mu\mu';\alpha\beta;\nu'\nu}({\rm i}p_s,{\rm i}p_m,{\rm i}p_s) =
   U_{\mu\mu';\alpha\beta;\nu'\nu}({\rm i}p_s,{\rm i}p_m,{\rm i}p_s) +
&
\nonumber \\
& +U_{\mu\mu';\alpha\beta';\xi\nu}({\rm i}p_s,{\rm i}p_m,{\rm i}p_s)
   G_{\beta'\delta}({\rm i}p_m)G_{\xi'\xi}({\rm i}p_s)
   \Gamma_{\delta\beta;\nu'\xi}({\rm i}p_m,{\rm i}p_s) +
&
\nonumber \\
& +U_{\mu\mu';\alpha\delta}({\rm i}p_s,{\rm i}p_m)
   G_{\delta\delta'}({\rm i}p_m)
   \Gamma_{\delta'\beta;\nu'\nu}({\rm i}p_m,{\rm i}p_s) +
&
\nonumber \\
& +U_{\delta'\mu';\alpha\alpha'}({\rm i}p_s,{\rm i}p_m)
   G_{\delta\delta'}({\rm i}p_s)G_{\alpha'\beta'}({\rm i}p_m)
\partial\Gamma_{\mu\delta;\beta'\beta;\nu'\nu}
             ({\rm i}p_s,{\rm i}p_m,{\rm i}p_s) ,
\label{eq2.38}
\end{eqnarray}
and
\begin{equation}
U_{\mu\mu';\alpha\beta;\nu'\nu}({\rm i}p_s,{\rm i}p_m,{\rm i}p_s) =
\delta U_{\nu'\mu';\alpha\beta}({\rm i}p_s,{\rm i}p_m) /
                \delta G_{\nu\mu}({\rm i}p_s)
\label{eq2.39}
\end{equation}
is irreducible six-vertex due to disorder scattering.
  Using Eq.~(\ref{eq2.24}) once again we may express
$\partial\Gamma$  through $\Gamma$, $U_{4}$ and  $U_{6}$
only and after straightforward but lengthy manipulations
we arrive at the result
\begin{eqnarray}
&  \partial\Gamma_{\mu\mu';\alpha\beta;\nu'\nu}
                  ({\rm i}p_s,{\rm i}p_m,{\rm i}p_s) =
   U_{\mu\mu';\alpha\beta;\nu'\nu}({\rm i}p_s,{\rm i}p_m,{\rm i}p_s) +
&
\nonumber \\
& +\Gamma_{\delta'\mu';\alpha\alpha'}({\rm i}p_s,{\rm i}p_m)
   G_{\delta\delta'}({\rm i}p_s)G_{\alpha'\beta'}({\rm i}p_m)
   U_{\mu\delta;\beta'\beta;\nu'\nu}({\rm i}p_s,{\rm i}p_m,{\rm i}p_s) +
&
\nonumber \\
& +U_{\mu\mu';\alpha\beta';\xi\nu}({\rm i}p_s,{\rm i}p_m,{\rm i}p_s)
   G_{\beta'\delta}({\rm i}p_m)G_{\xi'\xi}({\rm i}p_s)
   \Gamma_{\delta\beta;\nu'\xi}({\rm i}p_m,{\rm i}p_s) +
&
\nonumber \\
& +\Gamma_{\delta'\mu';\alpha\alpha'}({\rm i}p_s,{\rm i}p_m)
   G_{\delta\delta'}({\rm i}p_s)G_{\alpha'\beta'}({\rm i}p_m)
   U_{\mu\delta;\beta'\xi;\eta'\nu}({\rm i}p_s,{\rm i}p_m,{\rm i}p_s)
   \times
&
\nonumber \\
&  \times
   G_{\xi\xi'}({\rm i}p_m)G_{\eta\eta'}({\rm i}p_s)
   \Gamma_{\xi'\beta;\nu'\eta}({\rm i}p_m,{\rm i}p_s) +
&
\nonumber \\
& +\Gamma_{\mu\mu';\alpha\delta}({\rm i}p_s,{\rm i}p_m)
   G_{\delta\delta'}({\rm i}p_m)
   \Gamma_{\delta'\beta;\nu'\nu}({\rm i}p_m,{\rm i}p_s) .
&
\label{eq2.40}
\end{eqnarray}
Then substitution of (\ref{eq2.40}) into (\ref{eq2.35},\ref{eq2.36})
with the use of (\ref{eq2.28},\ref{eq2.29}) leads to final expression
for $\Sigma^{e-ph}[G,D]$ (see Figs.~5 and~6)
\begin{equation}
  \Sigma^{e-ph}_{\mu\nu}({\rm i}p_s) =
  \Sigma^{H}_{\mu\nu}({\rm i}p_s) +
  \Sigma^{F}_{\mu\nu}({\rm i}p_s) ,
\label{eq2.41}
\end{equation}
\begin{eqnarray}
& \displaystyle
   \Sigma^{F}_{\mu\nu}({\rm i}p_s) =  -T \sum_{{\rm i}p_m}
   D_{\gamma\gamma'}({\rm i}p_s - {\rm i}p_m)
&
\nonumber \\
& \displaystyle
   \Bigl\{
   M^{\gamma}_{\mu\alpha'}({\rm i}p_s,{\rm i}p_m)
   G_{\alpha'\beta'}({\rm i}p_m)
   M^{\gamma'}_{\beta'\nu}({\rm i}p_m,{\rm i}p_s) +
&
\nonumber \\
& \displaystyle
  +U_{\mu\mu';\delta\xi;\nu'\nu}({\rm i}p_s,{\rm i}p_m,{\rm i}p_s)
   G_{\mu'\alpha}({\rm i}p_s)
   M^{\gamma}_{\alpha\alpha'}({\rm i}p_s,{\rm i}p_m)
   G_{\alpha'\delta}({\rm i}p_m) \times
&
\nonumber \\
&  \times
   G_{\xi\beta}({\rm i}p_m)
   M^{\gamma'}_{\beta\beta'}({\rm i}p_m,{\rm i}p_s)
   G_{\beta'\nu'}({\rm i}p_s)
   \Bigr\}
\label{eq2.42}
\end{eqnarray}
\begin{eqnarray}
& \displaystyle
   \Sigma^{H}_{\mu\nu}({\rm i}p_s) =  T \sum_{{\rm i}p_m}
   D_{\gamma\gamma'}(0)
   \Bigl\{
   M^{\gamma}_{\mu\nu}
   G_{\alpha\beta}({\rm i}p_m)
   M^{\gamma'}_{\beta\alpha} +
&
\nonumber \\
& \displaystyle
 +\Gamma_{mu\mu';\delta'\delta}({\rm i}p_s,{\rm i}p_m)
  G_{\delta\beta}({\rm i}p_m)
  M^{\gamma}_{\beta\beta'}
  G_{\beta'\delta}({\rm i}p_m)
  G_{\mu'\nu'}({\rm i}p_s)
  M^{\gamma'}_{\nu'\nu} +
&
\nonumber \\
& \displaystyle
 +M^{\gamma}_{\mu\mu'}
  G_{\mu'\nu'}({\rm i}p_s)
  G_{\delta\beta}({\rm i}p_m)
  M^{\gamma'}_{\beta\beta'}
  G_{\beta'\delta'}({\rm i}p_m)
 \Gamma_{\nu'\nu;\delta'\delta}({\rm i}p_s,{\rm i}p_m) +
&
\nonumber \\
& \displaystyle
 +\Gamma_{\mu\mu';\delta\delta'}({\rm i}p_s,{\rm i}p_m)
  \Gamma_{\nu'\nu;\alpha'\alpha}({\rm i}p_s,{\rm i}p_m)
  \times
&
\nonumber \\
& \displaystyle
  \times
  G_{\mu'\xi}({\rm i}p_s)
  M^{\gamma}_{\xi\xi'}
  G_{\delta'\alpha'}({\rm i}p_m)
  G_{\epsilon\delta}({\rm i}p_m)
  G_{\alpha\epsilon'}({\rm i}p_m)
  M^{\gamma'}_{\epsilon'\epsilon}
  G_{\xi'\nu'}({\rm i}p_s) +
&
\nonumber \\
& \displaystyle
 +U_{\mu\mu';\alpha\beta;\nu'\nu}({\rm i}p_s,{\rm i}p_m,{\rm i}p_s)
  \times
&
\nonumber \\
& \displaystyle
  \times
  G_{\mu'\delta}({\rm i}p_s)
  M^{\gamma}_{\delta\delta'}
  G_{\delta'\nu'}({\rm i}p_s)
  G_{\beta\beta'}({\rm i}p_m)
  M^{\gamma'}_{\beta'\alpha'}
  G_{\alpha'\alpha}({\rm i}p_m) +
&
\nonumber \\
& \displaystyle
 +U_{\mu\xi;\epsilon\beta;\nu'\nu}({\rm i}p_s,{\rm i}p_m,{\rm i}p_s)
  G_{\xi\xi'}({\rm i}p_s)
  G_{\epsilon'\epsilon}({\rm i}p_m)
  \Gamma_{\xi'\mu';\alpha\epsilon'}({\rm i}p_s,{\rm i}p_m)
  \times
&
\nonumber \\
& \displaystyle
  \times
  G_{\mu'\delta}({\rm i}p_s)
  M^{\gamma}_{\delta\delta'}
  G_{\delta'\nu'}({\rm i}p_s)
  G_{\beta\beta'}({\rm i}p_m)
  M^{\gamma'}_{\beta'\alpha'}
  G_{\alpha'\alpha}({\rm i}p_m) +
&
\nonumber \\
& \displaystyle
 +U_{\mu\mu';\alpha\epsilon;\xi\nu}({\rm i}p_s,{\rm i}p_m,{\rm i}p_s)
  G_{\xi'\xi}({\rm i}p_s)
  G_{\epsilon\epsilon'}({\rm i}p_m)
  \Gamma_{\nu'\xi';\epsilon'\beta}({\rm i}p_s,{\rm i}p_m)
  \times
&
\nonumber \\
& \displaystyle
  \times
  G_{\mu'\delta}({\rm i}p_s)
  M^{\gamma}_{\delta\delta'}
  G_{\delta'\nu'}({\rm i}p_s)
  G_{\beta\beta'}({\rm i}p_m)
  M^{\gamma'}_{\beta'\alpha'}
  G_{\alpha'\alpha}({\rm i}p_m) +
&
\nonumber \\
& \displaystyle
 +U_{\mu\xi;\epsilon'\zeta';\rho\nu}({\rm i}p_s,{\rm i}p_m,{\rm i}p_s)
  G_{\xi\xi'}({\rm i}p_s)
  G_{\epsilon\epsilon'}({\rm i}p_m)
  \Gamma_{\xi'\mu';\alpha\epsilon}({\rm i}p_s,{\rm i}p_m)
  \times
&
\nonumber \\
& \displaystyle
  \times
  \Gamma_{\xi'\mu';\alpha\epsilon'}({\rm i}p_s,{\rm i}p_m)
  G_{\rho'\rho}({\rm i}p_s)
  G_{\zeta'\zeta}({\rm i}p_m)
  \times
&
\nonumber \\
& \displaystyle
  \times
  G_{\mu'\delta}({\rm i}p_s)
  M^{\gamma}_{\delta\delta'}
  G_{\delta'\nu'}({\rm i}p_s)
  G_{\beta\beta'}({\rm i}p_m)
  M^{\gamma'}_{\beta'\alpha'}
  G_{\alpha'\alpha}({\rm i}p_m) +
  \Bigr\} .
&
\label{eq2.43}
\end{eqnarray}
%%%%%%%%%%%%%%%%%%%%%%%%%%%%%%%%%%%%%%%%%%%%%%%%%%%%%%%%%%%%%%%%%%%%%%%%%%%%
\begin{figure}[htbp]
\vspace*{13pt}
\epsfysize=1.5truein
\centerline{\epsfbox{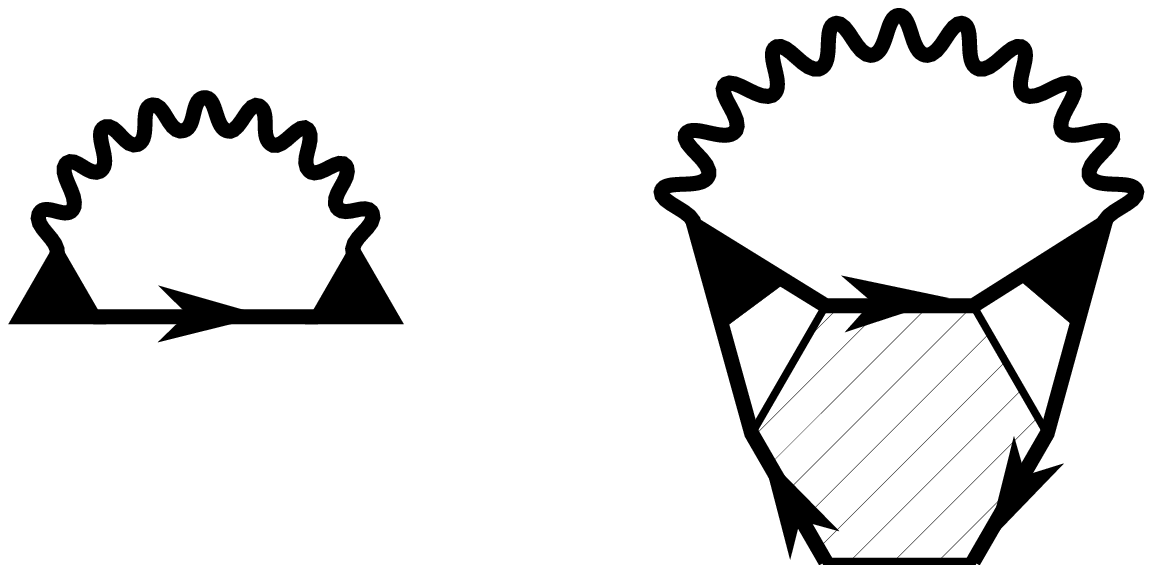}}
\vspace*{13pt}
\fcaption{Fock-type contributions to the
electron-phonon self-energy. Black triangle denotes the renormalized
electron-phonon three-leg vertex; dashed hexagon, the irreducible
disorder six-leg vertex. The other graphical elements are the same as
in previous Figures.}
\end{figure}
%%%%%%%%%%%%%%%%%%%%%%%%%%%%%%%%%%%%%%%%%%%%%%%%%%%%%%%%%%%%%%%%%%%%%%%%%%%%
%%%%%%%%%%%%%%%%%%%%%%%%%%%%%%%%%%%%%%%%%%%%%%%%%%%%%%%%%%%%%%%%%%%%%%%%%%%%
\begin{figure}[htbp]
\vspace*{13pt}
\epsfysize=6.0truein
\centerline{\epsfbox{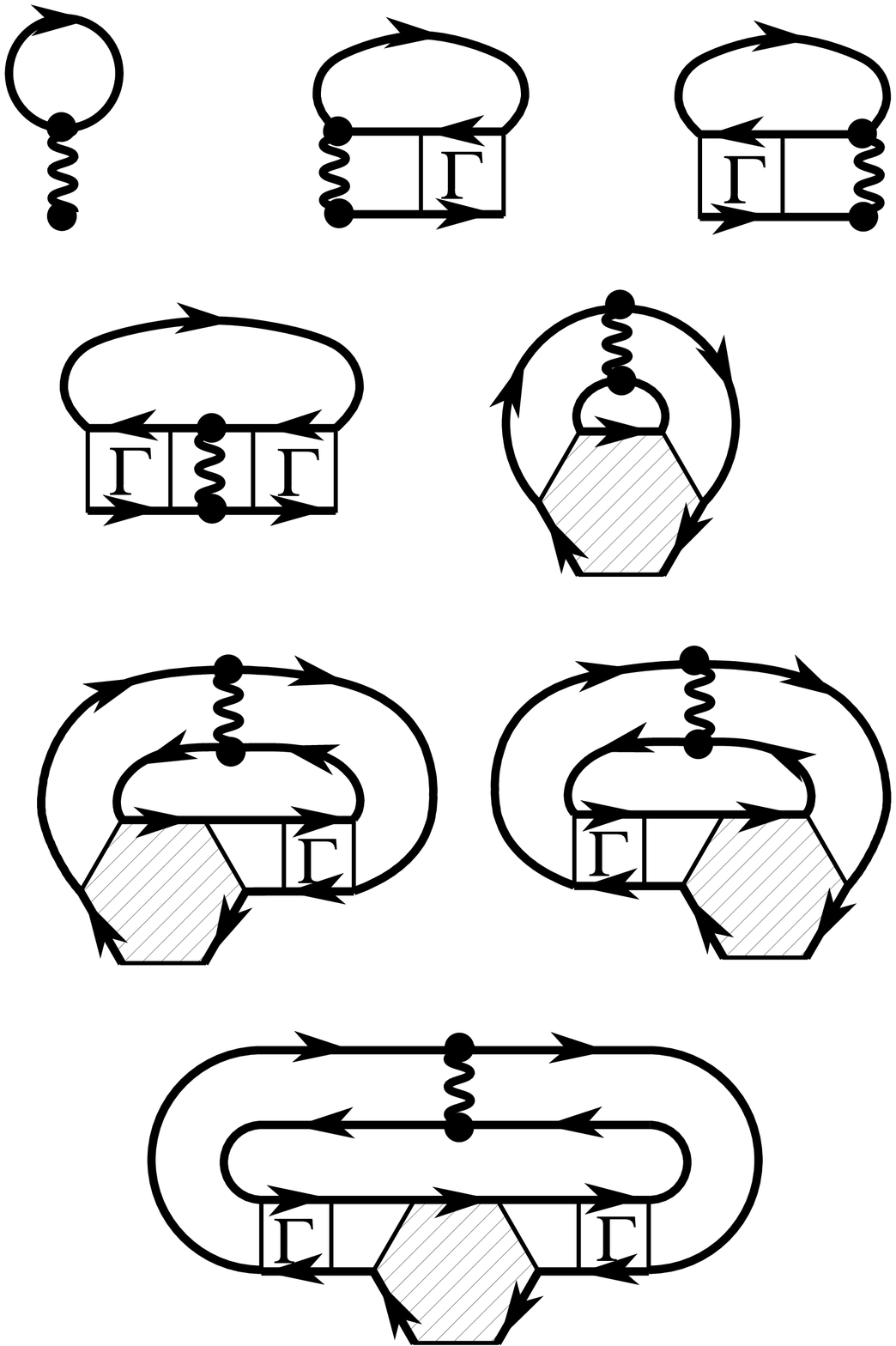}}
\vspace*{13pt}
\fcaption{Hartree-type contributions to the electron-phonon self-energy.}
\end{figure}
%%%%%%%%%%%%%%%%%%%%%%%%%%%%%%%%%%%%%%%%%%%%%%%%%%%%%%%%%%%%%%%%%%%%%%%%%
  The formulas (\ref{eq2.30}--\ref{eq2.33})
and (\ref{eq2.41}--\ref{eq2.43}) are
those  which  constitute
closed  set  of  self-consistent  equations  in  the  case  of  EPI
renormalizations  taken  into   account   consistently   with   the
approximate $\Sigma^{e}[G]$ given.

  To obtain  the  equations  for  the  superconductive  transition
temperature $T_{c}$ one should linearize the equation for $\Sigma[G,D]$
in anomalous part which is assumed to be small near $T_{c}$.  This  is
formally achieved by expanding $\Sigma [G]$ in a functional series
in  $\delta G$,
$\delta G$ being anomalous part of $G$, and then using  Dyson's
equation  to
express $\delta G$ back through $\delta\Sigma$, where $\delta\Sigma$
is the anomalous  contribution to $\Sigma[G,D]$. We have
\begin{eqnarray}
&  \displaystyle
   \Sigma[G_{n}+\delta G,D] = \Sigma[G_{n},D]
         + {\left.\frac{\delta\Sigma[G,D]}{\delta G}\right|}_{G=G_{n}}
           \delta G =
&
\nonumber \\
&  \displaystyle
   \Sigma[G_{n},D]
         +{\left.\frac{\delta\Sigma[G,D]}{\delta G}\right|}_{G=G_{n}}
         G_{n}\delta\Sigma G_{n} =
         \Sigma_{n} +\delta\Sigma
\label{eq2.44}
\end{eqnarray}
(subscript ``n'' means that the corresponding quantity is taken in the
normal state of superconductor) and equating contributions of order
$O(\delta\Sigma)$
in left- and right-hand sides of Eq.~(\ref{eq2.44})  we  obtain
formal equation
\begin{equation}
         \delta\Sigma   =
         {\left.\frac{\delta\Sigma[G,D]}{\delta G}\right|}_{G=G_{n}}
         G_{n} \delta\Sigma G_{n},
\label{eq2.45}
\end{equation}
and $T_{c}$ may be found from solubility condition  to  this  equations.
Note, that the solubility condition may be written as  equality  of
some formal determinant  to  zero,  but  we  omit  explicit  formal
expression of such  a  form.  The  diagrammatic  representation  of
(\ref{eq2.45}) in the case under discussion contains ninty seven
terms and is too cumbersome to be written down here.

  Hence, normal part of $\Sigma[G,D]$ near $T_{c}$ is still determined by
Eqs.~(\ref{eq2.32},\ref{eq2.33}) and (\ref{eq2.41}--\ref{eq2.43})
with one-particle  Green  function  $G$
taken in  normal  phase  of  superconductor,  and  we  have  formal
equation (\ref{eq2.45}), from which the expression
for $T_{c}$  may  be  derived
after suitably fixing approximation for $\Sigma^{e}[G]$.

  In the conclusion of this Section some remarks are to the point.

   Obviously, the approximation to $\Sigma[G,D]$  given  by
(\ref{eq2.32},\ref{eq2.33},\ref{eq2.41}--\ref{eq2.43}) is
conserving in a whole  by  construction.

   The expression for $W^{e-ph}[G,D]$ (\ref{eq2.27}) does not
actually depend on
particular alloy model since the assumptions on the alloy model
enter into this expression only implicitly through irreducible
four-leg vertex, so that the form of $W^{e-ph}[G,D]$ is independent
of the disorder type. We will demonstrate the correctness of the
statement above in subsequent Section for the substitutional disorder
of a general type carrying out configurational averaging in a formally
exact way.

   The equations
Eqs.~(\ref{eq2.32},\ref{eq2.33}) and (\ref{eq2.41}--\ref{eq2.43})
enable also  to  consider
``inconsistent'' renormalizations of EPI vertices. In  this  case  we
assume  that  $\Sigma^{e}[G]$ is calculated within certain conserving
approximation and renormalizations of EPI are taken into account
within another, but still conserving approximation for $\Sigma^{e}[G]$.
This  happens, for instance,
when  discussing  the  so-called  Anderson's
theorem where one assumes that disorder does  not  influence  the
EPI at all. The last, in turn, implies that the  approximation  for
$\Sigma^{e}[G]$ within which we renormalize EPI
vertices  is  $\Sigma^{e}[G]=0$,  and
therefore irreducible four-leg disorder vertex and of course
higher vertices vanishes identically.

\section{T--Matrix Approach to the Superconductivity
of Disordered Alloys}
\noindent
  In this Section we derive the equations~(\ref{eq2.30}--\ref{eq2.33})
  and~(\ref{eq2.41}--\ref{eq2.43})
for  the  case  of  formally   exact   configurational   (disorder)
averaging. Making use of multiple scattering theory, or  so  called
$T$--matrix approach,\cite{c20,c37} we establish relations
with the approach of the previous Section and demonstrate that  the
Eqs.~(\ref{eq2.30}--\ref{eq2.33}) and~(\ref{eq2.41}--\ref{eq2.43})
are actually the case for general  type
of substitutional disorder when the disorder averaging  is  carried
out exactly. However, if  the  matter is a specific  approximation
within certain model of substitutional alloy, it is more convenient
to formulate the approximation in terms of  generating  functional,
so that the approaches of Section~2 and  of  this
Section  complement each other.

  For the first time the application of $T$--matrix approach  to  the
problem of alloy superconductivity has been given in Refs.\cite{c6,c7,c10}
where the model  of  a  binary  alloy  was  considered  with  local
(single-site) BCS-type interaction and, hence,  with  purely  local
superconductive pairing. In  these  papers  the  equation  for  the
self-energy was derived within the framework of Coherent  Potential
Approximation (CPA) and  resulting  BCS  coupling  renormalized  by
disorder was presented in Ref.\cite{c7}.

  The results of this Section are a strightforward  generalization
of the approach of Refs.\cite{c6,c7,c8,c9,c10}
to the case of non-local
superconductive pairing owing to EPI with bare EPI  vertices  being
independent of disorder.

  For particular alloy configuration, consider one-particle Matsubara
Green's function ${\bf g}(\tau)$ with matrix elements
\begin{equation}
   {\left\{{\bf g}(\tau)\right\}}_{\alpha\beta} =
   -{<{\rm T}_{\tau}{\rm C}_{\alpha}(\tau){\rm C}^{\dag}_{\beta}(0)>}_{H} ,
\label{eq3.1}
\end{equation}
where $<\ldots>_{H}$ means usual Gibbs averaging, with $H$ being the
Hamiltonian (\ref{eq2.1}).

  After transforming to Matsubara frequencies and taking EPI  into
account one has for ${\bf g}^{-1}({\rm i}p_s)$ the expression
\begin{equation}
   {\bf g}^{-1}({\rm i}p_s)= {\rm i}p_s \tau_{0} -
           {\bf H}\tau_{3} -
           {\bf V}^{e-ph}({\rm i}p_s) ,
\label{eq3.2}
\end{equation}
with ${\bf H}$ being the matrix with elements (\ref{eq2.3});
${\bf V}^{e-ph}({\rm i}p_s)$ being the effective
interaction due  to  EPI  in  this  disorder  configuration,
possessing functional dependence on ${\bf g}({\rm i}p_s)$.

  To derive equations  required  it  is  sufficient  to  know  the
functional ${\bf V}^{e-ph}({\rm i}p_s)$ to the lowest  order
in  EPI  of  the  self-consistent
perturbation theory. To this order the expression for
${\bf V}^{e-ph}({\rm i}p_s)$ has the form
\begin{equation}
    {\bf V}^{e-ph}({\rm i}p_s) =
             {\bf V}^{H}({\rm i}p_s) +
             {\bf V}^{F}({\rm i}p_s)  ,
\label{eq3.3}
\end{equation}
\begin{equation}
   {\bf V}^{H}_{\alpha\beta}({\rm i}p_s) =
   T \sum_{{\rm i}p_m}
   D_{\gamma\gamma'}(0)
   M^{\gamma}_{\alpha\beta}
   g_{\alpha'\beta'}({\rm i}p_m)
   M^{\gamma'}_{\beta'\alpha'} ,
\label{eq3.4}
\end{equation}
\begin{equation}
   {\bf V}^{F}_{\alpha\beta}({\rm i}p_s) =
   -T \sum_{{\rm i}p_m}
   D_{\gamma\gamma'}({\rm i}p_s - {\rm i}p_m)
   M^{\gamma}_{\alpha\alpha'}g_{\alpha'\beta'}({\rm i}p_m)
   M^{\gamma'}_{\beta'\beta},
\label{eq3.5}
\end{equation}
where ${\bf V}^{H}({\rm i}p_s)$ and
${\bf V}^{F}({\rm i}p_s)$ are Hartree and Fock
contributions to the effective potential respectively.

  For the configurational averaging to be carried out we introduce
$T$-matrix. In the usual manner one has\cite{c20,c37}
\begin{eqnarray}
&  {\bf T}({\rm i}p_s) =
      \left({\bf V}({\rm i}p_s) - {\bf \Sigma}({\rm i}p_s) \right) +
      \left({\bf V}({\rm i}p_s) - {\bf \Sigma} ({\rm i}p_s) \right)
       {\bf G}({\rm i}p_s) {\bf T}({\rm i}p_s), &
\label{eq3.6} \\
&  {\bf T}({\rm i}p_s) =
      \left({\bf V}({\rm i}p_s) - {\bf \Sigma}({\rm i}p_s)\right) +
       {\bf T}({\rm i}p_s) {\bf G}({\rm i}p_s)
      \left({\bf V}({\rm i}p_s) - {\bf \Sigma}({\rm i}p_s)\right) , &
\label{eq3.7}
\end{eqnarray}
where
\begin{equation}
   {\bf V}({\rm i}p_s) = {\bf V}^{e}
    + {\bf V}^{e-ph}({\rm i}p_s),
\label{eq3.8}
\end{equation}
${\bf V}^{e}$ is random contribution to Hamiltonian (\ref{eq2.2}),
depending on disorder configuration, and
\begin{eqnarray}
  {\bf g}({\rm i}p_s) & = & {\bf G}({\rm i}p_s) +
      {\bf G}({\rm i}p_s){\bf T}({\rm i}p_s)
      {\bf G}({\rm i}p_s) ,
\label{eq3.9} \\
  <{\bf g}({\rm i}p_s)> & = & {\bf G}({\rm i}p_s) +
     {\bf G}({\rm i}p_s) <{\bf T}({\rm i}p_s)>
     {\bf G}({\rm i}p_s) ,
\label{eq3.10}
\end{eqnarray}
\begin{equation}
   {\bf G}({\rm i}p_s) = {\bf G}_{0}({\rm i}p_s) +
        {\bf G}_{0}({\rm i}p_s) {\bf \Sigma}({\rm i}p_s)
        {\bf G}({\rm i}p_s) .
\label{eq3.11}
\end{equation}
Here $<\ldots>$ stands for  configurational  averaging  with  particular
probability distribution and ${\bf G}_{0}({\rm i}p_s)$ is
completely  determined  by
non-random part of ${\bf H}$ (\ref{eq2.2}).

  For self-consistency ${\bf\Sigma}({\rm i}p_s)$ is fixed by the condition
\begin{equation}
   <{\bf T}({\rm i}p_s)>  = 0,
\label{eq3.12}
\end{equation}
and, hence, from (\ref{eq3.9},\ref{eq3.10}) and (\ref{eq3.12}) it follows
\begin{equation}
   <{\bf g}({\rm i}p_s)>  = {\bf G}({\rm i}p_s) .
\label{eq3.13}
\end{equation}

  To  carry  out  configurational  averaging  we  must   use   two
apparently different schemes of averaging: one for the  alloy  part
of the problem and another for the effective  scattering  owing  to
EPI (the latter scheme is reminiscent of Virtual Crystal Approximation
(VCA) for ordinary ``alloy'' problems). To this end we decompose
${\bf\Sigma}({\rm i}p_s)$ according to natural
partitioning of ${\bf V}({\rm i}p_s)$ into two
parts owing to disorder scattering and EPI respectively
\begin{equation}
   {\bf\Sigma}({\rm i}p_s) = {\bf\Sigma}^{e}({\rm i}p_s) +
          {\bf\Sigma}^{e-ph}({\rm i}p_s) ,
\label{eq3.14}
\end{equation}
and then rewrite Eqs.~(\ref{eq3.6},\ref{eq3.7}) for $T$-matrix
in the following form
\begin{eqnarray}
  {\bf T}({\rm i}p_s) & = & {\bf T}_{1}({\rm i}p_s) +
        {\bf T}_{2}({\rm i}p_s) ,
\label{eq3.15} \\
  {\bf T}_{1}({\rm i}p_s) & = & {\bf V}_{1}({\rm i}p_s)+
        {\bf V}_{1}({\rm i}p_s){\bf G}({\rm i}p_s)
        {\bf T}({\rm i}p_s)  ,
\label{eq3.16} \\
  {\bf T}_{2}({\rm i}p_s) & = & {\bf V}_{2}({\rm i}p_s)+
        {\bf V}_{2}({\rm i}p_s){\bf G}({\rm i}p_s)
        {\bf T}({\rm i}p_s)  ,
\label{eq3.17}
\end{eqnarray}
where
\begin{eqnarray}
  {\bf V}_{1}({\rm i}p_s) & = & {\bf V}^{e} -
        {\bf \Sigma}^{e}({\rm i}p_s) ,
\label{eq3.18} \\
  {\bf V}_{2}({\rm i}p_s) & = & {\bf V}^{e-ph}({\rm i}p_s) -
        {\bf \Sigma}^{e-ph}({\rm i}p_s) .
\label{eq3.19}
\end{eqnarray}
After introducing partial $T$-matrices ${\bf T}^{e}$
and ${\bf T}^{e-ph}$ by
\begin{eqnarray}
&  {\bf T}^{e}({\rm i}p_s) = {\bf V}_{1}({\rm i}p_s)+
        {\bf V}_{1}({\rm i}p_s){\bf G}({\rm i}p_s)
        {\bf T}^{e}({\rm i}p_s) ,  &
\label{eq3.20} \\
&  {\bf T}^{e-ph}({\rm i}p_s) = {\bf V}_{2}({\rm i}p_s)+
        {\bf V}_{2}({\rm i}p_s){\bf G}({\rm i}p_s)
        {\bf T}^{e-ph}({\rm i}p_s) ,  &
\label{eq3.21}
\end{eqnarray}
equations~(\ref{eq3.16},\ref{eq3.17}) for ${\bf T}_{1}$
and ${\bf T}_{2}$ transform as
\begin{eqnarray}
&  {\bf T}_{1}({\rm i}p_s) = {\bf T}^{e}({\rm i}p_s)+
        {\bf T}^{e}({\rm i}p_s){\bf G}({\rm i}p_s)
        {\bf T}_{2}({\rm i}p_s) ,  &
\label{eq3.22} \\
&  {\bf T}_{2}({\rm i}p_s) = {\bf T}^{e-ph}({\rm i}p_s)+
        {\bf T}^{e-ph}({\rm i}p_s){\bf G}({\rm i}p_s)
        {\bf T}_{1}({\rm i}p_s) . &
\label{eq3.23}
\end{eqnarray}
Bearing in mind the lowest order  in  EPI  let  us  find  the  full
$T$-matrix (\ref{eq3.6},\ref{eq3.7},\ref{eq3.15}) up to the first
order in ${\bf V}_{2}$.  In  this  case Eq.~(\ref{eq3.21}) becomes
\begin{equation}
  {\bf T}^{e-ph}({\rm i}p_s) = {\bf V}_{2}({\rm i}p_s) .
\label{eq3.24}
\end{equation}
  To this order in ${\bf V}_{2}$ the result
for the full $T$--matrix reads
\begin{eqnarray}
&  {\bf T}({\rm i}p_s) = {\bf T}^{e}({\rm i}p_s) +
        {\bf V}_{2}({\rm i}p_s) +
        {\bf V}_{2}({\rm i}p_s){\bf G}({\rm i}p_s)
        {\bf T}^{e}({\rm i}p_s)  &
\nonumber \\
&       + {\bf T}^{e}({\rm i}p_s){\bf G}({\rm i}p_s)
        {\bf V}_{2}({\rm i}p_s) +
        {\bf T}^{e}({\rm i}p_s){\bf G}({\rm i}p_s)
        {\bf V}_{2}({\rm i}p_s)
        {\bf G}({\rm i}p_s){\bf T}^{e}({\rm i}p_s) . &
\label{eq3.25}
\end{eqnarray}
Fixing up to now arbitrary quantity $\Sigma^{e}({\rm i}p_s)$
by the condition
\begin{equation}
   <{\bf T}^{e}({\rm i}p_s)> = 0 ,
\label{eq3.26}
\end{equation}
we obtain from (\ref{eq3.25}) the equation
for ${\bf\Sigma}^{e-ph}({\rm i}p_s)$:
\begin{eqnarray}
& {\bf \Sigma}^{e-ph}({\rm i}p_s)=
    <{\bf V}^{e-ph}({\rm i}p_s)>+  &
\nonumber \\
&
    <{\bf V}^{e-ph}({\rm i}p_s)
    {\bf G}({\rm i}p_s){\bf T}^{e}({\rm i}p_s)>
   + <{\bf T}^{e}({\rm i}p_s){\bf G}({\rm i}p_s)
    {\bf V}^{e-ph}({\rm i}p_s)> + &
\nonumber \\
&
  + <{\bf T}^{e}({\rm i}p_s){\bf G}({\rm i}p_s)
    {\bf V}_{2}({\rm i}p_s)
    {\bf G}({\rm i}p_s){\bf T}^{e}({\rm i}p_s)> , &
\label{eq3.27}
\end{eqnarray}
where the use was made of Eqs.~(\ref{eq3.12},\ref{eq3.19})
and~(\ref{eq3.26}).

  Introducing in a standard fashion\cite{c20,c37}
full four- and six-leg vertices
\begin{eqnarray}
  \Gamma_{\mu\mu';\nu\nu'}({\rm i}p_s,{\rm i}p_m) & = &
   <T^{e}_{\mu\mu'}({\rm i}p_s)T^{e}_{\nu\nu'}({\rm i}p_m)>,
\label{eq3.28}  \\
  \Gamma_{\alpha\alpha';\beta\beta';\gamma\gamma'}
   ({\rm i}p_s,{\rm i}p_m,{\rm i}p_s) & = &
   <T^{e}_{\alpha\alpha'}({\rm i}p_s)T^{e}_{\beta\beta'}({\rm i}p_m)
    T^{e}_{\gamma\gamma'}({\rm i}p_s)>,
\label{eq3.29}
\end{eqnarray}
writing down indices (the summation over  doubly  repeated
Greek indices is implicitly meant here and in  what  follows) and then
substituting (\ref{eq3.4},\ref{eq3.5}) and~(\ref{eq3.9},\ref{eq3.10})
into (\ref{eq3.27}) we obtain
\begin{eqnarray}
&
  \Sigma^{e-ph}_{\mu\nu}({\rm i}p_s)
   +\Gamma_{\mu\mu';\nu'\nu}({\rm i}p_s)({\rm i}p_s)
   G_{\mu'\alpha}({\rm i}p_s)
   \Sigma^{e-ph}_{\alpha\beta}({\rm i}p_s)
   G_{\beta\nu}({\rm i}p_s) =
\nonumber \\
&
 =\Sigma^{(1)}_{\mu\nu}({\rm i}p_s) +
  \Sigma^{(2)}_{\mu\nu}({\rm i}p_s) ,
\label{eq3.30}
\end{eqnarray}
where
\begin{eqnarray}
& \displaystyle
   \Sigma^{(1)}_{\mu\nu}({\rm i}p_s) =
   -T \sum_{{\rm i}p_m}
   D_{\gamma\gamma'}({\rm i}p_s - {\rm i}p_m) \Bigl\{
   M^{\gamma}_{\mu\alpha}G_{\alpha\beta}({\rm i}p_m)
   M^{\gamma'}_{\beta\nu} +
&
\nonumber \\
& \displaystyle
   \Gamma_{\mu\mu';\beta\beta'}({\rm i}p_s,{\rm i}p_m)
   G_{\mu'\alpha}({\rm i}p_s) M^{\gamma}_{\alpha\alpha'}
   G_{\alpha'\beta}({\rm i}p_m)
   G_{\beta'\nu'}({\rm i}p_m)M^{\gamma'}_{\nu'\nu} +
&
\nonumber \\
& \displaystyle
   M^{\gamma}_{\mu\mu'}G_{\mu'\alpha}({\rm i}p_m)
   G_{\alpha'\beta}({\rm i}p_m) M^{\gamma'}_{\beta\beta'}
   G_{\beta'\nu'}({\rm i}p_s)
   \Gamma_{\alpha\alpha';\nu'\nu}({\rm i}p_s,{\rm i}p_m) +
&
\nonumber \\
& \displaystyle
  +\Gamma_{\mu\mu';\beta\beta';\nu'\nu}({\rm i}p_s,{\rm i}p_m,{\rm i}p_s)
   \times
&
\nonumber \\
& \displaystyle
   \times
   G_{\mu'\alpha}({\rm i}p_s)M^{\gamma}_{\alpha\alpha'}
   G_{\alpha'\beta}({\rm i}p_m)
   G_{\beta'\delta}({\rm i}p_m)M^{\gamma'}_{\delta\delta'}
   G_{\delta'\nu'}({\rm i}p_s)
   \Bigr\}
&
\label{eq3.31}
\end{eqnarray}
and
\begin{eqnarray}
& \displaystyle
   \Sigma^{(2)}_{\mu\nu}({\rm i}p_s) =
   T \sum_{{\rm i}p_m}
   D_{\gamma\gamma'}(0) \Bigl\{
   M^{\gamma}_{\beta\alpha}
   G_{\alpha\beta}({\rm i}p_m)
   M^{\gamma'}_{\mu\nu} +
&
\nonumber \\
& \displaystyle
   \Gamma_{\mu\mu';\beta'\beta}({\rm i}p_s,{\rm i}p_m)
   G_{\beta\alpha}({\rm i}p_m) M^{\gamma}_{\alpha\alpha'}
   G_{\alpha'\beta'}({\rm i}p_m)
   G_{\mu'\nu'}({\rm i}p_s)M^{\gamma'}_{\nu'\nu} + &
\nonumber \\
& \displaystyle
   M^{\gamma}_{\mu\mu'}G_{\mu'\nu'}({\rm i}p_s)
   G_{\alpha\beta}({\rm i}p_m) M^{\gamma'}_{\alpha\alpha'}
   G_{\beta'\alpha'}({\rm i}p_m)
   \Gamma_{\nu'\nu;\beta\beta'}({\rm i}p_s,{\rm i}p_m) +
&
\nonumber \\
& \displaystyle
  +\Gamma_{\mu\mu';\beta\beta';\nu'\nu}({\rm i}p_s,{\rm i}p_m,{\rm i}p_s)
   \times
&
\nonumber \\
& \displaystyle
   \times
   G_{\mu'\alpha}({\rm i}p_s)
   M^{\gamma}_{\alpha\alpha'}
   G_{\alpha'\nu'}({\rm i}p_s)
   G_{\beta'\delta}({\rm i}p_m)
   M^{\gamma'}_{\delta\delta'}
   G_{\delta'\beta}({\rm i}p_m)
   \Bigr\} .
&
\label{eq3.32}
\end{eqnarray}

  The equations~(\ref{eq3.30}--\ref{eq3.32}) are essentially
a generalization of corresponding equations of Refs.\cite{c6,c7,c10}
to the case with EPI--induced
superconductive pairing and when configurational  averaging
is carried out exactly,  and  they  may  be  used  to  discuss  the
phenomenon of the superconductivity of alloys  on  the  grounds  of
certain approximate solution  for  ${\bf T}^{e}$ (\ref{eq3.20})
and  ${\bf\Sigma}^{e}({\rm i}p_s)$ (\ref{eq3.26}).
In particular, for the case of diagonal alloy disorder the replacing
${\bf T}^{e}$ with $t$ in (\ref{eq3.30}--\ref{eq3.32}),
where $t$ is the single-site $T$--matrix,
\begin{equation}
  t({\rm i}p_s)=\left(\varepsilon_{i}\tau_{3}-
                  \Sigma^{e}({\rm i}p_s)\right) +
                  \left(\varepsilon_{i}\tau_{3}-
                  \Sigma^{e}({\rm i}p_s)\right)G_{ii}({\rm i}p_s)
                  t({\rm i}p_s),
\label{eq3.33}
\end{equation}
retaining single-site EPI contributions only and using CPA
expression for  $\Sigma^{e}$,  one may obtain the equations
of Refs.\cite{c6,c7} with the exception that superconductive
pairing is due to EPI, not BCS interaction.

  However we prefer the equations for the self-energy
${\bf\Sigma}({\rm i}p_s)$ in the form, which does
not rely upon $T$--matrix and gives a possibility to analyze
arbitrary conserving single-site
approximation  for ${\bf\Sigma}({\rm i}p_s)$
within an unified scheme. Therefore we rederive the expressions
(\ref{eq2.41}--\ref{eq2.43}) for ${\bf\Sigma}^{e-ph}({\rm i}p_s)$
from (\ref{eq3.30}--\ref{eq3.32}) to demonstrate that the two
expressions are in fact equivalent in the case of exact
configurational averaging.

  In a standard way, the irreducible disorder four-leg vertex, $U_{4}$,
is  introduced  by  the  Bethe-Salpeter  equation\cite{c20,c37}
\begin{eqnarray}
&  \Gamma_{\mu\mu';\nu'\nu}({\rm i}p_s,{\rm i}p_m) =
    U_{\mu\mu';\nu'\nu}({\rm i}p_s,{\rm i}p_m)   &
\nonumber \\
&  + U_{\mu\alpha;\beta\nu}({\rm i}p_s,{\rm i}p_m)
     G_{\alpha\alpha'}({\rm i}p_s)G_{\beta'\beta}({\rm i}p_m)
    \Gamma_{\alpha'\mu';nu'\beta'}({\rm i}p_s,{\rm i}p_m) &
\label{eq3.34}
\end{eqnarray}
(note the inverse order of introducing of full and irreducible vertices
in this and in the previous Sections). Multiplying (\ref{eq3.30})
by  the  combination  ${\bf G}{\bf U}_{4}{\bf G}$
(where indices  must  be  identified  explicitly)  we  can  express
$<{\bf T}^{e}{\bf G}{\bf\Sigma}^{e-ph}{\bf G}{\bf T}^{e}>$ in the
left-hand side of (\ref{eq3.30}) through ${\bf U}_{4}$, ${\bf\Gamma}_{4}$,
${\bf\Gamma}_{6}$ and ${\bf G}$.
Substituting the result obtained back into
(\ref{eq3.30}) and collecting terms we then arrive
to the result which reads
\begin{equation}
  \Sigma^{e-ph}_{\mu\nu}({\rm i}p_s) =
  \Sigma^{H}_{\mu\nu}({\rm i}p_s) +
  \Sigma^{F}_{\mu\nu}({\rm i}p_s) ,
\label{eq3.35}
\end{equation}
\begin{eqnarray}
& \displaystyle
   \Sigma^{F}_{\mu\nu}({\rm i}p_s) =
   -T \sum_{{\rm i}p_m}
   D_{\gamma\gamma'}({\rm i}p_s - {\rm i}p_m) \Bigl\{
   M^{\gamma}_{\mu\alpha}G_{\alpha\beta}({\rm i}p_m)
   M^{\gamma'}_{\beta\nu} +
&
\nonumber \\
& \displaystyle
   \Gamma_{\mu\mu';\beta\beta'}({\rm i}p_s,{\rm i}p_m)
   G_{\mu'\alpha}({\rm i}p_s) M^{\gamma}_{\alpha\alpha'}
   G_{\alpha\beta}({\rm i}p_m)
   G_{\beta'\nu'}({\rm i}p_m)M^{\gamma'}_{\nu'\nu} +
\nonumber \\
& \displaystyle
   M^{\gamma}_{\mu\mu'}G_{\mu'\alpha}({\rm i}p_m)
   G_{\alpha'\beta}({\rm i}p_m) M^{\gamma'}_{\beta\beta'}
   G_{\beta'\nu'}({\rm i}p_s)
   \Gamma_{\alpha\alpha';\nu'\nu}({\rm i}p_m,{\rm i}p_s) +
&
\nonumber \\
& \displaystyle
  +\Theta_{\mu\mu';\beta\beta';\nu'\nu}({\rm i}p_s,{\rm i}p_m,{\rm i}p_s)
   \times
&
\nonumber \\
& \displaystyle
   \times
   G_{\mu'\alpha}({\rm i}p_s)
   M^{\gamma}_{\alpha\alpha'}
   G_{\alpha'\beta}({\rm i}p_m)
   G_{\beta'\delta}({\rm i}p_m)
   M^{\gamma'}_{\delta\delta'}
   G_{\delta'\nu'}({\rm i}p_s)
   \Bigr\}
&
\label{eq3.36}
\end{eqnarray}
\begin{eqnarray}
& \displaystyle
   \Sigma^{H}_{\mu\nu}({\rm i}p_s) =
   T \sum_{{\rm i}p_m}
   D_{\gamma\gamma'}(0) \Bigl\{
   M^{\gamma}_{\beta\alpha}
   G_{\alpha\beta}({\rm i}p_m)
   M^{\gamma'}_{\mu\nu} +
&
\nonumber \\
& \displaystyle
   \Gamma_{\mu\mu';\beta'\beta}({\rm i}p_s,{\rm i}p_m)
   G_{\beta\alpha}({\rm i}p_m) M^{\gamma}_{\alpha\alpha'}
   G_{\alpha'\beta'}({\rm i}p_m)
   G_{\mu'\nu'}({\rm i}p_s)M^{\gamma'}_{\nu'\nu} +
&
\nonumber \\
&  \displaystyle
   M^{\gamma}_{\mu\mu'}G_{\mu'\nu'}({\rm i}p_s)
   G_{\alpha\beta}({\rm i}p_m) M^{\gamma'}_{\alpha\alpha'}
   G_{\beta'\alpha'}({\rm i}p_m)
   \Gamma_{\nu'\nu;\beta\beta'}({\rm i}p_s,{\rm i}p_m) +
&
\nonumber \\
& \displaystyle
   +\Theta_{\mu\mu';\beta\beta';\nu'\nu}({\rm i}p_s,{\rm i}p_m,{\rm i}p_s)
   \times
&
\nonumber \\
& \displaystyle
   \times
   G_{\mu'\alpha}({\rm i}p_s)
   M^{\gamma}_{\alpha\alpha'}
   G_{\alpha'\nu'}({\rm i}p_s)
   G_{\beta'\delta}({\rm i}p_m)
   M^{\gamma'}_{\delta\delta'}
   G_{\delta'\beta}({\rm i}p_m)
   \Bigr\} ,
\label{eq3.37}
\end{eqnarray}
and
\begin{eqnarray}
&  \Theta_{\mu\mu';\delta\xi;\nu'\nu}({\rm i}p_s,{\rm i}p_m,{\rm i}p_s)=
   \Gamma_{\mu\mu';\delta\xi;\nu'\nu}
              ({\rm i}p_s,{\rm i}p_m,{\rm i}p_s) - &
\nonumber \\
&  U_{\mu\xi;\nu'\nu}({\rm i}p_s,{\rm i}p_s)
   G_{\xi\xi'}({\rm i}p_s)
   \Gamma_{\xi'\mu';\delta\delta'}({\rm i}p_s,{\rm i}p_m) - &
\nonumber \\
&  U_{\mu\mu';\xi'\nu}({\rm i}p_s,{\rm i}p_s)
   G_{\xi\xi'}({\rm i}p_s)
   \Gamma_{\delta\delta';\nu'\xi}({\rm i}p_m,{\rm i}p_s) - &
\nonumber \\
&  U_{\mu\xi;\kappa\nu}({\rm i}p_s,{\rm i}p_s)
   G_{\xi\xi'}({\rm i}p_s)
   G_{\kappa\kappa'}({\rm i}p_s)
   \Gamma_{\xi'\mu';\delta\delta';\nu\kappa}
                ({\rm i}p_s,{\rm i}p_m,{\rm i}p_s) . &
\label{eq3.38}
\end{eqnarray}
Here the expression (\ref{eq3.38}) is nothing but
a variational derivative  $\partial{\bf\Gamma}$
of ${\bf\Gamma}_{4}$ (\ref{eq3.28}) with respect to $G$.
Indeed, the variation of the $T$--matrix (\ref{eq3.20})
with respect to ${\bf G}$ gives
\begin{eqnarray}
&  \delta{\bf T}^{e}/\delta{\bf G} =
   - {\bf T}^{e} \delta\left({{\bf T}^{e}}^{-1}\right)
   /\delta{\bf G}{\bf T}^{e}  &
\nonumber \\
&  = -{\bf T}^{e} \left\{
              \delta\left(
                    \left( {\bf V}^{e}-{\bf\Sigma}^{e}\right)^{-1} - {\bf G}
                   \right) / \delta{\bf G}
                 \right\}{\bf T}^{e} &
\nonumber \\
&  ={\bf T}^{e} \left\{
        - \left( {\bf V}^{e}-{\bf\Sigma}^{e} \right)^{-1}
         \delta {\bf\Sigma}^{e}/\delta {\bf G}
         \left( {\bf V}^{e}-{\bf\Sigma}^{e} \right)^{-1}
         + \delta {\bf G}/\delta {\bf G}
          \right\}{\bf T}^{e} &
\nonumber \\
&  = - \left(1+{\bf T}^{e} {\bf G} \right)
       \delta {\bf\Sigma}^{e}/\delta {\bf G}
       \left(1+ {\bf G} {\bf T}^{e} \right) +
       {\bf T}^{e}
       \delta {\bf G}/\delta {\bf G}
       {\bf T}^{e} , &
\label{eq3.39}
\end{eqnarray}
and
\begin{equation}
   \frac{\delta G_{\alpha\beta}}{\delta G_{\gamma\delta}} =
   \delta_{\alpha\gamma}\delta_{\beta\delta} .
\label{eq3.40}
\end{equation}
Introducing
\begin{equation}
   \frac{
        \delta \Sigma^{e}_{\alpha\beta}({\rm i}p_s)
        }{\delta G_{\gamma\delta}({\rm i}p_s)
        } =
        U_{\alpha\gamma;\delta\beta}({\rm i}p_s,{\rm i}p_s) ,
\label{eq3.41}
\end{equation}
explicitly writing down indices in (\ref{eq3.39}) and then substituting
(\ref{eq3.39}) into the following expression
\begin{equation}
   \partial\Gamma_{\mu\mu';\delta\delta';\nu'\nu}
                  ({\rm i}p_s,{\rm i}p_m,{\rm i}p_s) =
      <\frac{
            \delta T^{e}_{\nu'\mu'}({\rm i}p_s)
            }{
            \delta G_{\nu\mu}({\rm i}p_s)
            }
            T^{e}_{\delta\delta'}({\rm i}p_m)> ,
\label{eq3.42}
\end{equation}
one obtains
\begin{eqnarray}
&  \partial\Gamma_{\mu\mu';\delta\xi;\nu'\nu}
                     ({\rm i}p_s,{\rm i}p_m,{\rm i}p_s) =
   \Gamma_{\mu\mu';\delta\xi;\nu'\nu}
                     ({\rm i}p_s,{\rm i}p_m,{\rm i}p_s) -
\nonumber \\
&  U_{\mu\xi;\nu'\nu}({\rm i}p_s,{\rm i}p_s)
   G_{\xi\xi'}({\rm i}p_s)
   \Gamma_{\xi'\mu';\delta\delta'}({\rm i}p_s,{\rm i}p_m) - &
\nonumber \\
&  U_{\mu\mu';\xi'\nu}({\rm i}p_s,{\rm i}p_s)
   G_{\xi\xi'}({\rm i}p_s)
   \Gamma_{\delta\delta';\nu'\xi}({\rm i}p_m,{\rm i}p_s) - &
\nonumber \\
&  U_{\mu\xi;\kappa\nu}({\rm i}p_s,{\rm i}p_s)
   G_{\xi\xi'}({\rm i}p_s)
   G_{\kappa\kappa'}({\rm i}p_s)
   \Gamma_{\xi'\mu';\delta\delta';\nu\kappa}
                     ({\rm i}p_s,{\rm i}p_m,{\rm i}p_s) , &
\label{eq3.43}
\end{eqnarray}
where we have used~(\ref{eq3.41}) and
definitions~(\ref{eq3.28}) and~(\ref{eq3.29}). The
four-leg vertex (\ref{eq3.41}) is completely determined by the
Bethe-Salpeter equation and, as shown in Appendix~A, the definition
(\ref{eq3.41}) is in fact the Ward's identity.

  Comparing Eqs.~(\ref{eq3.43}) and~(\ref{eq3.38}) one sees that
they coincide with each other and, consequently,
(\ref{eq3.35}--\ref{eq3.37}) and
(\ref{eq2.34}--\ref{eq2.36}) are identical.
In Appendix~A we will demonstrate that Eq.(\ref{eq3.26})
determining the self-energy ${\bf\Sigma}^{e}({\rm i}p_s)$
can be represented in the form~(\ref{eq2.33}).
Simple repetition of the transformations of the
previous Section (Eqs.~(\ref{eq2.34}--\ref{eq2.43})) then
leads to Eqs.~(\ref{eq2.41}--\ref{eq2.43}).
This completes the demonstration that two sets of
equations (\ref{eq2.33},\ref{eq2.41}--\ref{eq2.43})
and~(\ref{eq3.26},\ref{eq3.30}--\ref{eq3.32})
are equivalent for the case of exact configurational averaging.

  Note that the case of exact configurational  averaging  we
dealt with in this Section may be considered as an extreme case  of
conserving approximation in the sense that $\Sigma^{e}[G]$ coincides with the
exact value, and the condition of $\phi$-derivability of the
approximation\cite{c36}
\begin{equation}
   \frac{
         \delta \Sigma^{e}_{\alpha\beta}({\rm i}p_s)
         }{
         \delta G_{\gamma\delta}({\rm i}p_s)
         } =
   \frac{
         \delta \Sigma^{e}_{\delta\gamma}({\rm i}p_s)
         }{
         \delta G_{\beta\alpha}({\rm i}p_s)
         } ,
\label{eq3.44}
\end{equation}
can be readly verified (see Appendix~A). Therefore, as for
conserving approximations, formal results which will be obtained in
the rest of the paper are valid for this case too, and we will  not
discuss it separately in the following.

  Concluding this Section let us stress some important points.

  In contrast to Refs.\cite{c6,c7,c8,c9,c10},
where site-diagonal  contributions  to
the  effective  scattering  potential  (which  come  from  BCS-like
interaction) have been treated on an equal  footing  with  disorder
contributions, that is within the same approximation (in fact CPA),
we rather apply the same Virtual Crystal type  averaging  procedure
to both site-diagonal and site-off-diagonal  EPI  contributions  to
${\bf V}({\rm i}p_s)$. The decomposition
of $\Sigma[G,D]$, Eq.~(\ref{eq3.14}),  and,  hence,  of
the full $T$--matrix, Eqs.~(\ref{eq3.15}--\ref{eq3.17}),
gives a  possibility tocarry out
two different averaging procedures.  The very
content of the Migdal-Eliashberg theory, based on the addiabaticity
theorem,\cite{c44} dictates such the way of partitioning  the  self-energy
$\Sigma[G,D]$ into the parts owing to disorder
scattering and EPI, as well
as it dictates the virtual crystal type averaging procedure for the
EPI contributions. Then the Eliashberg-type equations  acquire  the
character of effective field equations with  two  effective  fields
$\Sigma^{e}[G]$ and $\Sigma^{e-ph}[G,D]$
coupled via Dyson's equation
for the one-particle electron Green's function $G$.

  Suppose for a moment that we  have  decomposed  the  full
self-energy into,
for example, three parts owing to disorder scattering,
site-diagonal    and    site-off-diagonal    EPI     contributions,
respectively. Then, if we use CPA-like, or  even  exact,  averaging
scheme for site-diagonal EPI contributions, this will give  results
which are formally beyond the  accuracy  of  the  Migdal-Eliashberg
theory as they contain,  besides  the  lowest  order  contribution,
higher order EPI contributions to $\Sigma^{e-ph}[G,D]$.

  Another important feature is that
the  decomposition  (\ref{eq3.8})  of
${\bf V}({\rm i}p_s)$ and $\Sigma[G,D]$, Eq.~(\ref{eq3.14}),
is the only possible for  both  the
Ward's identity (\ref{eq3.41}) and $\phi$-derivability
criterion (\ref{eq3.44})  for
$\Sigma^{e}[G]$ to hold. Rather different approaches of this  Section
and of Section~2 are then in tight analogy with each other: the form
of the equations are the same and detailed structure of  underlying
quantities  may  be  established  using  the  connections   between
disorder vertices and the $T$--matrix.

\section{Ward's Identities and the Anderson's Theorem}
\noindent
  As a first application of the approach being developed  in  this
paper we consider the so-called Anderson's theorem  which  concerns
the  influence  of  nonmagnetic  disorder  on  the  superconductive
transition temperature $T_{c}$.

  This theorem\cite{c1,c2,c4} was established within  BCS  model
and Virtual Crystal  Approximation  (VCA)  for  dilute  nonmagnetic
impurities\cite{c3} and  for  concentrated  nonmagnetic  impurities
taken into account within Coherent Potential Approximation (CPA).\cite{c47}
For the latter case  the  theorem  was  discussed  on  more
general grounds.\cite{c7}

  For isotropic superconductor where pairing  owing  to  EPI is
primarily of $s$-character there is a standard consideration in the
literature\cite{c4}  of  how  dilute nonmagnetic impurities
(essentially the case of weak alloy disordering) influence $T_{c}$.

  For the case where disorder scattering strength  is  arbitrary,
which was treated by  using  Average  $T$-matrix  Approximation  (the
concentration of the impurities is small none the less), and in the
absence of EPI renormalizations  this  question  was  discussed
in connection with one-particle  state  density  variations
near the Fermi  energy.\cite{c48} It was demonstrated that, while usual
cancellation of anomalous disorder-scattering contributions holds
in the equation for the gap function, the asymmetry of the
normal-phase one-particle state density near the chemical potential leads
to non-zero (in fact, non-constant) spectral shifts. In this  case,
equations  for  the  gap  function  and  for  the  spectral  weight
renormalization function $Z({\rm i}p_s)$
no longer decouple which results in the
disorder  contributions to  the  gap   function   via   explicit
dependence of the corresponding equation on $Z({\rm i}p_s)$.
Then,  as  was
concluded,\cite{c48} in the presence of  any
structure  of  the  density  of
states (DOS) near the Fermi level, the  Anderson's  theorem  breaks
down, which seems to be in contrast with the situation for the  BCS
model. This point needs further clarification.

Further we shall see that the account of the Fermi-surface energy
dependence and its electron-hole symmetry is not directly related
to the Anderson's theorem. It is strong ${\bf k}$--dependence
of the superconducting order parameter
($\Delta_{\bf k}({\rm i}p_s)$ is not constant over the Fermi surface)
only which may lead to the suppression of superconductivity by
non-magnetic impurities in the case of weak disorder. However
from the poit of view of the standard considerations\cite{c4}
this situation does not correspond to $s$-pairing (basis function
in the expansion in partial waves  for an anisotropic
Fermi surface  is unity for $s$-channel!). So, the term
``$s$-pairing'' will be understand hereafter as follows:
the gap function $\Delta_{\bf k}$ has the same point symmetry
as the ``bare''electronic spectrum $\epsilon_{\bf k}$, that is
$\Delta_{\bf k}=\Delta(\epsilon_{\bf k})$. One may expect that this holds
true for all conventional superconductors.

  Within the Bogolyubov's formulation  of  the  BCS-like  model,
that is for the model with  disorder-independent  attractive
four-fermionic interaction which is
almost of infinite range and  has  a
separable kernel, and after adding disorder-de\-pend\-ent two-fermionic
contributions, the Anderson's theorem may be understand basing on
qualitative considerations. It states that the  expression  for $T_{c}$
has usual BCS form with  the  normal  state  one-particle  spectral
density renormalized by the disorder, replacing bare  one.  Indeed,
by  analogy with the case where diorder is absent,\cite{c49}
one can construct an approximate
Hamiltonian of effective-field type, which gives asymptotically
exact (in the thermodynamical limit) solutions for the
thermodynamical quantities and for correlation functions  as  well.
For fixed disorder configuration the effective field (in fact,  the
superconductive gap) which enters the approximate  Hamiltonian,  is
spatially homogeneous and  can  be  determined  through  the
self-consistency conditions containing anomalous averages
of fermionic  operators  forming  Cooper  pairs  in  this  disorder
configuration. The anomalous averages are then expressed in terms
of the one-particle spectral density which is non-negative  and  is
the same for both spin-up and spin-down electron states because, as
was pointed out by Anderson, the states  of  electrons  forming
Cooper pair are time-reversal images  of  each  other.  Hence,  the
effective field  may  depend  on  disorder  via  the  corresponding
dependence of the spectral density, but the ``disorder'' fluctuations
of the effective field are strongly suppressed in the
thermodynamical limit because of infinite range of the
interaction, so that the effective field can be safely replaced  by
its configurationally averaged value in the self-consistency
equation. Residual disorder averaging touches only the spectral
density; the self-consistency equations become similar to those  of
usual BCS theory; however they contain  the  renormalized  spectral
density instead of bare one. The statement of  the  Anderson's
theorem then follows. Note, that for this particular model the
conclusion takes place irrespectively of the strength and type of
quenched substitutional disorder and also of whether renormalized
and/or bare state densities possess any structure.

  The situation for the model~(\ref{eq2.1}) is somewhat different.
In this case effective EPI-induced electron-electron interaction is rather
of short or intermediate range than of (infinitely) long  range,
``disorder'' fluctuations of the electron-phonon part of the
self-energy are not small in general and lead  to  renormalizations
of ``bare'' EPI  vertices.  But if one neglects completely these
renormalizations the statement may be established  which is quite
analogous to the Anderson's theorem for the BCS-like model.

  In this Section we demonstrate that, for  superconductive  alloy
with i\-so\-tropic $s$-type  pairing  owing  to  EPI  and  for  arbitrary
conserving approximation for the disorder self-energy $\Sigma^{e}[G]$,
quenched substitutional disorder influences superconductive transition
temperature $T_{c}$ only through normal-state one-particle  spectral
density  renormalized  by  all  the  interactions  in  the  system,
provided disorder does not change ``bare'' EPI vertices. Surely, this
makes the theorem to be rather a formal statement because the
effects of EPI-vertex renormalizations, as we will see, are not
small except for the case of weak alloy disordering (dilute
alloys). Note, also, the Anderson's theorem is not a  statement  of
mathematical regour but holds to certain  approximations  usually
accepted in conventional theory of superconductivity.

  The most general form for spatial Fourier transforms of
${\bf\Sigma}^{e}({\rm i}p_s)$ and ${\bf\Sigma}^{e-ph}({\rm i}p_s)$
is\cite{c4,c5}
\begin{eqnarray}
   \Sigma^{e}_{\bf k}({\rm i}p_s) & = &
       {\rm i}p_s[1-\gamma_{\bf k}({\rm i}p_s)] \tau_{0} +
       \chi^{e}_{\bf k}({\rm i}p_s) \tau_{3} +
       \phi^{e}_{\bf k}({\rm i}p_s) \tau_{1} ,
\label{eq4.1} \\
   \Sigma^{e-ph}_{\bf k}({\rm i}p_s) & = &
       {\rm i}p_s \gamma_{\bf k}({\rm i}p_s)
       [1 - Z_{\bf k}({\rm i}p_s)] \tau_{0} +
       \chi^{e-ph}_{\bf k}({\rm i}p_s) \tau_{3} +
       \phi^{e-ph}_{\bf k}({\rm i}p_s) \tau_{1} ,
\label{eq4.2}
\end{eqnarray}
where only $\tau_{1}$ contribution to the anomalous part is retained (in
the  absence  of  external  magnetic  fields  we  can  always  make
$\tau_{2}$-contributions vanish by suitable gauge fixing because usual
gauge invariance under phase transformations is preserved within
$\phi$-derivable approximations) and
\begin{equation}
   \gamma_{\bf k}({\rm i}p_s) = 1 -
         \frac{
               \Sigma^{e}_{{\bf k}\uparrow}({\rm i}p_s) -
               \Sigma^{e}_{-{\bf k}\downarrow}(-{\rm i}p_s)
              }{
               2 {\rm i}p_s
              } .
\label{eq4.3}
\end{equation}
  The representation~(\ref{eq4.2}) for
$\Sigma^{e-ph}_{\bf k}({\rm i}p_s)$ differs from the standard one\cite{c4}
in that the disorder contribution have been picked  out  explicitly
into the factor $\gamma_{\bf k}({\rm i}p_s)$
(which is nothing but  the  disorder  self-energy correction).

%%%%%%%%%%%%%%%%%%%%%%%%%%%%%%%%%%%%%%%%%%%%%%%%%%%%%%%%%%%%%%%%%%%%%%%%%%%%%
\begin{figure}[htbp]
\vspace*{13pt}
\epsfysize=0.7truein
\centerline{\epsfbox{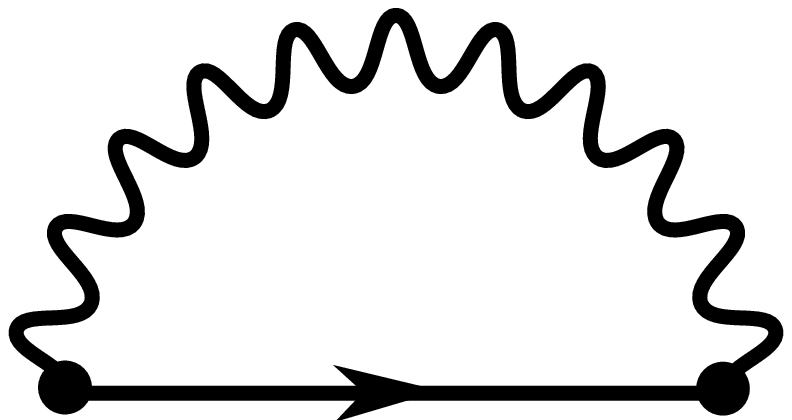}}
\vspace*{13pt}
\fcaption{Standard contribution to the normal
electron-phonon self-energy without disorder renormalizations.}
\end{figure}
%%%%%%%%%%%%%%%%%%%%%%%%%%%%%%%%%%%%%%%%%%%%%%%%%%%%%%%%%%%%%%%%%%%%%%%%%%%%%
%%%%%%%%%%%%%%%%%%%%%%%%%%%%%%%%%%%%%%%%%%%%%%%%%%%%%%%%%%%%%%%%%%%%%%%%%%%%%
\begin{figure}[htbp]
\vspace*{13pt}
\epsfysize=1.0truein
\centerline{\epsfbox{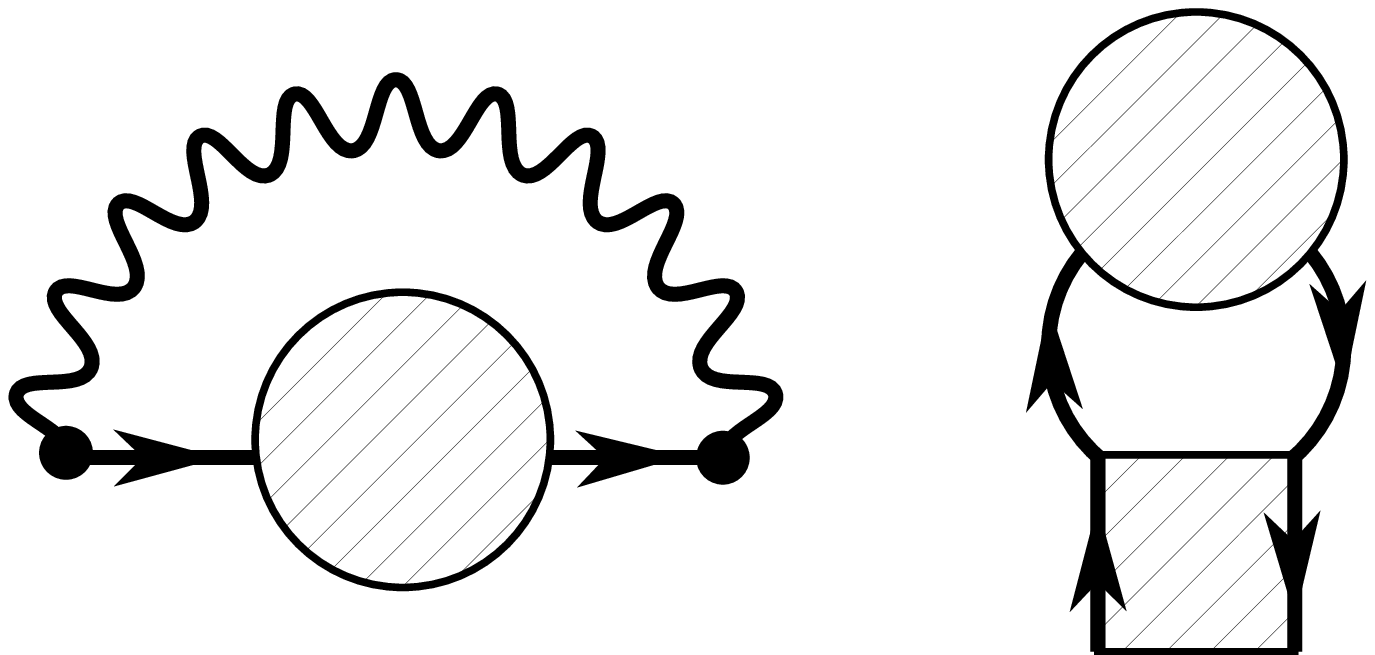}}
\vspace*{13pt}
\fcaption{Full contribution to the anomalous
self-energy. Dashed circle represents the anomalous self-energy;
dashed square, the irreducible disorder four-leg vertex. All the other
graphical elements are the same as before.}
\end{figure}
%%%%%%%%%%%%%%%%%%%%%%%%%%%%%%%%%%%%%%%%%%%%%%%%%%%%%%%%%%%%%%%%%%%%%%%%%%%%%
  Linearized Eliashberg-type equations without disorder
renormalization of EPI read (see Figs.~7 and~8 for the normal and anomalous
contributions to the self-energy)
\begin{equation}
   {\rm i}p_s \gamma_{\bf k}({\rm i}p_s)
                 \left[1 - Z_{\bf k}({\rm i}p_s)\right] =
        - T_{c} \sum_{{\bf q},{\rm i}p_m}
          \lambda_{\bf kq}({\rm i}p_s - {\rm i}p_m)
          \Psi_{\bf q}({\rm i}p_m)
          {\rm i}p_s \gamma_{\bf q}({\rm i}p_m)
          Z_{\bf q}({\rm i}p_m) ,
\label{eq4.4}
\end{equation}
\begin{equation}
   \chi^{e-ph}_{\bf k}({\rm i}p_s) =
        - T_{c} \sum_{{\bf q},{\rm i}p_m}
          \lambda_{\bf kq}({\rm i}p_s - {\rm i}p_m)
          \Psi_{\bf q}({\rm i}p_m)
          \varepsilon_{\bf q}({\rm i}p_m) ,
\label{eq4.5}
\end{equation}
\begin{eqnarray}
& \displaystyle
   \phi_{\bf k}({\rm i}p_s) =
        T_{c} \sum_{{\bf q},{\rm i}p_m}
        \lambda_{\bf kq}({\rm i}p_s - {\rm i}p_m)
          \Psi_{\bf q}({\rm i}p_m)
          \phi_{\bf q}({\rm i}p_m) + &
\nonumber \\
& \displaystyle
        + \sum_{\bf q} U^{\uparrow\downarrow}_{\bf kq}({\rm i}p_s)
          \Psi_{\bf q}({\rm i}p_s)
          \phi_{\bf q}({\rm i}p_s) , &
\label{eq4.6}
\end{eqnarray}
where
\begin{equation}
\phi_{\bf k}({\rm i}p_s)=
\phi^{e}_{\bf k}({\rm i}p_s)+\phi^{e-ph}_{\bf k}({\rm i}p_s)
\label{eq4.7}
\end{equation}
is the full anomalous contribution to $\Sigma_{\bf k}({\rm i}p_s)$,
$U^{\uparrow\downarrow}_{\bf kq}({\rm i}p_s)$
is proper irreducible four-leg vertex
related to anomalous contribution to $\Sigma^{e}({\rm i}p_s)$, and
\begin{equation}
   \Psi_{\bf k}({\rm i}p_s) =
      \left[
      \left( p_s\gamma_{\bf k}({\rm i}p_s)
             Z_{\bf k}({\rm i}p_s) \right)^{2} +
      \left( \varepsilon_{\bf k}({\rm i}p_s) \right)^{2}
      \right]^{-1} ,
\label{eq4.8}
\end{equation}
\begin{equation}
   \varepsilon_{\bf k}({\rm i}p_s) = t_{\bf k} +
          \chi^{e}_{\bf k}({\rm i}p_s)     +
          \chi^{e-ph}_{\bf k}({\rm i}p_s) ,
\label{eq4.9}
\end{equation}
with $t_{\bf k}$ being bare electron dispersion defined by non-random part of
$H^{e}$, Eqs.(\ref{eq2.2}, \ref{eq2.8}).
Quantities $\gamma^{e}_{\bf k}({\rm i}p_s)$ and
$\chi^{e}_{\bf k}({\rm i}p_s)$
describe  disorder scattering and satisfy formal equation:
\begin{equation}
   {\rm i}p_s[1-\gamma_{\bf k}({\rm i}p_s)] \tau_{0} +
   \chi^{e}_{\bf k}({\rm i}p_s) \tau_{3} =
      \left.
      \frac{
            \delta W^{e}[G]
           }{
            \delta G_{\bf k}({\rm i}p_s)
           }
      \right|_{T=T_{c}}
\label{eq4.10}
\end{equation}
Explicit form of this equation depends on the  approximation  used
for $W^{e}[G]$ or $\Sigma^{e}[G]$ and is not important
for further consideration. Note that we have also included Hartree
contributions to the renormalization of the chemical potential.

  At last, the expression for
$\lambda_{\bf kq}({\rm i}\omega_n)$ reads
\begin{equation}
   \lambda_{\bf kq}({\rm i}\omega_n) =
           \int_{0}^{\infty} d\Omega \alpha^{2}F_{\bf kq}(\Omega)
              \frac{
                    \Omega
                   }{
                    \omega^{2}_{n} + \Omega^{2}
                   } .
\label{eq4.11}
\end{equation}
Here $\alpha^{2}F_{\bf kq}(\Omega )$ coincides
with the usual Eliashberg function up to  a
factor of dimension energy and has the form
\begin{equation}
   \alpha^{2}F_{\bf kq}(\Omega) =
      \sum_{s} \left| M_{\bf kq}^{(s)} \right|^{2}
               \left\{
                 - \frac{1}{\pi}\,
                 {\rm Im} D^{(s)}_{{\bf k}-{\bf q}}(\Omega + {\rm i}0)
               \right\} .
\label{eq4.12}
\end{equation}

  Setting in (\ref{eqA.10}) of Appendix~A
\begin{eqnarray}
   G'=G_{ij,\uparrow}({\rm i}p_s), \quad
   G =G_{ji,\downarrow}(-{\rm i}p_s) ,
\label{eq4.13} \\
   \Sigma'=\Sigma^{e}_{ij \uparrow}({\rm i}p_s), \quad
   \Sigma =\Sigma^{e}_{ji \downarrow}(-{\rm i}p_s),
\label{eq4.14}
\end{eqnarray}
where all quantities are taken in the normal state,
and Fourier transforming one then obtains
\begin{equation}
   \Sigma^{e}_{{\bf k}\uparrow}({\rm i}p_s) -
     \Sigma^{e}_{-{\bf k}\downarrow}(-{\rm i}p_s) =
          \sum_{\bf q} U^{\uparrow\downarrow}_{\bf kq}({\rm i}p_s)
               \left( G_{{\bf q}\uparrow}({\rm i}p_s) -
                      G_{-{\bf q}\downarrow}(-{\rm i}p_s) \right),
\label{eq4.15}
\end{equation}
or, with account of the definition (\ref{eq4.3})
for $\gamma_{\bf k}({\rm i}p_s)$,
\begin{equation}
   \gamma_{\bf k}({\rm i}p_s) = 1 +
          \sum_{\bf q} U^{\uparrow\downarrow}_{\bf kq}({\rm i}p_s)
          \Psi_{\bf q}({\rm i}p_s)
          Z_{\bf q}({\rm i}p_s) \gamma_{\bf q}({\rm i}p_s) ,
\label{eq4.16}
\end{equation}
where the use was made  of
Eqs.~(\ref{eq4.1},\ref{eq4.2},\ref{eq4.8},\ref{eq4.9})  and  of  the
expression
\begin{equation}
   G^{-1}_{{\bf k}\sigma} = {\rm i}p_s - t_{\bf k} -
          \Sigma^{e}_{{\bf k}\sigma}({\rm i}p_s) -
          \Sigma^{e-ph}_{{\bf k}\sigma}({\rm i}p_s) ,
\label{eq4.17}
\end{equation}
and it is essential that the irreducible  four-leg  vertices  which
enter Eqs.~(\ref{eq4.6}) and (\ref{eq4.16}) are the same.

  Introducing in a standard fashion normal-state one-particle
spectral density
\begin{equation}
   A_{{\bf k}\sigma}(\epsilon)=
        -\frac{1}{\pi} {\rm Im} G_{{\bf k}\sigma}(\epsilon + {\rm i}0)
\label{eq4.18}
\end{equation}
and noting that
\begin{equation}
   A_{{\bf k}\uparrow}(\epsilon) =
          A_{-{\bf k}\downarrow}(\epsilon)
\label{eq4.19}
\end{equation}
(this is, in fact, a consequence of time reversal symmetry  of  the
problem), we can write
\begin{equation}
   \gamma_{\bf k}({\rm i}p_s)
    Z_{\bf k}({\rm i}p_s) \Psi_{\bf k}({\rm i}p_s) =
          \overline{\Psi}_{\bf k}({\rm i}p_s) ,
\label{eq4.20}
\end{equation}
\begin{equation}
   \overline{\Psi}_{\bf k}({\rm i}p_s) =
        \int d\epsilon \frac{
                              A_{\bf k}(\epsilon)
                            }{
                             p^{2}_{s} + \epsilon^{2}
                            }
\label{eq4.21}
\end{equation}
and
\begin{equation}
  \overline{Y}_{\bf k}({\rm i}p_s)=
  \varepsilon_{\bf k}({\rm i}p_s)\overline{\Psi}_{\bf k}({\rm i}p_s)=
        \int d\epsilon \frac{
                              \epsilon A_{\bf k}(\epsilon)
                            }{
                             p^{2}_{s} + \epsilon^{2}
                            } .
\label{eq4.22}
\end{equation}
Here we fix the zero of the energy self-consistently  at  the  true
chemical potential.

The system of the equations (\ref{eq4.4}--\ref{eq4.6}) then reads
\begin{equation}
   {\rm i}p_s \gamma_{\bf k}({\rm i}p_s)
   \left[1 - Z_{\bf k}({\rm i}p_s)\right] =
        - T_{c} \sum_{{\bf q},{\rm i}p_m}
          \lambda_{\bf kq}({\rm i}p_s - {\rm i}p_m)
          \overline{\Psi}_{\bf q}({\rm i}p_m)
          {\rm i}p_s ,
\label{eq4.23}
\end{equation}
\begin{equation}
   \chi^{e-ph}_{\bf k}({\rm i}p_s) =
        - T_{c} \sum_{{\bf q},{\rm i}p_m}
          \lambda_{\bf kq}({\rm i}p_s-{\rm i}p_m)
          \overline{Y}_{\bf k}({\rm i}p_s)
\label{eq4.24}
\end{equation}
and
\begin{eqnarray}
&  \displaystyle
   \gamma_{\bf k}({\rm i}p_s)
     Z_{\bf k}({\rm i}p_s) \Delta_{\bf k}({\rm i}p_s) =
        T_{c} \sum_{{\bf q},{\rm i}p_m}
          \lambda_{\bf kq}({\rm i}p_s - {\rm i}p_m)
          \overline{\Psi}_{\bf q}({\rm i}p_m)
          \Delta_{\bf q}({\rm i}p_m) + &
\nonumber \\
&  \displaystyle
  +\sum_{\bf q} U^{\uparrow\downarrow}_{\bf kq}({\rm i}p_s)
          \Psi_{\bf q}({\rm i}p_s)
           Z_{\bf q}({\rm i}p_s) \gamma_{\bf q}({\rm i}p_s)
          \Delta_{\bf q}({\rm i}p_s) , &
\label{eq4.25}
\end{eqnarray}
where
\begin{equation}
    \Delta_{\bf k}({\rm i}p_s) =
       \frac{
             \phi_{\bf k}({\rm i}p_s)
            }{
              Z_{\bf k}({\rm i}p_s) \gamma_{\bf k}({\rm i}p_s)
            }
\label{eq4.26}
\end{equation}
is, as usual, the gap function.

  Solving (\ref{eq4.23}) for $Z_{\bf k}({\rm i}p_s)$:
\begin{equation}
   Z_{\bf k}({\rm i}p_s) = 1 +
       \frac{1}{{\rm i}p_s}{\gamma_{\bf k}({\rm i}p_s)}
       T_{c} \sum_{{\bf q},{\rm i}p_m}
       \lambda_{\bf kq}({\rm i}p_s - {\rm i}p_m)
                   \overline{\Psi}_{\bf q}({\rm i}p_m) {\rm i}p_m
\label{eq4.27}
\end{equation}
and inserting the result into (\ref{eq4.25}) we rewrite the equation for
$\Delta_{\bf k}({\rm i}p_s)$ in the form
\begin{eqnarray}
& \displaystyle
  \sum_{\bf q} Q_{\bf kq}({\rm i}p_s)
       \gamma_{\bf q}({\rm i}p_s)
       \Delta_{\bf q}({\rm i}p_s) =
        T_{c} \sum_{{\bf q},{\rm i}p_m}
        \lambda_{\bf kq}({\rm i}p_s - {\rm i}p_m)
          \overline{\Psi}_{\bf q}({\rm i}p_m)
          \Delta_{\bf q}({\rm i}p_m) - &
\nonumber \\
& \displaystyle
  -\frac{\Delta_{\bf k}({\rm i}p_s)}{{\rm i}p_s}
       T_{c} \sum_{{\bf q},{\rm i}p_m}
       \lambda_{\bf kq}({\rm i}p_s - {\rm i}p_m)
                   \overline{\Psi}_{\bf q}({\rm i}p_m) {\rm i}p_m , &
\label{eq4.28}
\end{eqnarray}
where
\begin{equation}
   Q_{\bf kq}({\rm i}p_s) =
         \delta_{\bf kq}({\rm i}p_s) -
         U^{\uparrow\downarrow}_{\bf kq}({\rm i}p_s)
         Z_{\bf q}({\rm i}p_s)\Psi_{\bf q}({\rm i}p_s)
\label{eq4.29}
\end{equation}
and up to now no approximations has been made.

  Usual way to proceed further is  to  introduce  harmonics  which
constitute orthogonal and full set at some  ``bare''  Fermi  surface,
that is so called Fermi surface harmonics. We have\cite{c4,c50}
\begin{equation}
   \sum_{\bf k} F_{J}({\bf k}) F_{J'}({\bf k})
                  \delta(\epsilon - \varepsilon^{*}_{\bf k}) =
            N_{0}(\epsilon) \delta_{JJ'} ,
\label{eq4.30}
\end{equation}
\begin{equation}
   N_{0}(\epsilon) = \sum_{\bf k}
                     \delta(\epsilon - \varepsilon^{*}_{\bf k}) .
\label{eq4.31}
\end{equation}
The expansion over Fermi surface harmonics for various ${\bf k}$-dependent
quantities reads
\begin{equation}
   C_{\bf k} = \sum_{J} \int {\rm d}\epsilon
           \delta(\epsilon - \varepsilon^{*}_{\bf k})
           F_{J}({\bf k})C_{J}(\epsilon) ,
\label{eq4.32}
\end{equation}
\begin{equation}
   K_{\bf kq} = \sum_{JJ'} \int {\rm d}\epsilon {\rm d}\epsilon'
                      \delta(\epsilon - \varepsilon^{*}_{\bf k})
                      \delta(\epsilon' - \varepsilon^{*}_{\bf q})
                      F_{J}({\bf k})F_{J'}({\bf q})K_{JJ'}(\epsilon\epsilon'),
\label{eq4.33}
\end{equation}
\begin{equation}
   K_{\bf kq} = \sum_{J} \int {\rm d}\epsilon
                        \delta(\epsilon - \varepsilon^{*}_{\bf k})
                        F_{J}({\bf k})K_{J,\bf q}(\epsilon) ,
\label{eq4.34}
\end{equation}
\begin{equation}
   A_{\bf k}B_{\bf k} = \sum_{J} \int {\rm d}\epsilon
                            \delta(\epsilon - \varepsilon^{*}_{\bf k})
                            F_{J}(k)
                            \sum_{LL'} C_{JLL'}
                                 A_{L}(\epsilon) B_{L'}(\epsilon) ,
\label{eq4.35}
\end{equation}
\begin{equation}
   \sum_{\bf k}A_{\bf k}B_{\bf k}
                          = \int {\rm d}\epsilon N_{0}(\epsilon)
                            \sum_{L}A_{L}(\epsilon)B_{L}(\epsilon) ,
\label{eq4.36}
\end{equation}
where coefficients $C_{JLL'}$, being Clebsch-Gordan coefficients for the
harmonics, obey the property
\begin{equation}
   C_{0LL'} = C_{L0L'} = C_{LL'0} = \delta_{LL'},
\label{eq4.37}
\end{equation}
$\varepsilon^{*}_{\bf k}$ is ``bare'' dispersion and we choose
$\varepsilon^{*}_{\bf k}=t_{\bf k}$,  and  $t_{\bf k}$ is  bare
electron dispersion defined by nonrandom part of $H^{e}$, Eq.~(\ref{eq2.2}).

  With the use of (\ref{eq4.30}--\ref{eq4.36})
the equation (\ref{eq4.28}) can be represented as
\begin{eqnarray}
& \displaystyle
   \sum_{J'} \int {\rm d}\epsilon' N_{0}(\epsilon')
                Q_{JJ'}(\epsilon\epsilon';{\rm i}p_s)
                \sum_{LL'} C_{J'LL'}
                          \overline{\Psi}_{L}(\epsilon';{\rm i}p_s)
                          \Delta_{L'}(\epsilon';{\rm i}p_s) = &
\nonumber \\
& \displaystyle
   =  T_{c} \sum_{{\rm i}p_m} \sum_{J'}
             \int {\rm d}\epsilon' N_{0}(\epsilon')
                \lambda_{JJ'}(\epsilon\epsilon';{\rm i}p_s-{\rm i}p_m)
                \times
\nonumber \\
& \displaystyle
               \times
               \sum_{LL'} C_{J'LL'}
                          \overline{\Psi}_{L}(\epsilon';{\rm i}p_m)
                          \Delta_{L'}(\epsilon';{\rm i}p_m) -
   T_{c} \sum_{LL'} C_{JLL'}
             \frac{\Delta_{L}(\epsilon;{\rm i}p_s)}{{\rm i}p_s}
             \times &
\nonumber \\
& \displaystyle
             \times
             \sum_{J'}
             \int {\rm d}\epsilon' N_{0}(\epsilon')
                \lambda_{J'L'}(\epsilon\epsilon';{\rm i}p_s-{\rm i}p_m)
                          \overline{\Psi}_{L}(\epsilon';{\rm i}p_m)
                          {\rm i}p_m &
\label{eq4.38}
\end{eqnarray}
Since $s$-type pairing dominates for the case of isotropic
superconductor and, moreover, $\Delta_{L}(\epsilon ;{\rm i}p_s)$ is
almost independent of  $\epsilon$   in  a  narrow
energy region $\vert\epsilon\vert<\omega_{D}$ the
following approximation takes place
\begin{equation}
   \Delta_{L}(\epsilon;{\rm i}p_s) \longrightarrow
     \Delta_{0}(\epsilon^{*};{\rm i}p_s) = \Delta ({\rm i}p_s) .
\label{eq4.39}
\end{equation}
After substituting (\ref{eq4.39}) into (\ref{eq4.38})
and using (\ref{eq4.37}) we get
\begin{eqnarray}
& \displaystyle
    \sum_{\bf q} Q_{0,\bf q}(\epsilon^{*};{\rm i}p_s)
                \gamma_{\bf q}({\rm i}p_s)
       \Delta ({\rm i}p_s) =
       T_{c} \sum_{{\bf q},{\rm i}p_m}
              \lambda_{0,\bf q}(\epsilon^{*};{\rm i}p_s-{\rm i}p_m)
               \overline{\Psi}_{\bf q}({\rm i}p_m)
               \Delta ({\rm i}p_m) - &
\nonumber \\
& \displaystyle
  - \frac{\Delta ({\rm i}p_s)}{{\rm i}p_s}
       T_{c} \sum_{{\bf q},{\rm i}p_m}
             \lambda_{0,\bf q}(\epsilon^{*};{\rm i}p_s-{\rm i}p_m)
             \overline{\Psi}_{\bf q}({\rm i}p_m) {\rm i}p_m &
\label{eq4.40}
\end{eqnarray}
The Ward's identity (\ref{eq4.16}) can be rewritten as
\begin{equation}
   \sum_{\bf q} Q_{\bf  k\,q}({\rm i}p_s) \gamma_{\bf q}({\rm i}p_s) = 1 ,
\label{eq4.41}
\end{equation}
where $Q_{\bf  k\,q}({\rm i}p_s)$ is defined by (\ref{eq4.29}).
After carrying  out  expansion over Fermi surface harmonics,
the expression (\ref{eq4.41}) becomes
\begin{equation}
   \sum_{\bf q} Q_{J,\bf q}(\epsilon,{\rm i}p_s) \gamma_{\bf q}({\rm i}p_s) =
            \delta_{J0} ,
\label{eq4.42}
\end{equation}
and thus
\begin{equation}
   \sum_{\bf q} Q_{0,\bf q}(\epsilon,{\rm i}p_s)
                      \gamma_{\bf q}({\rm i}p_s) = 1 .
\label{eq4.43}
\end{equation}
The use of (\ref{eq4.43}) in the left-hand side of (\ref{eq4.40})
immediately yields
\begin{eqnarray}
& \displaystyle
   \Delta ({\rm i}p_s) \left\{ 1 +
       \frac{1}{{\rm i}p_s}
          T_{c} \sum_{{\bf q},{\rm i}p_m}
                \lambda_{0,\bf q}(\epsilon^{*};{\rm i}p_s-{\rm i}p_m)
                \overline{\Psi}_{\bf q}({\rm i}p_m) {\rm i}p_m
   \right\} = &
\nonumber \\
& \displaystyle
   =T_{c} \sum_{{\bf q},{\rm i}p_m}
         \lambda_{0,\bf q}(\epsilon^{*};{\rm i}p_s-{\rm i}p_m)
         \overline{\Psi}_{\bf q}({\rm i}p_m)\Delta ({\rm i}p_m) . &
\label{eq4.44}
\end{eqnarray}

  This final expression clearly shows that disorder  contributions
enter the gap-function equation only
through $\overline{\Psi}_{\bf k}({\rm i}p_s)$ and, thus,
through renormalized  normal-state  one-particle  spectral  density
$A_{\bf k}(\epsilon )$ (Eq.~(\ref{eq4.18})).
Disorder contributions to anomalous part of the
self-energy cancel the factor $\gamma_{\bf k}({\rm i}p_s)$
in the left hand side  of
(\ref{eq4.23}) owing to exact Ward's
identity (\ref{eq4.16}) and  we  may  conclude
that it is a type of Ward's cancellations which leads to
the Anderson's theorem.

\section{Reduced Equations for the Isotropic superconductor.
         The Abrikosov's Identity}
\noindent
    In  this  section  we  will  derive  a   set   of   linearized
Eliashberg-type equations within isotropic  approximation  for  the
gap function $\Delta_{\bf k}({\rm i}p_s)$,
that is for the case  where  ${\bf k}$-dependence  of
$\Delta_{\bf k}({\rm i}p_s)$ is rather weak and
may  be  completely  neglected  in  the
narrow energy  interval  $\pm\omega_{D}$ near the true chemical potential.
Analogously to the case where disorder contributions to EPI  matrix
elements have been neglected, we will demonstrate  that,  owing  to
exact Ward's identity, anomalous contributions to  the  self-energy,
which come from disorder scattering, may be eliminated from the full
equation for the gap function.

  With the use of (\ref{eq4.1}--\ref{eq4.3}) linearized
Eliashberg-type equations can be formally written as
\begin{equation}
   {\rm i}p_s \gamma_{\bf k}({\rm i}p_s)[1 - Z_{\bf k}({\rm i}p_s)] =
    -T_{c}\sum_{{\bf q},{\rm i}p_m}L^{(z)}_{\bf kq}({\rm i}p_s,{\rm i}p_m) ,
\label{eq5.1}
\end{equation}
\begin{equation}
   \chi^{e-ph}_{\bf k}({\rm i}p_s) =
   -T_{c}\sum_{{\bf q},{\rm i}p_m}L^{(\chi)}_{\bf kq}({\rm i}p_s,{\rm i}p_m) ,
\label{eq5.2}
\end{equation}
\begin{eqnarray}
&  \displaystyle
\phi_{\bf k}({\rm i}p_s) =
    T_{c}\sum_{{\bf q},{\rm i}p_m}L^{(1)}_{\bf kq}({\rm i}p_s,{\rm i}p_m)
          \Psi_{\bf q}({\rm i}p_m)
          \phi_{\bf q}({\rm i}p_m) +
&
\nonumber \\
& \displaystyle
  + T_{c}\sum_{{\bf q},{\rm i}p_m}L^{(2)}_{\bf kq}({\rm i}p_s,{\rm i}p_m)
          \Psi_{\bf q}({\rm i}p_s)
          \phi_{\bf q}({\rm i}p_s) +
        \sum_{\bf q} U^{\uparrow\downarrow}_{\bf kq}({\rm i}p_s)
          \Psi_{\bf q}({\rm i}p_s)
          \phi_{\bf q}({\rm i}p_s) ,
&
\label{eq5.3}
\end{eqnarray}
where we have denoted contributions to $Z_{\bf k}({\rm i}p_s)$
and $\chi_{\bf k}({\rm i}p_s)$
as $L^{(z)}_{\bf kq}({\rm i}p_s,{\rm i}p_m)$
and $L^{(\chi )}_{\bf kq}({\rm i}p_s,{\rm i}p_m)$ in
Eqs.~(\ref{eq5.1}) and~(\ref{eq5.2}) respectively.
Analogously, in (\ref{eq5.3})
$L^{(1)}_{\bf kq}({\rm i}p_s,{\rm i}p_m)$ means the
contributions where  anomalous  part
$\phi_{\bf k}({\rm i}p_s)$ is summed over both
quasimomentum  ${\bf q}$ and frequency ${\rm i}p_m$ , and
$L^{(2)}_{\bf kq}({\rm i}p_s,{\rm i}p_m)$,
the contributions which contain
$\phi_{\bf k}({\rm i}p_s)$ summed over the quasimomentum only.
Note, bytheway,  that while both Hartree- and Fock-type diagrams
contribute to $L^{(2)}_{\bf kq}({\rm i}p_s,{\rm i}p_m)$, only
Fock-type diagrams give non-zero contribution to
$L^{(1)}_{\bf kq}({\rm i}p_s,{\rm i}p_m)$.
The  last  term  in  (\ref{eq5.3})  is anomalous disorder scattering
contribution to $\phi_{\bf k}({\rm i}p_s)$, and we  have
also explicitly written down sums over internal  quasimomentum  and
frequency.

  Making use of Eq.~(\ref{eq4.19})
we can express (\ref{eq5.3}) in terms  of  the  one-particle
spectral density. A little manipulation yields
\begin{eqnarray}
& \displaystyle
  Z_{\bf k}({\rm i}p_s)\gamma_{\bf k}({\rm i}p_s)\Delta_{\bf k}({\rm i}p_s) =
     T_{c}\sum_{{\bf q},{\rm i}p_m}L^{(1)}_{\bf kq}({\rm i}p_s,{\rm i}p_m)
          \overline{\Psi}_{\bf q}({\rm i}p_m)
          \Delta_{\bf q}({\rm i}p_m) +
&
\nonumber \\
& \displaystyle
   +T_{c} \sum_{{\bf q},{\rm i}p_m}L^{(2)}_{\bf kq}({\rm i}p_s,{\rm i}p_m)
          \overline{\Psi}_{\bf q}({\rm i}p_s)
          \Delta_{\bf q}({\rm i}p_s) +
        \sum_{\bf q} U^{\uparrow\downarrow}_{\bf kq}({\rm i}p_s)
          \overline{\Psi}_{\bf q}({\rm i}p_s)
          \Delta_{\bf q}({\rm i}p_s) ,
&
\label{eq5.4}
\end{eqnarray}
and
\begin{equation}
   \gamma_{\bf k}({\rm i}p_s) =
        \sum_{\bf q} U^{\uparrow\downarrow}_{\bf kq}({\rm i}p_s)
          \overline{\Psi}_{\bf q}({\rm i}p_s) .
\label{eq5.5}
\end{equation}
Again, the gap function $\Delta_{\bf k}({\rm i}p_s)$ has been
introduced, see Eq.~(\ref{eq4.26}).

  Formal solution to the equation (\ref{eq5.1})
for $Z_{\bf k}({\rm i}p_s)$ has the form
\begin{equation}
   Z_{\bf k}({\rm i}p_s) = 1 +
     \frac{1}{{\rm i}p_s \gamma_{\bf k}({\rm i}p_s)}
     T_{c}\sum_{{\bf q},{\rm i}p_m}L^{(z)}_{\bf kq}({\rm i}p_s,{\rm i}p_m).
\label{eq5.6}
\end{equation}
The substitution of (\ref{eq5.6}) and (\ref{eq4.16})
into (\ref{eq5.4}) gives
\begin{eqnarray}
& \displaystyle
  \Delta_{\bf k}({\rm i}p_s) \left\{ 1 +
       \frac{1}{{\rm i}p_s}
          T_{c} \sum_{{\bf q},{\rm i}p_m}
                L^{(z)}_{\bf kq}({\rm i}p_s,{\rm i}p_m)
          \right\} +
&
\nonumber \\
& \displaystyle
   +  \Delta_{\bf k}({\rm i}p_s)
        \sum_{\bf q} U^{\uparrow\downarrow}_{\bf kq}({\rm i}p_s)
          \overline{\Psi}_{\bf q}({\rm i}p_s) =
&
\nonumber \\
& \displaystyle
  = T_{c} \sum_{{\bf q},{\rm i}p_m} \left\{
               L^{(1)}_{\bf kq}({\rm i}p_s,{\rm i}p_m)
               \overline{\Psi}_{\bf q}({\rm i}p_m)\Delta_{\bf q}({\rm i}p_m) +
               \right.
&
\nonumber \\
& \displaystyle
             \left.
             + L^{(2)}_{\bf kq}({\rm i}p_s,{\rm i}p_m)
               \overline{\Psi}_{\bf q}({\rm i}p_s)\Delta_{\bf q}({\rm i}p_s)
               \right\} +
           \sum_{\bf q} U^{\uparrow\downarrow}_{\bf kq}({\rm i}p_s)
               \overline{\Psi}_{\bf q}({\rm i}p_s) \Delta_{\bf q}({\rm i}p_s)
&
\label{eq5.7}
\end{eqnarray}
and, after expanding over Fermi surface harmonics (\ref{eq4.30}) and  using
(\ref{eq4.32}--\ref{eq4.36}), one obtains
\begin{eqnarray}
& \displaystyle
  \sum_{LL'}\Delta_{L}(\epsilon;{\rm i}p_s)
   \left\{\delta_{LJ} + C_{JLL'}\frac{1}{{\rm i}p_s}T_{c}
          \sum_{{\bf q},{\rm i}p_m}
              L^{(z)}_{L',\bf q}(\epsilon;{\rm i}p_s,{\rm i}p_m)
   \right\} +
&
\nonumber \\
& \displaystyle
  + \sum_{LL'}C_{JLL'}\Delta_{L}(\epsilon;{\rm i}p_s)
             \sum_{\bf q}U^{\uparrow\downarrow}_{L',\bf q}(\epsilon;{\rm i}p_s)
                     \overline{\Psi}_{\bf q}({\rm i}p_s) =
&
\nonumber \\
& \displaystyle
  = T_{c}\sum_{{\rm i}p_m}\sum_{J'}\sum_{LL'}
        \int d\epsilon' N_{0}(\epsilon')C_{J'LL'} \times
&
\nonumber \\
& \displaystyle
            \times
            \left\{
               L^{(1)}_{JJ'}(\epsilon\epsilon';{\rm i}p_s,{\rm i}p_m)
               \overline{\Psi}_{L}(\epsilon';{\rm i}p_m)
               \Delta_{L'}(\epsilon';{\rm i}p_m)+
             \right.
&
\nonumber \\
& \displaystyle
            \left.
            +  L^{(2)}_{JJ'}(\epsilon\epsilon';{\rm i}p_s,{\rm i}p_m)
               \overline{\Psi}_{L}(\epsilon';{\rm i}p_s)
               \Delta_{L'}(\epsilon';{\rm i}p_s)
             \right\} +
&
\nonumber \\
& \displaystyle
 + \sum_{J'}\sum_{LL'}\int d\epsilon' N_{0}(\epsilon')C_{J'LL'}
            U^{\uparrow\downarrow}_{JJ'}(\epsilon\epsilon'{\rm i}p_s)
            \overline{\Psi}_{L}(\epsilon';{\rm i}p_s)
            \Delta_{L'}(\epsilon';{\rm i}p_s) .
&
\label{eq5.8}
\end{eqnarray}
As in the previous section, a weak dependence  of
$\Delta_{J}(\epsilon ;{\rm i}p_s)$  on  $\epsilon$  within the region
$\vert\epsilon\vert\leq\omega_{D}$ near $E_{F}$ is supposed and we assume
that the pairing is mainly of $s$-type, that is the harmonics
with $J=0$ give dominant
contribution  into  (\ref{eq5.8}). Substituting (\ref{eq4.39})
into  (\ref{eq5.8}) and
recalling (\ref{eq4.32}--\ref{eq4.36}) we then have
\begin{eqnarray}
& \displaystyle
   \Delta ({\rm i}p_s) \left\{ 1 +
       \frac{1}{{\rm i}p_s}
          T_{c} \sum_{{\bf q},{\rm i}p_m}
                L^{(z)}_{0,\bf q}(\epsilon^{*};{\rm i}p_s,{\rm i}p_m)
   \right\} =
&
\nonumber \\
& \displaystyle
 = T_{c} \sum_{{\bf q},{\rm i}p_m} \left\{
               L^{(1)}_{0,\bf q}(\epsilon^{*};{\rm i}p_s,{\rm i}p_m)
               \overline{\Psi}_{\bf q}({\rm i}p_m)\Delta ({\rm i}p_m) +
               \right.
&
\nonumber \\
& \displaystyle
            \left.
            +  L^{(2)}_{0,\bf q}(\epsilon^{*};{\rm i}p_s,{\rm i}p_m)
               \overline{\Psi}_{\bf q}({\rm i}p_s)\Delta ({\rm i}p_s)
               \right\} ,
&
\label{eq5.9}
\end{eqnarray}
and this equation does not contain anomalous  contribution  due  to
disorder scattering explicitly.

  If we introduce an auxiliary quantity $\tilde{\Sigma}({\rm i}p_s)$
by the expression
\begin{equation}
   \tilde{\Sigma}({\rm i}p_s) =
       {\rm i}p_s[1-\tilde{Z}({\rm i}p_s)] \tau_{0} +
       \chi({\rm i}p_s) \tau_{3} +
       \tilde{Z}({\rm i}p_s)\Delta({\rm i}p_s) \tau_{1} ,
\label{eq5.10}
\end{equation}
we then can write for the isotropic  superconductor  the  following
set of linearized Eliashberg-type equations
\begin{equation}
   {\rm i}p_s[1 - \tilde{Z}({\rm i}p_s)] =
        - T_{c} \sum_{{\bf q},{\rm i}p_m}
                     L^{(z)}_{0,\bf q}({\rm i}p_s,{\rm i}p_m) ,
\label{eq5.11}
\end{equation}
\begin{equation}
   \chi({\rm i}p_s) =
        - T_{c} \sum_{{\bf q},{\rm i}p_m}
                 L^{(\chi)}_{0,\bf q}({\rm i}p_s,{\rm i}p_m) ,
\label{eq5.12}
\end{equation}
\begin{eqnarray}
& \displaystyle
   \tilde{Z}({\rm i}p_s)\Delta({\rm i}p_s) =
        T_{c} \sum_{{\bf q},{\rm i}p_m}
               L^{(1)}_{0,\bf q}(\epsilon^{*};{\rm i}p_s,{\rm i}p_m)
               \overline{\Psi}_{\bf q}({\rm i}p_m)\Delta ({\rm i}p_m) +
&
\nonumber \\
& \displaystyle
    + T_{c} \sum_{{\bf q},{\rm i}p_m}
               L^{(2)}_{0,\bf q}(\epsilon^{*};{\rm i}p_s,{\rm i}p_m)
               \overline{\Psi}_{\bf q}({\rm i}p_s)\Delta ({\rm i}p_s) .
&
\label{eq5.13}
\end{eqnarray}
Being considered as equations on auxiliary
self-energy  $\tilde{\Sigma}({\rm i}p_s)$ these equations
may by depicted in a form formally coinciding with  that  shown  in
Figs.~5 and~6, except that picking out of isotropic
contributions is  meant  implicitly.  We  see  that  the  anomalous
disorder-scattering  contribution  to  the   self-energy   can   be
eliminated owing to the Ward's identity for the  disorder  part  of
the self-energy similarly to  the  case  studied  in  the  previous
Section and the remaining explicite dependence on disorder  is  due
to renormalized EPI vertices involved in the equation for auxiliary
self-energy $\tilde{\Sigma}({\rm i}p_s)$.

  The fact of the Ward's cancellations of anomalous  contributions
coming  from   disorder   scattering   and   disorder   self-energy
contributions to the gap function  manifests  itself  also  in  the
so-called Abrikosov's identity.\cite{c47} To establish this
identity, we consider the linearized
$\phi^{e}_{\bf k}({\rm i}p_s)$ contribution,
\begin{equation}
   \phi^{e}_{\bf k}({\rm i}p_s)=
        \sum_{\bf q}U^{\uparrow\downarrow}_{\bf kq}({\rm i}p_s)
                \Psi_{\bf q}({\rm i}p_s)Z_{\bf q}({\rm i}p_s)
                \gamma_{\bf q}({\rm i}p_s)\Delta_{\bf q}({\rm i}p_s) ,
\label{eq5.14}
\end{equation}
to the function $\phi_{\bf k}({\rm i}p_s)$.
Bearing in mind isotropic  $s$-type  pairing
and making use of the expansion over
harmonics (\ref{eq4.30}) one obtains
\begin{equation}
   \phi^{e}_{\bf k}({\rm i}p_s)=
        \sum_{\bf q}U^{\uparrow\downarrow}_{\bf kq}({\rm i}p_s)
                \Psi_{\bf q}({\rm i}p_s)Z_{\bf q}({\rm i}p_s)
                \gamma_{\bf q}({\rm i}p_s)\Delta ({\rm i}p_s) ,
\label{eq5.15}
\end{equation}
where we have retained a residual ${\bf k}$-dependence of the irreducible
four-leg vertex. Recalling (\ref{eq4.16}) we have
\begin{equation}
   \frac{
         \phi^{e}_{\bf k}({\rm i}p_s)
        }{
        \Delta ({\rm i}p_s)
        }= \gamma_{\bf k}({\rm i}p_s)-1 .
\label{eq5.16}
\end{equation}
Substituting in this expression the definition (\ref{eq4.3})
for  $\gamma_{\bf k}({\rm i}p_s)$
and using the time reversal symmetry we arrive at  the  Abrikosov's
identity
\begin{equation}
   \frac{
         \phi^{e}_{\bf k}({\rm i}p_s)
        }{
        \Delta ({\rm i}p_s)
        }=-\frac{
                {\rm i}\,{\rm Im}\Sigma_{\bf k}({\rm i}p_s)
                }{
                {\rm i}p_s
                }
\label{eq5.17}
\end{equation}
in the form proposed by Lustfield,\cite{c47} so that this identity is not
specific for BCS-like models but holds  also  for  the  case  where
pairing is due to EPI and irrespectively of the strength  and  type
of quenched substitutional disorder.

\section{Qualitative Analysis of the Reduced Isotropic Equations}
\noindent
  General formal equations (\ref{eq2.33},\ref{eq2.41}--\ref{eq2.43})
from which one may obtain the explicit expression for the superconductive
transition temperature $T_c$, are too cumbersome to work with directly.
However,  if  certain approximations are made
when calculating isotropic contributions to
$\tilde{\Sigma}({\rm i}p_s)$, qualitative analysis of
the reduced equations  (\ref{eq5.11}--\ref{eq5.13})  becomes
possible. Confining ourselves hereafter by the binary alloy model~(\ref{eq2.3})
introduced in Section~2 we present here general formulas for $T_{c}$ in
the case  of  weak  and  intermediate  EPI  coupling  and  for  the
renormalization of the EPI vertices by disorder scattering.

  To pick out isotropic contributions to $\tilde{\Sigma}({\rm i}p_s)$,
let us consider
terms of order $M$ of the self-consistent perturbational series in $V$
contributing to $\Sigma ({\rm i}p_s)$. The isotropic part resulting from a
particular $M$'th-order diagram can be represented
in terms of the functions
$G_{J}(\epsilon;{\rm i}p_s)$ and
$\lambda_{JJ'}(\epsilon\epsilon';{\rm i}\omega_n)$ which are the
coefficients  of  the
expansions over Fermi surface harmonics of the one-particle Green's
function $G_{\bf k}({\rm i}p_s)$ and of the function
$\lambda_{\bf kq}({\rm i}\omega_n)$ (defined by~(\ref{eq4.11}))
respectively.

  There arise also combinations of the form
\begin{eqnarray}
   N^{-1}_{0}(\epsilon) \sum_{{\bf k},{\bf p},\bf q}
           F_{J_{1}}({\bf k})
           F_{J_{2}}({\bf p})
           F_{J_{3}}({\bf q})
           F_{J_{3}}({\bf k-p+q})& &
\nonumber \\
&          \hspace*{-10em}\delta (\epsilon-t_{\bf k})
           \delta (\epsilon_{1}-t_{\bf p})
           \delta (\epsilon_{2}-t_{\bf q})
           \delta (\epsilon_{3}-t_{\bf k-p+q})&
\label{eq6.1}
\end{eqnarray}
and the like containing more complex combinations of $\delta$-functions
and Fermi surface harmonics $F_{J}({\bf k})$.
Quite convincing  check  of  the
statement made is to write down the expression  for  the  isotropic
part of $\tilde{\Sigma}({\rm i}p_s)$ within simplest alloy
approximation,  that  is  VCA,
but corresponding details are mainly of auxiliary character and  we
do not present them here.

  Our further approximations are formulated as follows.

  Firstly, we suppose that approximation adopted for
$\Sigma^{e}_{\bf k}({\rm i}p_s)$  is
such as not to disturb drastically the ${\bf k}$-dependence
of $G_{\bf k}({\rm i}p_s)$
comparatively with that of the bare one-particle Green's  function,
at least for the narrow energy region $\vert\epsilon\vert<\omega_{D}$
near $E_{F}$, what  implies
that $\Sigma_{\bf k}({\rm i}p_s)$ is almost ${\bf k}$-independent
in this region. This  holds
exactly for all energies within arbitrary single-site approximation
to $\Sigma^{e}_{\bf k}({\rm i}p_s)$ and approximately
for $\Sigma^{e-ph}_{\bf k}({\rm i}p_s)$, to a good extent
however. More exactly, we may suppose that
$\Sigma^{e}_{\bf k}({\rm i}p_s)\approx\Sigma(t_{\bf k};{\rm i}p_s)$,
with $t_{\bf k}$ being ``bare'' electron spectrum, and in particular
we may have $\Sigma^{e}_{\bf k}({\rm i}p_s)=\Sigma({\rm i}p_s)$
within an arbitrary single-site approximation.
It is also assumed that dominant contribution to
$\lambda_{JJ'}(\epsilon\epsilon';{\rm i}\omega_n)$ results
from harmonics with $J=J'=0$. Hence, we have the
following substitutions:
\begin{eqnarray}
   G_{J}(\epsilon;{\rm i}p_s)  & \rightarrow &
       G_{0}(\epsilon;{\rm i}p_s) \delta_{J0} ,
\label{eq6.2} \\
   \lambda_{JJ'}(\epsilon\epsilon';{\rm i}\omega_m) & \rightarrow &
         \lambda_{00}(\epsilon\epsilon';{\rm i}\omega_m)
         \delta_{J0} \delta_{J'0} .
\label{eq6.3}
\end{eqnarray}
In this situation combination (\ref{eq6.1}) simplifies
\begin{equation}
   N^{-1}_{0}(\epsilon) \sum_{{\bf k},{\bf p},\bf q}
           \delta (\epsilon-t_{\bf k})
           \delta (\epsilon_{1}-t_{\bf p})
           \delta (\epsilon_{2}-t_{\bf q})
           \delta (\epsilon_{3}-t_{\bf k-p+q}) ,
\label{eq6.4}
\end{equation}
and so do the like.

  Secondly, the dependence of
$\lambda_{00}(\epsilon\epsilon';{\rm i}\omega_n)$ on $\epsilon$ and
$\epsilon'$ is
supposed to be sufficiently weak in the region
$\vert\epsilon\vert ,\vert\epsilon'\vert<\omega_{D}$ and
therefore may be completely neglected. This leads to the
replacement
\begin{equation}
   \lambda_{00}(\epsilon\epsilon';{\rm i}\omega_m) \rightarrow
      \lambda_{00}(\epsilon^{*}\epsilon^{*};{\rm i}\omega_m) =
      \lambda({\rm i}\omega_m) ,
\label{eq6.5}
\end{equation}
where $\epsilon^{*}$ is some point within the energy interval specified.

  Thirdly, we decouple expressions of the type (\ref{eq6.4}) and the  like
in a simple manner
\begin{eqnarray}
& \displaystyle
   N^{-1}_{0}(\epsilon) \sum_{{\bf k},{\bf p},\bf q}
           \delta (\epsilon    -t_{\bf k})
           \delta (\epsilon_{1}-t_{\bf p})
           \delta (\epsilon_{2}-t_{\bf q})
           \delta (\epsilon_{3}-t_{\bf k-p+q}) \rightarrow &
\nonumber \\
& \displaystyle
   N^{-1}_{0}(\epsilon) \sum_{{\bf k},{\bf p},\bf q}
           \delta (\epsilon    -t_{\bf k})
           \delta (\epsilon_{1}-t_{\bf p})
           \delta (\epsilon_{2}-t_{\bf q}) \left\{ \sum_{\bf p'}
           \delta (\epsilon_{3}-t_{\bf k+q-p'})
           \right\}  &
\nonumber \\
&
     = N_{0}(\epsilon_{1})N_{0}(\epsilon_{2}) N_{0}(\epsilon_{3}) . &
\label{eq6.6}
\end{eqnarray}
In these approximations one-particle Green's  function  enters  the
analytical expressions for diagrams in the combination
\begin{equation}
   \int {\rm d}\epsilon N_{0}(\epsilon) G_{0}(\epsilon;{\rm i}p_s) =
        \sum_{\bf k} G_{\bf k}({\rm i}p_s) =
        \frac{1}{N}  \sum_{i} G_{ii}({\rm i}p_s) =
        G({\rm i}p_s)
\label{eq6.7}
\end{equation}
where the use was made of  (\ref{eq4.30}--\ref{eq4.36}).
What  happens when accepting these
simplifications is that the vertices become purely
local quantities. They contain only site-diagonal contributions,
so that these approximations may be justified for the case  where
site-off-diagonal contributions carrying all remaining
${\bf k}$-dependence of the vertices are not important.
The latter holds true, in
particular, when the localization of electrons by disorder is
absent or at least weak for the energies within the energy
interval of interest.

  In any respect, these simplifications are fully transparent for
single-site approximations to the alloy self-energy where all
irreducible vertices are purely local quantities and as for more
complex non-single-site approximations we suppose the validity of
them, of course with the remarks above in mind.
Note also, that the other way of introducing the approximations
proposed is to use the limit of infinite spartial dimensionality
$d=\infty$,\cite{c51,c52,c53} which has become very popular
last time when discussing problems of the theory of strongly
correlated systems.\cite{c29,c30,c31,c32,c33}
The problem of the renormalization of EPI by disorder
may then be formulated in terms of the Holstein model\cite{c54}
in the limit $d=\infty$\cite{c55} and with the disorder contributions
added. This problem apeares to be a single-site problem\cite{c51,c52,c53}
and the ${\bf k}$-dependence of both the self-energy and the vertex
corrections is absent from the very begining.

  And at last fourthly, the expression (\ref{eq6.7}) can be represented as
\begin{eqnarray}
   G({\rm i}p_s) & =  & \displaystyle
          - \sum_{\bf k} \int {\rm d}\epsilon
            \frac{A_{\bf k}(\epsilon)}{p^{2}_{s}+\epsilon^{2}}
            \left\{
            {\rm i}p_s\tau_{0}+\epsilon\tau_{3}+\Delta({\rm i}p_s)\tau_{1}
            \right\}
\nonumber \\
    & = & \displaystyle
          - \int {\rm d}\epsilon
            \frac{N(\epsilon)}{p^{2}_{s}+\epsilon^{2}}
            \left\{
            {\rm i}p_s\tau_{0}+\epsilon\tau_{3}+\Delta({\rm i}p_s)\tau_{1}
            \right\}
\label{eq6.8}
\end{eqnarray}
where we have expanded $G({\rm i}p_s)$ up to the first order in
$\Delta ({\rm i}p_s)$
bearing in mind equation for $T_{c}$ and introduced renormalized
one-particle state density
\begin{equation}
   N(\epsilon) = \sum_{\bf k} A_{\bf k}(\epsilon)
\label{eq6.9}
\end{equation}
and $A_{\bf k}(\epsilon)$ is defined by (\ref{eq4.18},\ref{eq4.19}).
Choosing then $E_{F}=0$ and
assuming weak dependence of $N(\epsilon)$ on $\epsilon$ in the interval
$|\epsilon|<\omega_{D}$ we can write
\begin{equation}
   G({\rm i}p_s) = - {\rm i}\pi N(0) {\rm Sign}(p_s)\tau_{0}
             -\frac{\Delta({\rm i}p_s)}{\vert p_s\vert}\pi N(0)\tau_{1} .
\label{eq6.10}
\end{equation}
It is this simple frequency dependence of normal part of $G({\rm i}p_s)$
which enables, along with approximations (\ref{eq6.2}--\ref{eq6.6}),
to analyze qualitatively the equations (\ref{eq5.11}--\ref{eq5.13}).
Key idea of such analysis is this.
After using approximations (\ref{eq6.2}--\ref{eq6.6}) and (\ref{eq6.10}),
decoupling of dependence on frequency and disorder potential $V$  and
concentration $c$ proves to be possible order by order in the
perturbational expansion. The equations (\ref{eq5.11}--\ref{eq5.13})
can then be represented as
\begin{equation}
   T_{c} \sum_{{\rm i}p_m} \sum_{\alpha\beta}
              f_{\alpha\beta}(c,\pi N(0)V)
              \pi N(0)\lambda ({\rm i}p_s-{\rm i}p_m)
              \zeta_{\alpha}({\rm i}p_s) \zeta_{\beta}({\rm i}p_m) ,
\label{eq6.11}
\end{equation}
where ${\rm i}p_s$-dependence of $\zeta_{\alpha}({\rm i}p_s)$
is known and $\zeta_{\alpha}({\rm i}p_s)$  possesses  one
of the following simple forms
\begin{equation}
   1,                                       \quad
   {\rm Sign}(p_s),                         \quad
   \frac{{\rm Sign}(p_s)}{\vert p_s\vert},  \quad
   \frac{\Delta({\rm i}p_s)}{\vert p_s\vert} ,
\label{eq6.12}
\end{equation}
and the
factors $f_{\alpha\beta}(c,\pi N(0)V)$ in (\ref{eq6.11})
are wholly determined by the approximation
used to calculate $\Sigma^{e}({\rm i}p_s)$, and $\alpha,\beta= 1,\ldots,4$.

  We consider this possibility in detail
for Fock contributions to the self-energy first.

  Characterize diagrams of order $M$ by three numbers $N$, $L$, $Q$,
with $N$, $L$ and $Q$ being the numbers of successive disorder scattering
processes before, between and after two  successive  EPI  processes
respectively (as we shall see shortly only a parity of $N$, $L$ and $Q$
is of importance). Obviously, the number of full electron lines  in
the $M$'th-order diagram equals $M+1$, the number of $\tau_{3}$ Pauli
matrices is $M+2$ and $N+L+Q=M$. We also refer to the subset
of graphs with triple $NLQ$ fixed as to $NLQ$-family.

%%%%%%%%%%%%%%%%%%%%%%%%%%%%%%%%%%%%%%%%%%%%%%%%%%%%%%%%%%%%%%%%%%%%%%%%%%%%
\begin{figure}[htbp]
\vspace*{13pt}
\epsfysize=1.5truein
\centerline{\epsfbox{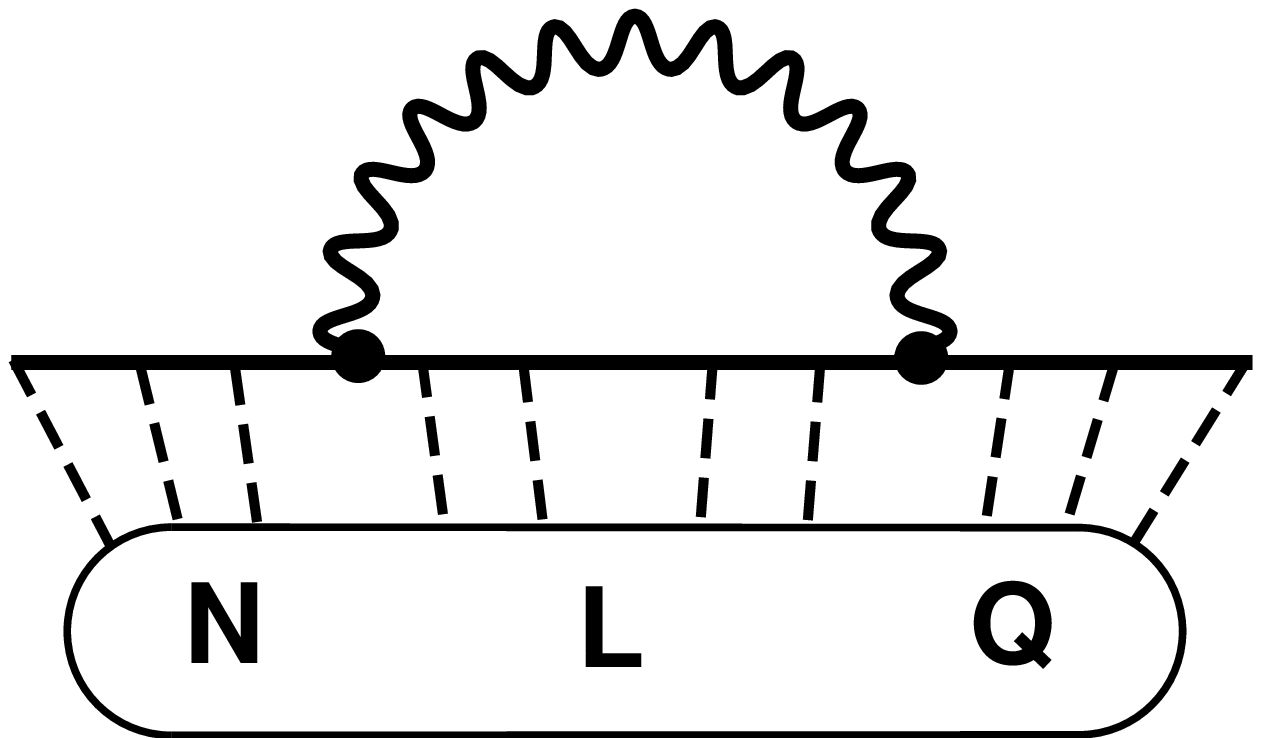}}
\vspace*{13pt}
\fcaption{General structure of the Fock-type
diagram. Scheme of \protect{$NLQ$}-partitioning. Broken line represents
a single-site scattering potential; the oval, the quantity
\protect{$B^{(D)}_{NLQ}(c)$}. For meaning of the other elements see
previous Figures.}
\end{figure}
%%%%%%%%%%%%%%%%%%%%%%%%%%%%%%%%%%%%%%%%%%%%%%%%%%%%%%%%%%%%%%%%%%%%%%%%%%%%

  Within the approximations~(\ref{eq6.2}--\ref{eq6.6})
and~(\ref{eq6.10}) the  contribution of $M$'th-order self-energy
diagram reads (see Fig.~9)
\begin{equation}
   B^{(D)}_{NLQ}V^{M} T_c
   \sum_{{\rm i}p_m}
   \lambda ({\rm i}p_s-{\rm i}p_m)\tau_{3}
   \left(G({\rm i}p_s)\tau_{3}\right)^{N}
   \left(G({\rm i}p_m)\tau_{3}\right)^{L+1}
   \left(G({\rm i}p_s)\tau_{3}\right)^{Q}
\label{eq6.13}
\end{equation}
where $B^{(D)}_{NLQ}(c)$ is a product of the renormalized cumulants for the
diagram ($D$) given.

  To proceed with the calculation of normal part of
$\tilde{\Sigma}({\rm i}p_s)$ we note
that diagrams with $M$ being even or odd contribute to
$\tilde{Z}({\rm i}p_s)$ or $\chi({\rm i}p_s)$ respectively.
Indeed, graphs with even (odd) $M$
contain even (odd) number of $\tau_{3}$-matrices,
normal contribution to $G({\rm i}p_s)$ is proportional
to unity $\tau_{0}$-matrix (see expression (\ref{eq6.10})), therefore
the contribution of such graphs to $\tilde{\Sigma}({\rm i}p_s)$
is proportional to $\tau_{0}$ ($\tau_{3}$)-matrix.

  In turn, we further classify graphs contributing
to $\tilde{Z}({\rm i}p_s)$ with
the parity of $N$, $L$ and $Q$. There are only four combinations for the
decomposition of $M$ even on $N$, $L$ and $Q$ with definite parities:
\begin{equation}
    eee,\quad
    oeo,\quad
    eoo,\quad
    ooe,
\label{eq6.14}
\end{equation}
and $e$ ($o$) means even (odd).

Consider for example $eee$-type
contributions. After inserting the normal part of $G({\rm i}p_s)$,
which  has the form $-{\rm i}\pi N(0){\rm Sign}(p_s)$,
general expression (\ref{eq6.13}) reduces to
\begin{equation}
   B^{(D)}_{NLQ}(c)\left(-{\rm i}V\right)^{M} \pi T_{c}\sum_{{\rm i}p_s}
         \tilde{\lambda}({\rm i}p_s-{\rm i}p_m)
         \tau_{3}^{M+2}{\rm Sign}^{N+Q}(p_s){\rm Sign}^{L+1}(p_m) ,
\label{eq6.15}
\end{equation}
where
\begin{equation}
   \tilde{V}=\pi N(0)V,\qquad
   \tilde{\lambda}({\rm i}\omega_n)= N(0)\lambda({\rm i}\omega_n) .
\label{eq6.16}
\end{equation}
Then, for $eee$-type contribution one has
\begin{equation}
   -{\rm i}B^{(D)}_{NLQ}(c)\left(-{\rm i}\tilde{V}\right)^{M}
         \pi T_{c}\sum_{{\rm i}p_s}
                  \tilde{\lambda}({\rm i}p_s-{\rm i}p_m)
                  {\rm Sign}(p_m) .
\label{eq6.17}
\end{equation}
Analogously, for other parity combinations, we have
\begin{eqnarray}
& \displaystyle
  -{\rm i}B^{(D)}_{NLQ}(c)\left(-{\rm i}\tilde{V}^{M}\right)
         \pi T_{c}\sum_{{\rm i}p_s}
                  \tilde{\lambda}({\rm i}p_s-{\rm i}p_m)
                  {\rm Sign}(p_m), &
   \quad (oeo)
\label{eq6.18} \\
& \displaystyle
  -{\rm i}B^{(D)}_{NLQ}(c)\left(-{\rm i}\tilde{V}^{M}\right)
         \pi T_{c}\sum_{{\rm i}p_s}
                  \tilde{\lambda}({\rm i}p_s-{\rm i}p_m)
                  {\rm Sign}(p_m), &
   \quad (eoo,\/ooe)
\label{eq6.19}
\end{eqnarray}
and, as we see, the precise form of frequency  dependence  for  the
expressions under sums is completely determined by  the  parity  of
the numbers $N$, $L$ and $Q$.

Putting everything together over all orders
in $V$ we then have the contribution of the Fock diargams to the
equation on $\tilde{Z}({\rm i}p_s)$ of the following form
\begin{eqnarray}
& \displaystyle
   - {\rm i} \left\{ f_{eee}(c,\tilde{V})+f_{oeo}(c,\tilde{V}) \right\}
                   \pi T_{c} \sum_{{\rm i}p_m}
                   \tilde{\lambda}({\rm i}p_s-{\rm i}p_m)
                   {\rm Sign}({\rm i}p_m) - &
\nonumber \\
& \displaystyle
   - {\rm i} \left\{ f_{eoo}(c,\tilde{V})+f_{ooe}(c,\tilde{V}) \right\}
                   \pi T_{c} \sum_{{\rm i}p_m}
                   \tilde{\lambda}({\rm i}p_s-{\rm i}p_m)
                   {\rm Sign}({\rm i}p_s), &
\label{eq6.20}
\end{eqnarray}
and the coefficient functions $f_{\alpha\beta\gamma}(c,\tilde{V})$ are
\begin{equation}
   f_{\alpha\beta\gamma}(c,\tilde{V}) =
      \sum_{NLQ=\{\alpha\beta\gamma\}}
           \sum_{\{D\}} B^{(D)}_{NLQ}(c)
                \left(-{\rm i}\tilde{V}\right)^{N+L+Q}.
\label{eq6.21}
\end{equation}
Here $B^{(D)}_{NLQ}(c)$ is a product of cumulants for a particular graph $(D)$,
inner sum runs over a set $\{{\bf D}\}$ of the Fock-type graphs
with triple $NLQ$ fixed, that
is over the $NLQ$-family, and outer sum is over all $NLQ$-families with
definite parity $\{\alpha\beta\gamma\}$ of $NLQ$ triple.

  Similarly one obtains the Fock contribution to $\chi({\rm i}p_s)$:
\begin{eqnarray}
& \displaystyle
   - {\rm i} \left\{ f_{ooo}(c,\tilde{V})+f_{eoe}(c,\tilde{V}) \right\}
                   \pi T_{c} \sum_{{\rm i}p_m}
                   \tilde{\lambda}({\rm i}p_s-{\rm i}p_m) - &
\nonumber  \\
& \displaystyle
   - {\rm i} \left\{ f_{eeo}(c,\tilde{V})+f_{oee}(c,\tilde{V}) \right\}
                   {\rm Sign}({\rm i}p_s)
                   \pi T_{c} \sum_{{\rm i}p_m}
                   \tilde{\lambda}({\rm i}p_s-{\rm i}p_m)
                   {\rm Sign}({\rm i}p_m). &
\label{eq6.22}
\end{eqnarray}

  Consider the anomalous contributions to $\tilde{\Sigma}({\rm i}p_s)$.
A particular $M$'th-order graph for the normal part
of $\tilde{\Sigma}({\rm i}p_s)$ generates $M+1$ graphs
contributing by successively replacing each normal  state  electron
line with anomalous part of~(\ref{eq6.10}) which is proportional to
$\tau_{1}$-matrix. These graphs may be further partitioned into two
complementary sets. The first set contains graphs where anomalous
part of $G({\rm i}p_s)$ is between two successive EPI processes; the
second, all the graphs remained.

For the first set we have
\begin{eqnarray}
& \displaystyle
   B^{(D)}_{NLQ}(c)\left(-{\rm i}\tilde{V}\right)^{M}
        \pi T_{c}\sum_{{\rm i}p_s}
        \tilde{\lambda}({\rm i}p_s-{\rm i}p_m)
        {\rm Sign}^{N+Q}(p_s) \times &
\nonumber \\
& \displaystyle
        \times
        {\rm Sign}^{L}(p_m)
        \frac{\Delta({\rm i}p_m)}{|p_m|}
        \left\{
              \sum_{S=1}^{L+1}\tau_{3}^{N+S-1}\tau_{1}\tau_{3}^{L+Q+1-S}
        \right\} , &
\label{eq6.23}
\end{eqnarray}
and the expression of the form
\begin{eqnarray}
&  \displaystyle
   B^{(D)}_{NLQ}(c)\left(-{\rm i}\tilde{V}\right)^{M}
        \pi T_{c}\sum_{{\rm i}p_s}
        \tilde{\lambda}({\rm i}p_s-{\rm i}p_m)
        {\rm Sign}^{N+Q-1}(p_s)
        {\rm Sign}^{L+1}(p_m)
        \frac{\Delta({\rm i}p_s)}{|p_s|} \times &
\nonumber \\
&  \displaystyle
        \times
        \left\{
              \sum_{S=1}^{N}\tau_{3}^{S-1}\tau_{1}\tau_{3}^{N+L+Q+1-S} +
              \sum_{S=1}^{Q}\tau_{3}^{N+L+S}\tau_{1}\tau_{3}^{Q-S}
        \right\} &
\label{eq6.24}
\end{eqnarray}
for the second set.

  Let now $M$ be odd. Pauli matrices enter such diagrams in
combinations
\begin{equation}
   \left(\tau_{3}\right)^{o}\tau_{1}\left(\tau_{3}\right)^{e} =
          {\rm i}\tau_{2} ,
\label{eq6.25}
\end{equation}
\begin{equation}
   \left(\tau_{3}\right)^{e}\tau_{1}\left(\tau_{3}\right)^{o} =
          -{\rm i}\tau_{2}
\label{eq6.26}
\end{equation}
(here $e$ ($o$) means even (odd) power respectivelly), leading formally
to non-zero $\tau_{2}$ contributions to anomalous part of
$\tilde{\Sigma}({\rm i}p_s)$. However
full contribution of such graphs vanishes identically owing to
usual gauge symmetry under phase transformations. This may be
established on rather general grounds for the equations
(\ref{eq2.30}--\ref{eq2.33}, \ref{eq2.41}--\ref{eq2.43}) without
any computational simplifications accepted
when evaluating contribution of a particular graph provided the
approximation to $\Sigma^{e}_{k}({\rm i}p_s)$ used possesses a property
of being conserving. But the last is precisely what we deal with.

  Now  we  check   that   within   computational   simplifications
(\ref{eq6.2}--\ref{eq6.6},\ref{eq6.10}) vanishing of
$\tau_{2}$-contributions is preserved. Consider
a particular graph of asymmetric $NLQ$-family, $N\ne Q$. There is always
a graph in the $QLN$-family which is symmetric to initially given one
by  reflection  with  respect  to  a  vertical  line   since   full
contribution  to  $\tilde{\Sigma}({\rm i}p_s)$ possesses this symmetry.
Both graphs
generate $M+1$ contributions to the anomalous part but with reversed
order of Pauli matrices relativelly to each other. Therefore whole
conribution of two asymmetric $NLQ$-  and  $QLN$-  families  vanishes
identically (recall Eqs.~(\ref{eq6.25},\ref{eq6.26})).
Analogously, this vanishing takes
place for the symmetric $NLQ$-family, $N=Q$, but within this  family,
because graphs with the symmetry specified belong now to  the  same
family: anomalous graphs of the first set cancel within the set and
so do graphs of the second set. Thus we conclude that such
contributions vanish identically order by order in the
pertubational series.

  For even $M$, after partitioning on  $NLQ$  we  have  for  the
anomalous part of $\tilde{\Sigma}({\rm i}p_s)$ the contribution
which for the first set reads
\begin{eqnarray}
& \displaystyle
  B^{(D)}_{NLQ}(c)\left(-{\rm i}\tilde{V}\right)^{M}\pi T_{c}\sum_{{\rm i}p_s}
        \tilde{\lambda}({\rm i}p_s-{\rm i}p_m)
        \frac{\Delta({\rm i}p_m)}{|p_m|}, &
        \qquad
        (eee)
\label{eq6.27} \\
& \displaystyle
  -B^{(D)}_{NLQ}(c)\left(-{\rm i}\tilde{V}\right)^{M}
        \pi T_{c}\sum_{{\rm i}p_s}
        \tilde{\lambda}({\rm i}p_s-{\rm i}p_m)
        \frac{\Delta({\rm i}p_m)}{|p_m|}, &
        \qquad
        (oeo)
\label{eq6.28}
\end{eqnarray}
and for the second set
\begin{eqnarray}
  \displaystyle
  B^{(D)}_{NLQ}(c)\left(-{\rm i}\tilde{V}\right)^{M}
        \pi T_{c}\sum_{{\rm i}p_s}
        \tilde{\lambda}({\rm i}p_s-{\rm i}p_m)
        \frac{2\Delta({\rm i}p_s)}{|p_s|}
        {\rm Sign}(p_s)
        {\rm Sign}(p_m)
        \quad
        (eee) &
\label{eq6.29} \\
  \displaystyle
  B^{(D)}_{NLQ}(c)\left(-{\rm i}\tilde{V}\right)^{M}
        \pi T_{c}\sum_{{\rm i}p_s}
        \tilde{\lambda}({\rm i}p_s-{\rm i}p_m)
        \frac{\Delta({\rm i}p_s)}{|p_s|} .
        \quad
        (eoo,\/ooe) &
\label{eq6a.30}
\end{eqnarray}
All other pariry combinations from (\ref{eq6.14}) give zero contributions
to the anomalous self-energy.

  Again,  summing  anomalous  contributions  over  all  orders  of
perturbational  expansion  we  obtain
the contribution of the Fock diagrams to the  anomalows part
of the self-energy $\tilde{\Sigma}({\rm i}p_s)$:
\begin{eqnarray}
& \displaystyle
    \left\{ f_{eee}(c,\tilde{V}) - f_{oeo}(c,\tilde{V})\right\}
    \pi T_{c} \sum_{{\rm i}p_m}
              \tilde{\lambda}({\rm i}p_s - {\rm i}p_m)
              \frac{\Delta({\rm i}p_m)}{|p_m|} +
\nonumber \\
& \displaystyle
   +\left\{ f_{eoo}(c,\tilde{V}) + f_{ooe}(c,\tilde{V})\right\}
   \frac{\Delta({\rm i}p_s)}{|p_s|}
   \pi T_{c} \sum_{{\rm i}p_m}
             \tilde{\lambda}({\rm i}p_s - {\rm i}p_m) +
\nonumber \\
& \displaystyle
   + 2 f_{eee}(c,\tilde{V})
   \frac{\Delta({\rm i}p_s)}{|p_s|}
   \pi T_{c} \sum_{{\rm i}p_m}
             \tilde{\lambda}({\rm i}p_s - {\rm i}p_m)
             {\rm Sign}({\rm i}p_s)
             {\rm Sign}({\rm i}p_m),
\label{eq6.31}
\end{eqnarray}
where functions $f_{\alpha\beta\gamma}$ are
defined by Eq.(\ref{eq6.21}).

Now we briefly consider evaluation of the Hartree-type contributions
to the self-energy $\tilde{\Sigma}({\rm i}p_s)$. The evaluation of these
contributions follow closely same roots as above, and minor modifications
are only necessary.

General Hartree-type diagram is shown in Fig.~10.
%%%%%%%%%%%%%%%%%%%%%%%%%%%%%%%%%%%%%%%%%%%%%%%%%%%%%%%%%%%%%%%%%%%%%%%%%%%%
\begin{figure}[htbp]
\vspace*{13pt}
\epsfysize=1.8truein
\centerline{\epsfbox{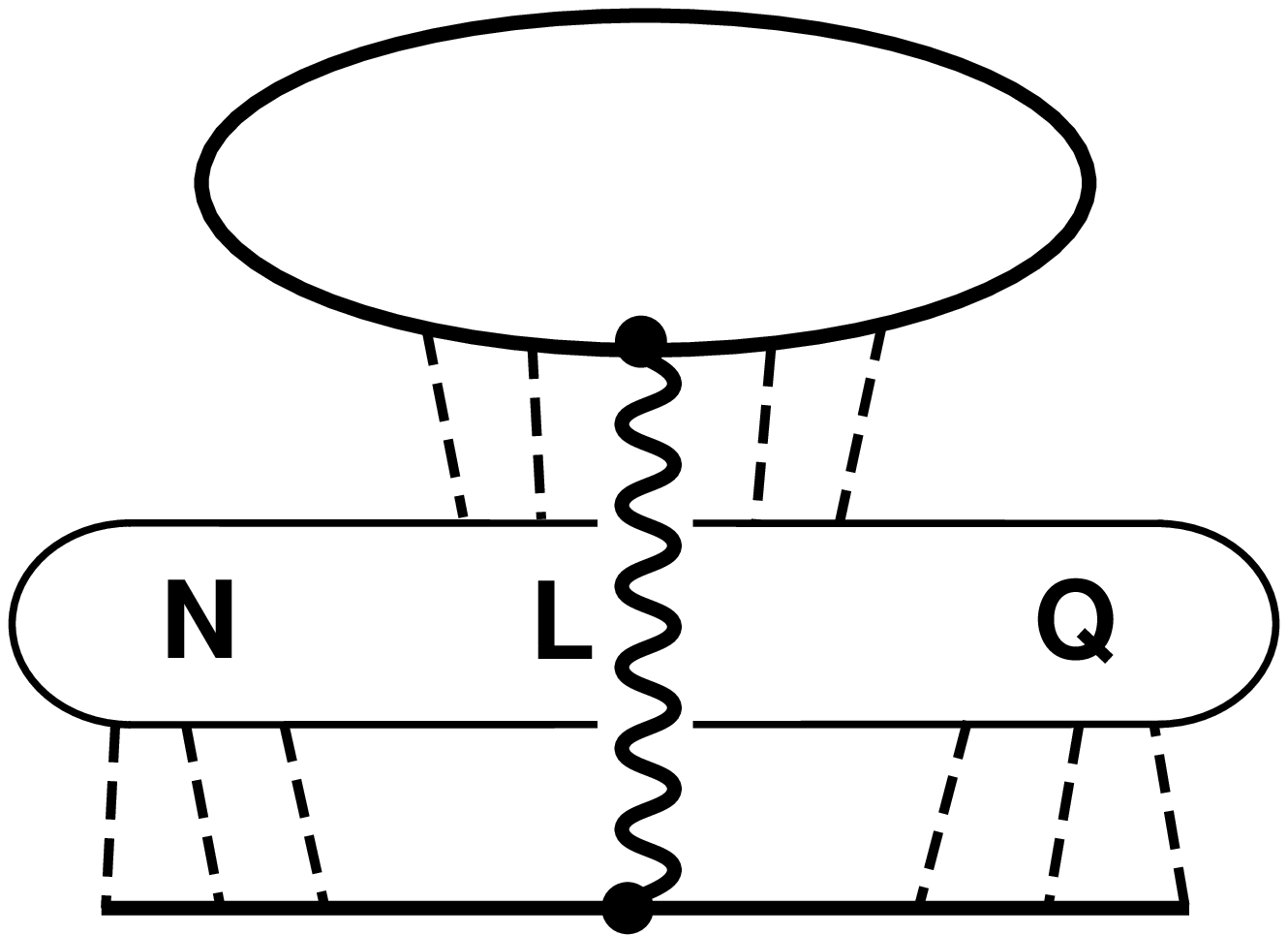}}
\vspace*{13pt}
\fcaption{General structure of the Hartree-type
diagram. Scheme of \protect{$NLQ$}-partitioning.}
\end{figure}
%%%%%%%%%%%%%%%%%%%%%%%%%%%%%%%%%%%%%%%%%%%%%%%%%%%%%%%%%%%%%%%%%%%%%%%%%%%%

We characterize
Hartree-type diagrams by three numbers $N$, $L$ and $Q$, where
$N$ and $Q$ have the same meaning as before and $L$ is the number
of successive disorder scattering processes on a fermionic loop.
Analitical expression for Hartree-type contributions has the form
\begin{equation}
 - B^{(D)}_{NLQ}V^{M} T_c
   \sum_{{\rm i}p_m}
   \lambda (0)\tau_{3}
   \left(G({\rm i}p_s)\tau_{3}\right)^{N}
   {\rm Sp}\left\{\left(G({\rm i}p_m)\tau_{3}\right)^{L+1}\right\}
   \left(G({\rm i}p_s)\tau_{3}\right)^{Q},
\label{eq6.32}
\end{equation}
$B^{(D)}_{NLQ}(c)$ is the product of the cumulants for particular diagram
$(D)$ with the triple $NLQ$ fixed, and diagram $(D)$ belongs now to the
set $\{{\bf D}\}$ of Hartree-type diagrams.
Then making the approximation~(\ref{eq6.10}) we have the
contributions to the equations on $\tilde{Z}({\rm i}p_s)$
and $\chi({\rm i}p_s)$ of the form
\begin{equation}
     2{\rm i}
     \left\{ f_{eoo}(c,\tilde{V})+f_{ooe}(c,\tilde{V}) \right\}
                   \tilde{\lambda}(0) {\cal U}
                   {\rm Sign}({\rm i}p_s),
\label{eq6.33}
\end{equation}
\begin{equation}
     2{\rm i}
     \left\{ f_{eoe}(c,\tilde{V})+f_{ooo}(c,\tilde{V}) \right\}
                   \tilde{\lambda}(0) {\cal U}
\label{eq6.34}
\end{equation}
and to the anomalous part of the self-energy:
\begin{equation}
    -2 \left\{ f_{eoo}(c,\tilde{V}) + f_{ooe}(c,\tilde{V})\right\}
    \tilde{\lambda}(0) {\cal U}
              \frac{\Delta({\rm i}p_s)}{|p_s|}.
\label{eq6.35}
\end{equation}
Here
\begin{equation}
  {\cal U}=\pi T_c \sum_{|p_s|\le W} 1,
\label{eq6.36}
\end{equation}
and we have introduced the cut-off $W$, with $W$ being of order
of the bandwidth. Note also that the functions $f_{\alpha\beta\gamma}$
in the expressions~(\ref{eq6.33}--\ref{eq6.35}) are defined
by Eq.(\ref{eq6.21}) where inner sum runs over Hartree set of diagrams.
However the results of the summation for both Hartree- and Fock-type diagrams
appear to be the same as may be demonstrated by strightforward manipulations
using the method of generating functions. Moreover, it is not even necessary
to calculate the coefficient functions~(\ref{eq7.21}) for Hartree-type diagrams
because Hartree-type diagrams do not explicitly contribute to the final
$T_c$ expression within the approximations used in this Section.

Finally, taking into account~(\ref{eq6.20},\ref{eq6.22},\ref{eq6.31})
and~(\ref{eq6.33}--\ref{eq6.35}) one obtains the following set of the
equations:
\begin{eqnarray}
& \displaystyle
   {\rm i}p_s[1-\tilde{Z}({\rm i}p_s)] =
     2{\rm i}
     \left\{ f_{eoo}(c,\tilde{V})+f_{ooe}(c,\tilde{V}) \right\}
                   \tilde{\lambda}(0) {\cal U}
                   {\rm Sign}({\rm i}p_s)
&
\nonumber \\
& \displaystyle
   - {\rm i} \left\{ f_{eee}(c,\tilde{V})+f_{oeo}(c,\tilde{V}) \right\}
                   \pi T_{c} \sum_{{\rm i}p_m}
                   \tilde{\lambda}({\rm i}p_s-{\rm i}p_m)
                   {\rm Sign}({\rm i}p_m)
&
\nonumber \\
& \displaystyle
   - {\rm i} \left\{ f_{eoo}(c,\tilde{V})+f_{ooe}(c,\tilde{V}) \right\}
                   \pi T_{c} \sum_{{\rm i}p_m}
                   \tilde{\lambda}({\rm i}p_s-{\rm i}p_m)
                   {\rm Sign}({\rm i}p_s)
&
\label{eq6.37}
\end{eqnarray}

\begin{eqnarray}
& \displaystyle
   \chi({\rm i}p_s)] =
     2{\rm i}
     \left\{ f_{eoe}(c,\tilde{V})+f_{ooo}(c,\tilde{V}) \right\}
                   \tilde{\lambda}(0) {\cal U}
&
\nonumber \\
& \displaystyle
   - {\rm i} \left\{ f_{ooo}(c,\tilde{V})+f_{eoe}(c,\tilde{V}) \right\}
                   \pi T_{c} \sum_{{\rm i}p_m}
                   \tilde{\lambda}({\rm i}p_s-{\rm i}p_m)
&
\nonumber \\
& \displaystyle
   - {\rm i} \left\{ f_{eeo}(c,\tilde{V})+f_{oee}(c,\tilde{V}) \right\}
                   {\rm Sign}({\rm i}p_s)
                   \pi T_{c} \sum_{{\rm i}p_m}
                   \tilde{\lambda}({\rm i}p_s-{\rm i}p_m)
                   {\rm Sign}({\rm i}p_m)
&
\label{eq6.38}
\end{eqnarray}

\begin{eqnarray}
& \displaystyle
   \tilde{Z}({\rm i}p_s)\Delta({\rm i}p_s) =
    -2 \left\{ f_{eoo}(c,\tilde{V}) + f_{ooe}(c,\tilde{V})\right\}
    \tilde{\lambda}(0) {\cal U}
              \frac{\Delta({\rm i}p_s)}{|p_s|}
&
\nonumber \\
& \displaystyle
         \left\{ f_{eee}(c,\tilde{V}) - f_{eee}(c,\tilde{V})\right\}
         \pi T_{c} \sum_{{\rm i}p_m}
                   \tilde{\lambda}({\rm i}p_s - {\rm i}p_m)
                   \frac{\Delta({\rm i}p_m)}{|p_m|} +
&
\nonumber \\
& \displaystyle
   \left\{ f_{eoo}(c,\tilde{V}) + f_{ooe}(c,\tilde{V})\right\}
   \frac{\Delta({\rm i}p_s)}{|p_s|}
   \pi T_{c} \sum_{{\rm i}p_m}
             \tilde{\lambda}({\rm i}p_s - {\rm i}p_m) +
&
\nonumber \\
& \displaystyle
   2 f_{oeo}(c,\tilde{V})
   \frac{\Delta({\rm i}p_s)}{|p_s|}
   \pi T_{c} \sum_{{\rm i}p_m}
             \tilde{\lambda}({\rm i}p_s - {\rm i}p_m)
             {\rm Sign}({\rm i}p_s)
             {\rm Sign}({\rm i}p_m)
&
\label{eq6.39}
\end{eqnarray}
Having been solved,
the Eq.(\ref{eq6.37}) yields for $\tilde{Z}({\rm i}p_s)$ the expression
\begin{eqnarray}
& \displaystyle
   \tilde{Z}({\rm i}p_s) =  1 -
     2\left\{ f_{eoo}(c,\tilde{V})+f_{ooe}(c,\tilde{V}) \right\}
           \tilde{\lambda}(0) {\cal U}
           \frac{1}{|p_s|}
&
\nonumber \\
& \displaystyle
   + \left\{ f_{eee}(c,\tilde{V})+f_{oeo}(c,\tilde{V}) \right\}
         \frac{1}{|p_s|}
          \pi T_{c} \sum_{{\rm i}p_m}
          \tilde{\lambda}({\rm i}p_s-{\rm i}p_m)
          {\rm Sign}({\rm i}p_m){\rm Sign}({\rm i}p_s)
&
\nonumber \\
& \displaystyle
   + \left\{ f_{eoo}(c,\tilde{V})+f_{ooe}(c,\tilde{V}) \right\}
        \frac{1}{|p_s|}
        \pi T_{c} \sum_{{\rm i}p_m}
        \tilde{\lambda}({\rm i}p_s-{\rm i}p_m) .
&
\label{eq6.40}
\end{eqnarray}
Substitution of (\ref{eq6.40}) into (\ref{eq6.39}) then
leads to the  equation  for
the gap function:
\begin{eqnarray}
& \displaystyle
   \Delta({\rm i}p_s) =
         \pi T_{c} \sum_{{\rm i}p_m}
                   \lambda^{eff}({\rm i}p_s - {\rm i}p_m)
                   \frac{\Delta({\rm i}p_m)}{|p_m|}
&
\nonumber \\
& \displaystyle
  -\frac{\Delta({\rm i}p_s)}{|p_s|}
   \pi T_{c} \sum_{{\rm i}p_m}
             \lambda^{eff}({\rm i}p_s - {\rm i}p_m)
             {\rm Sign}({\rm i}p_s)
             {\rm Sign}({\rm i}p_m) ,
&
\label{eq6.41}
\end{eqnarray}
where
\begin{eqnarray}
   \lambda^{eff}({\rm i}\omega_n) & = &
       \kappa(c,\tilde{V})N(0) \lambda({\rm i}\omega_n)
\nonumber \\
     & = &
       \kappa(c,\tilde{V}) \tilde{\lambda}({\rm i}\omega_n),
\label{eq6.42}
\end{eqnarray}
\begin{equation}
   \kappa(c,\tilde{V}) =
         f_{eee}(c,\tilde{V}) - f_{eoe}(c,\tilde{V}) ,
\label{eq6.43}
\end{equation}
with $\lambda^{eff}({\rm i}\omega_n)$ being
EPI kernel renormalized by disorder scattering
and $\kappa(c,\tilde{V})$ being the renormalization factor depending
on the approximation used for disorder self-energy.
As we have stated earlier Hartree-type contributions do not explicitly
appear in the equation for the gap function~(\ref{eq6.41}), which is
the result of compensation between Hartree contributions to the
function $\tilde{Z}({\rm i}p_s)$ and those of to the
anomalous self-energy $\phi({\rm i}p_s)$. Of course, such compensation
does not take place in general\cite{c16,c17,c18}, and is rather due to
the character of the computational simplifications used, of which
the local approximation (e.g., the limit $d=\infty$) is principal.

  Comparing (\ref{eq6.41}) with the standard result for
$\Delta({\rm i}p_s)$  which reads\cite{c4}
\begin{eqnarray}
& \displaystyle
   \Delta({\rm i}p_s) =
         \pi T_{c} \sum_{{\rm i}p_m}
                   \tilde{\lambda}({\rm i}p_s - {\rm i}p_m)
                   \frac{\Delta({\rm i}p_m)}{|p_m|} -
&
\nonumber \\
& \displaystyle
   \frac{\Delta({\rm i}p_s)}{|p_s|}
   \pi T_{c} \sum_{{\rm i}p_m}
             \tilde{\lambda}({\rm i}p_s - {\rm i}p_m)
             {\rm Sign}({\rm i}p_s)
             {\rm Sign}({\rm i}p_m)
&
\label{eq6.44}
\end{eqnarray}
we see that the two equations differ in the only respect:
the first contains $\lambda^{eff}({\rm i}\omega_n)$ and
the second, $\tilde{\lambda}({\rm i}\omega_n)$;
the form of the equation remains intact. Then using square-well
approximation\cite{c4} for $\lambda_{00}({\rm i}p_s-{\rm i}p_m)$
in the first term of Eq.(\ref{eq6.41})
\begin{equation}
   \lambda_{00}({\rm i}p_s-{\rm i}p_m)\longrightarrow
   \lambda\theta(\omega_{1}-|p_s|)\theta(\omega_{1}-|p_m|)
\label{eq6.45}
\end{equation}
and a slightly different version in the  contribution  which  comes
from the expression for $\tilde{Z}({\rm i}p_s)$
\begin{equation}
   \lambda_{00}({\rm i}p_s-{\rm i}p_m)\longrightarrow
   \lambda\theta(\omega_{1}-|p_s-p_m|)
\label{eq6.46}
\end{equation}
we immediately obtain for $T_{c}$
\begin{equation}
   T_{c} = 1.13 \omega_{D}
               \exp\left\{ -\frac{1+\lambda^{eff}}{\lambda^{eff}} \right\}
\label{eq6.47}
\end{equation}
where
\begin{equation}
   \lambda^{eff} = \kappa(c,\tilde{V})N(0)\lambda_{b}, \quad
   \lambda_{b} = 2 \int^{\infty}_{0} {\rm d}\Omega
        \frac{\alpha^{2}F_{00}(\epsilon^{*}\epsilon^{*};\Omega)}{\Omega}
\label{eq6.48}
\end{equation}

  The Eq.(\ref{eq6.41}) enables one to consider  more  general  situation
where, along with EPI mechanism, another pairing  mechanism  exists
which will be called, for definiteness, ``excitonic''.  Here  we
treat only the case of weak coupling in both  EPI  and ``excitonic''
channels.  The  corresponding  interaction   may   originate   from
electron-electron interaction, but should not constitute  its  main
part. For instance, one may  think  of  the  Weber  model\cite{c56} where
strong Mott-Hubbard correlations in the $d$-system  and  pairing  via
crystal-field excitations are considered.

  In this situation ``excitonic'' channel leads to the appearance of
the second scale in the effective interaction  kernel  and  to  the
logarithmic corrections to $\lambda^{eff}$.
Within square-well model  for  the
interaction kernel, $\lambda_{00}({\rm i}p_s-{\rm i}p_m)$
may be approximated as
\begin{equation}
   \lambda_{00}({\rm i}p_s-{\rm i}p_m)\longrightarrow
   \lambda_{1}\theta(\omega_{1}-|p_s|)\theta(\omega_{1}-|p_m|)+
   \lambda_{2}\theta(\omega_{2}-|p_s|)\theta(\omega_{2}-|p_m|) ,
\label{eq6.49}
\end{equation}
where $\omega_{1}$ and $\omega_{2}$ ($\omega_{1}<\omega_{2}$)
are  characteristic  phonon  and  exciton
energy scales, $\lambda_{1}$ and $\lambda_{2}$ are
corresponding coupling  parameters  of
dimension energy. Substituting (\ref{eq6.49}) into (\ref{eq6.41}) we obtain
\begin{equation}
   T_{c}=1.13\omega_{1}\exp\left\{-\frac{1}{\lambda^{eff}}\right\}
\label{eq6.50}
\end{equation}
with $\lambda^{eff}$ defined by an expression of the form
\begin{equation}
    \lambda^{eff}=\tilde{\lambda_{1}}+
        \frac{
             \tilde{\lambda_{2}}
             }{
             1-\tilde{\lambda_{2}}\ln|\omega_{2}/\omega_{1}|
             }
\label{eq6.51}
\end{equation}
and
\begin{equation}
   \tilde{\lambda_{1}} = \kappa(c,\tilde{V})N(0)\lambda_{1},
   \qquad
   \tilde{\lambda_{2}} = \kappa(c,\tilde{V})N(0)\lambda_{2}.
\label{eq6.52}
\end{equation}
The  expression (\ref{eq6.51}) retains its usual form, but up to a
renormalization of bare couplings by disorder scattering, which is
absorbed in the factor $\kappa(c,\tilde{V})$.

  Let us consider now disorder renormalizations of dilute
paramagnetic impurity contributions to the gap function and $T_c$
equations. In this case the equation (\ref{eq6.41}) has to be changed to
\begin{eqnarray}
& \displaystyle
   \Delta({\rm i}p_s) =
         \pi T_{c} \sum_{{\rm i}p_m}
                   \lambda^{eff}({\rm i}p_s - {\rm i}p_m)
                   \frac{\Delta({\rm i}p_m)}{|p_m|} -
&
\nonumber \\
& \displaystyle
   \frac{\Delta({\rm i}p_s)}{|p_s|}
   \pi T_{c} \sum_{{\rm i}p_m}
             \lambda^{eff}({\rm i}p_s - {\rm i}p_m)
             {\rm Sign}({\rm i}p_s)
             {\rm Sign}({\rm i}p_m)-
   2\gamma_{P}^{eff}
   \frac{\Delta({\rm i}p_s)}{|p_s|} ,
&
\label{eq6.53}
\end{eqnarray}
where
\begin{equation}
   \gamma_{P}=c_{P}V_{P}^{2}S(S+1),
   \qquad
   \tilde{\gamma_{P}}= N(0)\gamma_{P} ,
\label{eq6.54}
\end{equation}
\begin{equation}
   \gamma^{eff}=\kappa(c,\tilde{V})\tilde{\gamma_{P}} ,
\label{eq6.55}
\end{equation}
with $c_{P}$ being the concentration of the paramagnetic impurities,
$V_{P}$ being paramagnetic impurity scattering  potential,
$S$  being  the
impurity spin, and some details of the derivation of (\ref{eq6.53})
are given in Appendix~B.
We see that,  analogously  to  the  situations
without paramagnetic impurities, effects of  non-magnetic  disorder
lead to the replacement of bare couplings by the renormalized  ones
in the gap function equation and the renormalization factor is  the
same for all bare couplings. The gap function equation
preserves its form (compare with the case where non-magnetic
disorder is absent), and the results for the critical concentration
of the paramagnetic impurities and for the $T_{c}$ suppression  may  be
obtained by standard means.\cite{c4}
Thus we  have  the  equation
determining the critical concentration $n_{c}$ of the paramagnetic
impurities
\begin{equation}
   \gamma_{c}=(1+\lambda^{eff})T_{c0}/2.26 ,
\label{eq6.56}
\end{equation}
where $T_{c0}$ is the critical temperature
with $\gamma^{eff}=0$, $\gamma_{c}=n_{c}V_{P}^{2}S(S+1)$,
and an approximate expression
\begin{equation}
   \ln\left(\frac{T_{c0}}{T_{c}}\right)=
       \Psi(\alpha+1/2)-
       \Psi(1/2) ,
\label{eq6.57}
\end{equation}
gives the $T_{c}$ suppression by dilute paramagnetic impurities,
\begin{equation}
  \alpha=\frac{\gamma^{eff}_{P}}{\pi T_{c}(1+\lambda^{eff})}
        =0.14\frac{\gamma^{eff}_{P}}{\gamma_{c}}
             \frac{T_{c0}}{T_{c}}
\label{eq6.58}
\end{equation}
is  pair-breaking parameter and $\Psi (z)$, the digamma function.

\section{Calculations of the  Renormalization  Factor  within  Conserving
         Single-site Approxima\-tions to the Alloy Self-energy.}
\noindent
In this section we demonstrate that within an arbitrary conserving single-site
approximation determining the self-energy $\Sigma^{e}({\rm i}p_s)$ in the
normal phase of a superconductor the renormalization factor
$\kappa(c,\tilde{V})$ (\ref{eq6.43}) can be expressed in terms of
suitably defined generating  function  for $\Sigma^{e}({\rm i}p_s)$.
To this end we recall that the coefficient $B^{(D)}_{NLQ}(c)$
in (\ref{eq6.21}) is nothing but a product of all the  cumulants  resulting
from disorder-vertex contributions for the diagram ($D$) given, and these
cumulants are the same which enter $\Sigma^{e}({\rm i}p_s)$,
for the EPI renormalizations
are consistent with the approximation to $\Sigma^{e}({\rm i}p_s)$
by construction.

  Let us introduce $Q$-function, generating function connected  with
the number of successive  disorder  scattering  processes,  by  the
following expression
\begin{equation}
   Q(z_1,z_2,z_3) =
        \sum_{NLQ} \sum_{\{D\}} B^{(D)}_{NLQ}(c) z^N_1 z^L_2 z^Q_3 .
\label{eq7.1}
\end{equation}
Here $B^{(D)}_{NLQ}(c)$ is the product of the cumulants
for the Fock-type graph ($D$) having the triple $NLQ$ fixed,
inner sum runs over all graphs of $NLQ$-family
and  outer  sum  over  all  positive  integers $NLQ$.  Clearly,  the
functions $f_{\alpha\beta\gamma}(c,\tilde{V})$
can be expressed through $Q(z_1,z_2,z_3)$ by picking
out even and odd parts of the latter with respect to  each  of  its
three arguments, that is
\begin{equation}
   f_{\alpha\beta\gamma}(z_1,z_2,z_3) = \frac{1}{2^3}
        \sum_{\eta_1\eta_2\eta_3}
         \eta^{\alpha}_1
         \eta^{\beta}_2
         \eta^{\gamma}_3
         Q(\eta_1 z_1,\eta_2 z_2,\eta_3 z_3) ,
\label{eq7.2}
\end{equation}
where $\eta_i=\pm1, \quad (i=1,2,3)$, and
\begin{equation}
   f_{\alpha\beta\gamma}(c,\tilde{V}) =
     \left.
         f_{\alpha\beta\gamma}(z_1,z_2,z_3)
     \right|_{z_1=z_2=z_3= -{\rm i}\tilde{V}} .
\label{eq7.3}
\end{equation}
According to its definition the function $Q(z_1,z_2,z_3)$ is the product
of analogous functions for vertex corrections
\begin{equation}
   Q(z_1,z_2,z_3)=  Q_{3}(z_{1},z_{2})Q_{3}(z_{2},z_{3})
                    \left\{
                          1 + Q_{6}(z_{1},z_{2},z_{3})
                    \right\} ,
\label{eq7.4}
\end{equation}
where we have introduced $Q$-functions  for  irreducible  three-  and
six-leg  vertices. $Q_3(z_1,z_2)$  can  be  further  expressed  through
$Q_4(z_1,z_2)$ function, generating $Q$-function for four-leg vertex, as
\begin{equation}
   Q_{3}(z_{1},z_{2}) = 1 + Q_{3}(z_{1},z_{2})Q_{4}(z_{1},z_{2})  ,
\label{eq7.5}
\end{equation}
\begin{equation}
   Q_{3}(z_{1},z_{2}) =
      \left(
          1 - Q_{4}(z_{1},z_{2})
      \right)^{-1} .
\label{eq7.6}
\end{equation}

  Now we find the relations between the  disorder  self-energy
$\Sigma^{e}({\rm i}p_s)$ and generating functions introdiced above.
From Eqs.(\ref{eqC.4},\ref{eqC.10},\ref{eqC.12})
of Appendix~C we have
\begin{eqnarray}
   \Sigma^{e}[G] & = & V P_2(VG) ,
\label{eq7.7} \\
   U[G,G'] & = & V^{2} P_{4}(VG,VG') ,
\label{eq7.8} \\
   U[G,G',G''] & = & V^{3} P_{6}(VG,VG',VG'') ,
\label{eq7.9}
\end{eqnarray}
where
\begin{equation}
   P_2(z) = \sum_{n=0}^{\infty}q_n(c)z^n ,
\label{eq7.10}
\end{equation}
and $q_n(c)$ are renormalized cumulants for the single-site
approximation being used. (Note that we reffere to the generating
functions introduced in Appendix~C as to generating $P$-functions.)

Generating functions for four- and six-leg vertices can be
expressed as
\begin{eqnarray}
   P_{4}(z_{1},z_{2}) & = & \displaystyle
         \frac{1}{2\pi {\rm i}}
         \int_{C} {\rm d}z
              \frac{P_{2}(z)}{(z-z_{1})(z-z_{2})}
\nonumber \\
         & = & \displaystyle
         \frac{P_{2}(z_{1}) - P_{2}(z_{2})}{z_{1} - z_{2}} ,
\label{eq7.11} \\
   P_{6}(z_{1},z_{2},z_{3}) & =  & \displaystyle
         \frac{1}{2\pi {\rm i}}
         \int_{C} {\rm d}z
              \frac{P_{2}(z)}{(z-z_{1})(z-z_{2})(z-z_{3})} ,
\label{eq7.12}
\end{eqnarray}
where contours $C$ in the integrals encircle only the peculiarities
of the denominator of integrands. Noting that the irreducible
vertex of $2n$'th order has $n$ potential lines more than full electron
lines and using (\ref{eq7.8},\ref{eq7.9}) one obtains the relations:
\begin{eqnarray}
   Q_4(z_1,z_2) & = & z_1 z_2 P_4(z_1,z_2) ,
\label{eq7.13} \\
   Q_6(z_1,z_2,z_3) & = & z_1 z_2 z_3 P_6(z_1,z_2,z_3) .
\label{eq7.14}
\end{eqnarray}
Finally, we write down the expression for $\kappa(c,\tilde{V})$
once  more for reader convinience:
\begin{equation}
\kappa(c,\tilde{V})= f_{eee}(c,\tilde{V})-f_{oeo}(c,\tilde{V}) .
\label{eq7.15}
\end{equation}
So, we see that the knowledge of $P$-generating function for $\Sigma^{e}[G]$
is sufficient  to  calculate  renormalization  factor $\kappa(c,\tilde{V})$.

  The  expression (\ref{eq7.15}) for $\kappa(c,\tilde{V})$ is not
quite convenient to work with directly and may be transformed further.
To this end, we introduce, in tight analogy with the Bethe-Salpeter
equation (\ref{eq2.24}), the generating function $\Gamma_4(z_1,z_2)$ by
\begin{equation}
  \Gamma_4(z_1,z_2)=P_4(z_1,z_2)+z_1 z_2 P_4(z_1,z_2)\Gamma_4(z_1,z_2) .
\label{eq7.16}
\end{equation}
Then from (\ref{eq7.5},\ref{eq7.6},\ref{eq7.13})
and (\ref{eq7.16}), it follows
\begin{equation}
   Q_3(z_1,z_2)=1 + z_1 z_2 \Gamma_4(z_1,z_2)  ,
\label{eq7.17}
\end{equation}
and
\begin{equation}
   Q_3(z_1,z_2) = \frac{\Gamma_4(z_1,z_2)}{P_4(z_1,z_2)} .
\label{eq7.18}
\end{equation}
Using (\ref{eq7.14},\ref{eq7.17}) we can
rewrite expression (\ref{eq7.4}) for $Q(z_1,z_2,z_3)$ as
\begin{equation}
 Q(z_1,z_2,z_3)=1+z_1 z_2\Gamma_4(z_1,z_2)+z_2 z_3\Gamma_4(z_2,z_3)
                 +z_1 z_2 z_3 \delta Q(z_1,z_2,z_3) ,
\label{eq7.19}
\end{equation}
where
\begin{equation}
 \delta Q(z_1,z_2,z_3)=z_2\Gamma_4(z_1,z_2)\Gamma_4(z_2,z_3)
                      +Q_3(z_1,z_2)P_6(z_1,z_2,z_3)Q_3(z_2,z_3) .
\label{eq7.20}
\end{equation}
To proceed further we use  the  Ward-type  identity  which  relates
$P$-functions for vertices of different order.
From  (\ref{eqC.17},\ref{eqC.19}) of
Appendix~C the function $P_6(z_1,z_2,z_3)$ can be expressed as
\begin{equation}
   P_6(z_1,z_2,z_3)=\frac{P_4(z_1,z_2)-P_4(z_3,z_2)}{z_1-z_3} .
\label{eq7.21}
\end{equation}
In its turn, from (\ref{eq7.16}) one has
\begin{equation}
   P_4(z_1,z_2)=\left(\Gamma_4^{-1}(z_1,z_2) + z_1 z_2\right)^{-1}
\label{eq7.22}
\end{equation}
and using then (\ref{eq7.16},\ref{eq7.22}) one can obtain
\begin{eqnarray}
& \displaystyle
P_4(z_1,z_2)-P_4(z_3,z_2)=
  \frac{
       P_4(z_1,z_2)P_4(z_3,z_2)
       }{
       \Gamma_4(z_1,z_2)\Gamma_4(z_3,z_2)
       } \times &
\nonumber \\
& \displaystyle
       \times
       \left(
       \partial\Gamma(z_1,z_2,z_3)-z_2\Gamma_4(z_1,z_2)\Gamma_4(z_3,z_2)
       \right)(z_1-z_2) , &
\label{eq7.23}
\end{eqnarray}
where
\begin{equation}
   \partial\Gamma(z_1,z_2,z_3)=
           \frac{
                \Gamma_4(z_1,z_2)-\Gamma_4(z_3,z_2)
                }{
                z_1-z_3
                } .
\label{eq7.24}
\end{equation}
Substituting now the expression (\ref{eq7.23}) into (\ref{eq7.21}) one has
\begin{eqnarray}
& \displaystyle
 P_6(z_1,z_2,z_3)=
 \frac{
      P_4(z_1,z_2)P_4(z_3,z_2)
      }{
      \Gamma_4(z_1,z_2)\Gamma_4(z_3,z_2)
      } \times &
\nonumber \\
& \displaystyle
      \left(
      \partial\Gamma(z_1,z_2,z_3)-z_2\Gamma_4(z_1,z_2)\Gamma_4(z_3,z_2)
      \right) . &
\label{eq7.25}
\end{eqnarray}
Recalling (\ref{eq7.18}) and Eq.(\ref{eq7.20}) for $\delta Q(z_1,z_2,z_3)$,
we can write
\begin{equation}
  \delta Q(z_1,z_2,z_3) =\partial\Gamma(z_1,z_2,z_3) ,
\label{eq7.26}
\end{equation}
so that $Q(z_1,z_2,z_3)$ becomes
\begin{equation}
  Q(z_1,z_2,z_3)=1+z_1 z_2\Gamma_4(z_1,z_2)+z_2 z_3\Gamma_4(z_2,z_3)
                  +z_1 z_2 z_3\partial\Gamma(z_1,z_2,z_3) .
\label{eq7.27}
\end{equation}
As for the combination  $f_{eee}(z_1,z_2,z_3)-f_{oeo}(z_1,z_2,z_3)$,
a  little algebra with the use of (\ref{eq7.2}) yields
\begin{equation}
\frac{1}{4}
\left\{
      Q(z_1,z_2,-z_3)+Q(-z_1,-z_2,z_3)+Q(z_1,-z_2,-z_3)+Q(-z_1,z_2,z_3)
\right\} .
\label{eq7.28}
\end{equation}
Noting the symmetry $Q(z_1,z_2,z_3)=Q(z_3,z_2,z_1)$ one has
\begin{eqnarray}
\kappa(c,\tilde{V})&=&\left.\frac{1}{2}\{Q(z,-z,-z)+
                      Q(-z,z,z)\}\right|_{z=-{\rm i}\tilde{V}}
\label{eq7.29} \\
                   &=&\left.\frac{1}{2}\{Q(z,-z,-z)+
                      Q(z,z,-z)\}\right|_{z=-{\rm i}\tilde{V}}
\label{eq7.30} \\
                   &=&\left.\frac{1}{2}\{Q(z_1,z_2,z_2)+
                      Q(z_1,z_1,z_2)\}\right|_{z_1=-z_2=-{\rm i}\tilde{V}} .
\label{eq7.31}
\end{eqnarray}
Substitution of (\ref{eq7.27}) into (\ref{eq7.31}) leads
\begin{eqnarray}
&
\kappa(c,\tilde{V})=1+\frac{1}{2(z_1-z_2)^2}
    \bigl\{z_1^3(z_1-z_2)\Gamma_4(z_1,z_1)-z_2^3\Gamma_4(z_2,z_2)+ &
\nonumber \\
&   +2z_1 z_2(z_1-z_2)^2\Gamma_4(z_1,z_2)\bigr\}
     \left.\right|_{z_1=z_2=-{\rm i}\tilde{V}}  &
\label{eq7.32}
\end{eqnarray}
and on a line $z_1=-z_2$  this expression further reduces to
\begin{equation}
\kappa(c,\tilde{V})=1+
   \frac{
        z_1^4\Gamma_4(z_1,z_1)-2z_1^2 z_2^2\Gamma_4(z_1,z_2) +
        z_2^4\Gamma_4(z_2,z_2)
        }{
        (z_1-z_2)^2
        }
\left.\right|_{z_1=-z_2=-{\rm i}\tilde{V}} ,
\label{eq7.33}
\end{equation}
\begin{equation}
\kappa(c,\tilde{V})=1+
         {\tilde{V}}^2{\rm Im}_{z_1}{\rm Im}_{z_2}\Gamma_4(z_1,z_2)
      \left.\right|_{z_1=z_2={\rm i}\tilde{V}} ,
\label{eq7.34}
\end{equation}
where ${\rm Im}_z$  means picking out ``imaginary'' part with respect to a
complex variable $z$, which may be written as
\begin{equation}
  {\rm Im}_z f(\ldots,z,\ldots)=
  \frac{1}{2{\rm i}}\left(f(\ldots,z,\ldots)-f(\ldots,z^*,\ldots)\right)
\label{eq7.35}
\end{equation}
and points stand for variables other than $z$, and we retain explicit
dependence on $z_1$ and $z_2$ in (\ref{eq7.33},\ref{eq7.34})
for further use when analyzing the renormalizations within CPA.

  Now everything is ready to evaluate $\kappa(c,\tilde{V})$
within certain widely
known single site approximations. To be concrete we consider  three
such approximations: VCA, ATA and CPA.

\subsection{Virtual Crystal Approximation and EPI Renormalizations.}
\noindent
  For the case of VCA the self-energy reads\cite{c37}
\begin{equation}
     \Sigma^{e}_{VCA}({\rm i}p_s)=cV+c(1-c)VG({\rm i}p_s)
\label{eq7.36}
\end{equation}
and cumulants $q_n(c)$ are
\begin{equation}
    q_0(c)=c,\qquad  q_1(c)=c(1-c),\qquad q_n(c)=0\quad n>1 .
\label{eq7.37}
\end{equation}
The generating function $P_2(c)$ within this approximation has the
form
\begin{equation}
      P_2(z) = c + c(1-c) z
\label{eq7.38}
\end{equation}
and the use of (\ref{eq7.11}) gives
\begin{equation}
   P_4(z_1,z_2) = c(1-c) .
\label{eq7.39}
\end{equation}
Then, from (\ref{eq7.16}) and within the VCA accuracy, we have
\begin{equation}
   \Gamma_4(z_1,z_2) = P_4(z_1,z_2) = c(1-c)
\label{eq7.40}
\end{equation}
and from (\ref{eq7.33},\ref{eq7.34}) it follows
\begin{equation}
    \kappa(c,\tilde{V}) = 1 .
\label{eq7.41}
\end{equation}
So, there is no disorder contributions to  the  renormalization  of
EPI coupling, except for usual  ones  through  one  particle  state
density.

\subsection{Renormalizations of EPI within Average T-matrix Approximation.}
\noindent
  For ATA approximation the expression for $\Sigma^{e}_{ATA}({\rm i}p_s)$
has  the  form\cite{c37}
\begin{equation}
\Sigma^{e}_{ATA}({\rm i}p_s)=\frac{cV}{1-(1-c)VG({\rm i}p_s)} .
\label{eq7.42}
\end{equation}
General expression for $q_n(c)$ within ATA reads
\begin{equation}
       q_n(c)=c(1-c)^n,\qquad n\ge 1
\label{eq7.43}
\end{equation}
and the generating function $P_2(c)$ is
\begin{equation}
   P_2(z) =c\left[1 - (1-c)z\right]^{-1} .
\label{eq7.44}
\end{equation}
Then, similarly to the previous case, we obtain
\begin{equation}
   P_4(z_1,z_2) =c(1-c)
       \left[
       1 - (1-c)z_1
       \right]^{-1}
       \left[
       1 - (1-c)z_2
       \right]^{-1}
\label{eq7.45}
\end{equation}
and to required accuracy
\begin{equation}
   \Gamma_4(z_1,z_2) = P_4(z_1,z_2) ,
\label{eq7.46}
\end{equation}
so that in this case the result for $\kappa(c,\tilde{V})$ reads
\begin{equation}
 \kappa(c,\tilde{V})=1+
        \frac{
             c(1-c)^3{\tilde{V}}^4
             }{
             (1+(1-c)^2{\tilde{V}}^2)^2
             } .
\label{eq7.47}
\end{equation}

  Let us consider now the so called split band limit, $V\rightarrow\infty$.
Supposing that the chemical potential lies in the lower  A-subband,
that is
\begin{equation}
   \lim_{\tilde{V}\to\infty}N_B(0) = 0 .
\label{eq7.48}
\end{equation}
Introducing projected one  particle  state  densities  for  A-
and B-type atoms
\begin{equation}
     N(\epsilon) = (1-c)N_A(\epsilon)+cN_B(\epsilon) ,
\label{eq7.49}
\end{equation}
we find in this limit that
\begin{equation}
  \kappa(c,\infty)=\lim_{\tilde{V}\to\infty}\kappa(c,\tilde{V})=(1-c)^{-1}
\label{eq7.50}
\end{equation}
and effective EPI coupling becomes
\begin{equation}
  \lambda^{eff}=\kappa(c,\infty)N(0)\lambda=\lambda N_A(0) .
\label{eq7.51}
\end{equation}
Such a limiting behavior is fully transparent and might be expected
on physical grounds. Indeed, suppose $\tilde{V}$ is so large  that
electrons
can hop and  interact  with  phonons  only  on  A-type  atoms  (the
probability of finding an electron on B-type atom is  exponentially
small in this  case,  so  that  one  may  completely  neglect  this
process). Then two effects arise. The first is the  renormalization
of the electron state density  for  A-type  atoms  by  both  B-type
impurities and EPI. The second is the renormalization  of  chemical
potential which is due to the fact that B-type atoms are no  longer
accessible for the electron motion. The electrons can now propagate
in those states which belong to the renormalized  band  for  A-type
atoms and the pairing occurs for these states only. The temperature
$T_{c}$  depends solely on the properties of A-band states but not
on the
(macroscopical) size of superconductive cluster  and  this  is  the
reason why $\lambda^{eff}$ obeys the limiting
expression of  the  type (\ref{eq7.51})
where the dependence on the concentration of B-type impurities  and
on the strength of disorder scattering  enters  implicitly  through
the renormalized electron  state  density  $N_{A}(0)$  and,  of  course,
through the renormalized chemical potential. However the self-consistent
version of ATA used  here  fails  in  discribing  normal
state properties of an alloy (see Ref.\cite{c37} and references therein
for critique on this approximation), so that more complex
approximations have to be considered.

\subsection{Coherent Potential Approximation and EPI Renormalizations.}
\noindent
  For CPA the equation for a self-energy $\Sigma^{e}_{CPA}({\rm i}p_s)$
reads\cite{c37,c57}
\begin{equation}
   <t({\rm i}p_s)>=0
\label{eq7.52}
\end{equation}
where
\begin{equation}
  t({\rm i}p_s)=(V\eta_i-\Sigma^{e}_{CPA}) +
  (V\eta_i-\Sigma^{e}_{CPA})G_{ii}({\rm i}p_s)t({\rm i}p_s) .
\label{eq7.53}
\end{equation}
We may introduce $P_2(z)$ generating function by\cite{c58}
\begin{equation}
   <t(z)>=0 ,
\label{eq7.54}
\end{equation}
\begin{equation}
   t(z)=(\eta - P_2(z))+(\eta - P_2(z))zt(z) ,
\label{eq7.55}
\end{equation}
where $\eta$ is a random variable with the distribution
\begin{equation}
  {\rm Prob}\{\eta=1\} =c,\qquad   {\rm Prob}\{\eta=0\}=1-c .
\label{eq7.56}
\end{equation}
To calculate $P_4(z_1,z_2)$ we use the
Ward-type identity (\ref{eq7.11}). Then  a
kind of transformations used in Appendix~A, Eq.(\ref{eqA.5}), leads
\begin{equation}
     P_4(z_1,z_2)=\frac{
                  <t(z_1)t(z_2)>
                  }{
                  1+z_1 z_2<t(z_1)t(z_2)>
                  } ,
\label{eq7.57}
\end{equation}
and from Bethe-Salpeter type equation (\ref{eq7.16}) one has
\begin{equation}
   \Gamma_4(z_1,z_2) = <t(z_1)t(z_2)> .
\label{eq7.58}
\end{equation}
Using (\ref{eq7.34}) we can write $\kappa(c,\tilde{V})$ in the form
\begin{equation}
  \kappa(c,\tilde{V})=1+{\tilde{V}}^2 <({\rm Im} t({\rm i}\tilde{V}))^2> .
\label{eq7.59}
\end{equation}

  A more familiar form arises if one uses general  expression  for
$\kappa (c,\tilde{V})$ in the form (\ref{eq7.33}).
To this end we rewrite  the  expression
(\ref{eq7.33}) in the form
\begin{equation}
  \kappa(c,\tilde{V})=
        \frac{
             <(z_1+z_1^2 t(z_1)-z_2-z_2^2 t(z_2))^2>
             }{
             (z_1-z_2)^2
             }
     \left.\right|_{z_1=-z_2=-{\rm i}\tilde{V}} ,
\label{eq7.60}
\end{equation}
where the use was made of Eqs.(\ref{eq7.53}) and (\ref{eq7.57}),
and finally
\begin{equation}
  \kappa(c,\tilde{V}) =
        \frac{
             <[{\rm Im}(z+z^2t(z))]^2>
             }{
             ({\rm Im} z)^2
             }
     \left.\right|_{z=-{\rm i}\tilde{V}} .
\label{eq7.61}
\end{equation}
Within the approximation (\ref{eq6.10}) one has
\begin{equation}
  {\rm Re} G(+{\rm i}0)=0,\qquad {\rm Im} G(+{\rm i}0)= -{\rm i}\pi N(0) ,
\label{eq7.62}
\end{equation}
with $G(E+{\rm i}0)$ being analytical continuation of
$G({\rm i}p_s)$ onto the upper
half-plane of the complex energy plane, and the
expression (\ref{eq7.61}) becomes
\begin{equation}
   \kappa(c,\tilde{V})= \frac{<N_{\eta}^2(0)>}{<N_{\eta}(0)>^2} ,
\label{eq7.63}
\end{equation}
where in the course of transformations we have introduced projected
electron state densities as
\begin{equation}
  N_{\eta}(E)=-\frac{1}{\pi}{\rm Im} F_{\eta}(E),\qquad
  F_{\eta}(E)=G(E)+G(E)t_{\eta}(E+{\rm i}0)G(E) ,
\label{eq7.64}
\end{equation}
with $t(E+{\rm i}0)$ being analytical continuation of
$t({\rm i}p_s)$, Eq.(\ref{eq7.53}),
onto the upper half-plane and $G(E)$ being  on-site  one-particle
Green's function.

  As one may see the formula (\ref{eq7.63}) is exactly the same as in
Refs.\cite{c5,c7} and obeys the expected limiting behavior
Eq.(\ref{eq7.51}) (the reasons for this are the same as
in the case of ATA).

\section{Conclusions}
\noindent
  In this paper we have studied the electron system which includes
both   quenched   substitutional   disorder   and   electron-phonon
interaction. We  have  developed  a  rather  general  approach  for
constructing Eliashberg-type equations  describing  superconductive
transition in such systems and have analyzed a  linearized  version
of the equations in a number of widely known ``alloy'' approximations
to find explicit expressions for $T_{c}$.
The examples are Eq.(\ref{eq7.63})
for CPA, which coincide with  that  given  in  Refs.\cite{c5,c7},  and  also
Eq.(\ref{eq7.47}) for ATA (along with general
expression (\ref{eq6.47}) for $T_{c}$ and
Eq.(\ref{eq6.43}) for the renormalization
factor $\kappa(c,\tilde{V})$ of course).

  Analogously,  one  may  derive   rather   simple   and   compact
expressions for $T_{c}$ in any other ``alloy''
approximation  for  normal
phase and within more general models of quenched disorder including
off-diagonal one. The structure of the  equations
(\ref{eq2.30}--\ref{eq2.33},\ref{eq2.41}--\ref{eq2.43})
remains intact irrespectively of the type of quenched
substitutional disorder as well as the approximation for the
``alloy'' self-energy used (obviously, the latter must obey
$\phi$-derivability criterion, Eq.(\ref{eq3.44})).

  We want also mark a general conclusion on extreme importance  of
EPI  vertex  renormalization  by  alloy  disordering  and  on   the
Anderson's theorem for arbitrary type of quenched disorder  treated
within conserving approximations. As for  results
(\ref{eq6.50}--\ref{eq6.52})  and
(\ref{eq6.55}--\ref{eq6.58}) concerning the
influence of disordering on  logarithmic
corrections  to  the  effective  EPI   coupling   and   on   dilute
paramagnetic impurity  suppression of $T_c$ respectivelly,  to  our
knowledge they have been presented in such a general form  for  the
first time here.

  It would  be  interesting  enough  to  generalize  the  approach
presented to the case of non-singlet types of pairing, which  seems
to be quite possible. In this case, as one  knows,  the  Anderson's
theorem does not hold and the effects of the disorder influence  on
$T_{c}$  are much stronger.\cite{c4}

  Let us discuss briefly the applicability of the results for $T_{c}$
obtained to the ``alloy analogy'' for the Hubbard model.
Seemingly, they are not valid literally because of  very  different
structure of diagrams  for  quenched  disorder  and  for  effective
disorder within static approximation for the Hubbard  model,  where
disorder  is  not  quenched,  but  annealed.  In  this   case   the
``concentrations'' of ``alloy'' components themselves are  functionally
dependent on one-particle Green's function  and  should  be  varied
too. At a diagrammatic level of discussion the difference among two
types of the disorder  is  the  appearance  of  graphs  with  loops
contributing for the case of annealed disorder, while  such  graphs
are absent for quenched disorder. But even  if  one  neglects  this
difference (which may be quite reasonable, at  least  for  not  too
strong on-site repulsion when distributions for fluctuating  fields
are weakly  renormalized  by  electrons)  there  is  still  another
reason for why the present results are not directly  applicable
to the ``alloy analogy'' for the Hubbard model.  The  matter
is that in this case fluctuating fields  contain  both  charge  and
spin components, so that the effective  disorder  is  by  no  means
nonmagnetic.

  One can also see that simple approximations of a
type used by us earlier\cite{c28} are rather dangerous since a type of
Ward's cancellations may take place just as in the case of
Anderson's theorem and careless generalization of the present
approach to the ``alloy analogy'' within functional integral
treatments of the Hubbard model is required.

Let us stress once more the difference between our treatment
and the approach by Belitz.\cite{c16,c17,c18} Belitz, following
ideas due to Anderson,\cite{c1,c2} used the superconducting
order parameter
$\Delta_{\nu}\propto C^{\dag}_{\nu\uparrow}
C^{\dag}_{\overline{\nu}\downarrow}$
where $\nu$ and $\overline{\nu}$ are the states connected
by the time reversal symmetry for a particular disorder
configuration. At the same time we use ${\bf k}$-dependent
order parameter in a translationally invariant effective
medium. Therefore both approaches are not equivalent from
formal point of view: diagrams of the same form in $\nu$-
and ${\bf k}$-representation need not to give the same contribution and,
hence, the same result for $T_c$. In particular, in the approach by
Belitz the Hartree-type diagrams do not contribute to the anomalous
self-energy unlike present approach.

To establish the equivalence
one has to solve very  difficult, yet not investigated, problem
whether the random quantity $\Delta_{\nu}$ is a self-averaging one.
On the other hand, the order parameter used here seems to be directly
related to the quasiparticle spectrum in the superconducting phase  and
consequently to observables. To our opinion,
treatment of such important problems
as, for instance, order parameter anisotropy in high $T_c$ superconductors
is more  adequate within our approach. In any case, the order parameter used
here is physically satisfactory, provided the quasimomentum remains
well defined, that is, far from the localization transition. As for
the essentially strong disorder (near the Anderson's transition), both
the approaches apparently do not work and the choice of the order parameter
in the intermediate region may be done empirically.

\section{Acknowledgments}
\noindent
  We are grateful to  D.M.~Edwards  and  R.L.~Jacobs  for  helpful
discussions and critical remarks on our  paper    which  initiated
this  work.  We  are  also  indebted  to  V.Yu.~Irkhin   for   many
stimulating conversations on the subject of present paper.

\appendix{\ \ The Ward identities}
\noindent
  In this Appendix we present the derivation of Ward's  identities
(\ref{eq3.41}) and (\ref{eq4.15}). The  identities  of  this  type  was
established in  Ref.\cite{c59}  within  the  CPA  in  connection  with  the
calculation of transport properties of binary alloys. The relations
between self-energy $\Sigma^{e}[G]$, as given by CPA, and irreducible four-leg
vertex presented in Ref.\cite{c59} was also studied in Ref.\cite{c60}
with the  use of diagrammatic technique.

  To begin with, we establish the connection between variation  of
$\Sigma[G]$ caused by finite change in $G$. Such an identity was derived
within certain approximation and used when discussing the
problem of localization by disorder.\cite{c61}

  Given arbitrary but finite variation in one-particle Green's function $G$
\begin{equation}
   G'=G+\Delta G,  \qquad \mbox{$\Delta G$ --- finite!}
\label{eqA.1)}
\end{equation}
we consider
\begin{equation}
   <\Delta T>=<T'^{e}-T^{e}>,
\label{eqA.2}
\end{equation}
where $T^{e}$ is given by
\begin{equation}
   T^{e}({\rm i}p_s)=
        \left(
             V^{e}  -\Sigma^{e}({\rm i}p_s)
        \right)+
        \left(
             V^{e}  -\Sigma^{e}({\rm i}p_s)
        \right)G({\rm i}p_s)T^{e}({\rm i}p_s),
\label{eqA.3}
\end{equation}
$T'^{e}$ is defined by similar expression, but with $G$ and $\Sigma$
replaced by $G'$  and $\Sigma'$ respectively,
\begin{equation}
     \Sigma'=\Sigma^{e}+\Delta\Sigma ,
\label{eqA.4}
\end{equation}
and we impose the self-consistency condition $<T >=<T' >=0$ for $\Sigma$ to
acquire proper, albeit implicit, functional dependence on $G$.

The following chain of equalities holds:
\begin{eqnarray}
&  0 = <\Delta T^{e}> = < {T'}^{e}[G',\Sigma']
             - T^{e}[G, \Sigma]>
\nonumber \\
&  = - <{T'}^{e}\left(({T'}^{e})^{-1} - (T^{e})^{-1}\right)T^{e}>
\nonumber \\
&  = - <{T'}^{e}\left((V^{e}-{\Sigma'}^{e})^{-1})
        -(V^{e} - \Sigma^{e})^{-1})
        - (G'-G)\right)T^{e}>
\nonumber \\
&  = - <(1 + {T'}^{e}G')\Delta\Sigma
        (1 + GT^{e})-{T'}^{e}\Delta GT^{e}>
\nonumber \\
&  = <{T'}^{e}\Delta GT^{e}> -
     <(1+{T'}^{e}G')\Delta\Sigma(1+GT^{e})> - \Delta\Sigma .
\label{eqA.5}
\end{eqnarray}
In explicit index notation the last line of Eq.(\ref{eqA.5}) reads
\begin{equation}
   \Delta\Sigma_{\mu\nu} + \Gamma_{\mu\alpha;\beta\nu}[G',G]G_{\alpha\alpha'}
                           \Delta\Sigma_{\alpha'\beta'}G_{\beta'\beta} =
   \Gamma_{\mu\alpha;\beta\nu}[G',G]\Delta G_{\alpha\beta},
\label{eqA.6}
\end{equation}
where, analogously to (\ref{eq3.28}), we have introduced full disorder
four-leg vertex
\begin{equation}
   \Gamma_{\mu\alpha;\beta\nu}[G',G] =
      <{T'}^{e}_{\mu\alpha}{T}^{e}_{\beta\nu}> .
\label{eqA.7}
\end{equation}
The Bethe-Salpeter equation gives the irreducible four-leg vertex
\begin{equation}
   \Gamma_{\mu\alpha;\beta\nu}[G',G] = U_{\mu\alpha;\beta\nu}[G',G] +
   U_{\mu\mu';\nu'\nu}[G',G]{G'}_{\mu'\alpha'}G_{\beta'\nu'}
   \Gamma_{\alpha'\alpha;\beta\beta'}[G',G] .
\label{eqA.8}
\end{equation}
Multiplying (\ref{eqA.6}) by the combination
$G'U[G',G]G$ and using the Bethe-Solpeter equation (\ref{eqA.8})
we can write
\begin{eqnarray}
&
   \Gamma_{\alpha\alpha';\beta'\beta}[G',G]
   {G'}_{\alpha'\mu}
   G_{\nu\beta'}
   \Delta\Sigma_{\mu\nu} =
&
\nonumber \\
&
    =U_{\alpha\alpha';\beta'\beta}[G',G]{G'}_{\alpha'\mu'}G_{\nu'\beta'}
   \Gamma_{\mu'\delta;\gamma\nu'}[G',G]
   \Delta G_{\delta\gamma} .
&
\label{eqA.9}
\end{eqnarray}
After substituting this expression into (\ref{eqA.6}) and using
Eq.(\ref{eqA.8}) once again one obtains the Ward's identity
\begin{equation}
   \Delta\Sigma_{\mu\nu} =
    U_{\mu\alpha;\beta\nu}[G',G]\Delta G_{\alpha\beta} .
\label{eqA.10}
\end{equation}

Putting now $G'=G + \Delta G$, where $\Delta G$ tends to zero
we immediately have
\begin{equation}
   \frac{\delta\Sigma_{\mu\nu}}{\delta G_{\alpha\beta}} =
      U_{\mu\alpha;\beta\nu}[G,G] .
\label{eqA.11}
\end{equation}
  By the definition (\ref{eqA.7}) the vertex function $\Gamma[G',G]$
has the property
\begin{equation}
   \Gamma_{\mu\alpha;\beta\nu}[G',G] = \Gamma_{\beta\nu;\mu\alpha}[G,G']
\label{eqA.12}
\end{equation}
which induces corresponding property for irreducible four-leg vertex
\begin{equation}
   U_{\mu\alpha;\beta\nu}[G',G] = U_{\beta\nu;\mu\alpha}[G,G'] .
\label{eqA.13}
\end{equation}
Then the use of (\ref{eqA.11}) leads to the condition of $\phi$-derivability,
and the Eqs.~(\ref{eq3.26}) which implicitly determine the self-energy
$\Sigma^{e}[G]$ can be represented in the form stated in Eq.~(\ref{eq2.33})
with some functional $W^{e}[G]$.

  Note also that for any conserving approximation for $\Sigma^{e}[G]$ the
condition of $\phi$-derivability holds automatically and the identity of
the type (\ref{eqA.10}) may be established along the lines  of  Appendix  of
Ref.\cite{c61} and making  use  of  the  fact  that  complete  families  of
self-energy terms contribute to $\Sigma^{e}[G]$  in this case.

  Formal identity (\ref{eqA.10}) may be used to derive various relations
between alloy self-energy $\Sigma^{e}({\rm i}p_s)$ and one
particle Green's function $G({\rm i}p_s)$. For example, substituting
\begin{equation}
   \Sigma' = \Sigma^{e}_{\sigma}({\rm i}p_s), \quad
   \Sigma  = \Sigma^{e}_{\sigma}({\rm i}p_m)
\label{eqA.14}
\end{equation}
and
\begin{equation}
   G' = {G'}_{\sigma}({\rm i}p_s), \quad
   G  =  G_{\sigma}({\rm i}p_m)
\label{eqA.15}
\end{equation}
into (\ref{eqA.10}) we obtain
\begin{equation}
   \Sigma^{e}_{ij\sigma}({\rm i}p_s) - \Sigma^{e}_{ij\sigma}({\rm i}p_m)=
   U_{il;nj}({\rm i}p_s,{\rm i}p_m) \left\{
             G_{ln\sigma}({\rm i}p_s) - G_{ln\sigma}({\rm i}p_m)
             \right\}
\label{eqA.16}
\end{equation}
which after Fourier transforming and using
\begin{equation}
   {\rm Im}\Sigma^{e}_{{\bf k}\sigma}(E) = \frac{1}{2{\rm i}}
           \left\{
                 \Sigma^{e}_{{\bf k}\sigma}(E + {\rm i}\delta) -
                 \Sigma^{e}_{{\bf k}\sigma}(E - {\rm i}\delta)
           \right\}, \quad
                     \delta\to+0,
\label{eqA.17}
\end{equation}
\begin{equation}
   {\rm Im} G_{{\bf k}\sigma}(E) = \frac{1}{2{\rm i}}
           \left\{
                 G_{{\bf k}\sigma}(E + {\rm i}\delta) -
                 G_{{\bf k}\sigma}(E - {\rm i}\delta)
           \right\}, \quad
                     \delta\to+0,
\label{eqA.18}
\end{equation}
takes the form
\begin{equation}
   {\rm Im}\Sigma^{e}_{{\bf k}\sigma}(E) =
     \sum_{\bf q} U^{\sigma\sigma}_{\bf kq}(E^{+},E^{-}){\rm Im} G_{{\bf k}\sigma}(E),
     \quad
     E^{\pm} = E\pm {\rm i}0.
\label{eqA.19}
\end{equation}
This is nothing but an analogue of the unitarity condition for
alloy type problems we deal with in the present paper.

\appendix{\ \ Disorder-Induced Renormalization of Paramagnetic Impurity
              \protect{${\bf T_c}$}--Suppression}
\noindent
  In this Appendix we treat the problem of how $T_c$--suppression by
dilute paramagnetic impurities is renormalized by non-magnetic disorder.
Using the approach of Section~3 we derive the expression
for the self-energy correction which is due to
the scattering on the paramagnetic impurities, and then we apply
the analysis  of Section~5 and~6.

   Three assumptions are made. Firstly, paramagnetic impurities are
distributed randomly with the concentration $c_P$ over  the  host  and
independently of the non-magnetic ones. Secondly, the disorder due to
paramagnetic impurities is purely site-diagonal. Thirdly, both the
concentration of paramagnetic impurities and the scattering
intensity $V_P$ on them are small, so that the relevant small
parameter is $c_P V_P^2$.

  The contribution to the Hamiltonian (\ref{eq2.1})  due  to  paramagnetic
impurities reads
\begin{equation}
   H_P = \sum_{i} V^{P}_{i}\left\{
               S^{z}_{i}{\bf C}^{\dag}_{i}\tau_{0}{\bf C}_{i} +
               S^{+}_{i}{\bf C}^{\dag}_{i}\tau_{-}{\bf C}_{i} +
               S^{-}_{i}{\bf C}^{\dag}_{i}\tau_{+}{\bf C}_{i}
         \right\} ,
\label{eqB.1}
\end{equation}
where ${\bf C}^{\dag}_{i}$ and ${\bf C}_{i}$ are
row and column spinors (\ref{eq2.10}) on a site  $\bf i$, $S^{z}_{i}$ is
$z$-component of the impurity spin operator of spin $S$,
the matrices $\tau_{+}$ and $\tau_{-}$ are $1/2(\tau_{1}+{\rm i}\tau_{2})$
and $1/2(\tau_{1}-{\rm i}\tau_{2})$ respectively, $S^{+}_{i}$ and $S^{-}_{i}$
have similar definitions, and
\begin{eqnarray}
&  V^{P}_{i}= V_{P}\zeta_i, &
\label{eqB.2} \\
&  {\rm Prob}\{\zeta_i=0\}=(1-c_P),\qquad  {\rm Prob}\{\zeta_i=1\}=c_P, &
\nonumber
\end{eqnarray}
with $\zeta_i$ being independent random variables. Up to the lowest order
in $c_P V_P^2$, one has the correction to the effective potential
${\bf V}({\rm i}p_s)$  Eq.(\ref{eq3.3}),
\begin{equation}
  \delta V_{\alpha\beta}({\rm i}p_s)=
      cV_{P}^2 S(S+1)g_{\alpha\beta}({\rm i}p_s)
      \delta_{i_{\alpha},i_{\beta}} ,
\label{eqB.3}
\end{equation}
and we have neglected uninteresting  contribution  of  the  first
order in $V_P$ which gives a shift of the chemical potential.

  The expression (\ref{eqB.3}) can be written in the form
analogous to the  effective scattering potential caused by EPI
\begin{equation}
   \delta V_{\alpha\beta}({\rm i}p_s)=
     S_{\gamma\gamma'}K^{\gamma}_{\alpha\alpha'}
     g_{\alpha'\beta'}({\rm i}p_s)
     K^{\gamma'}_{\beta'\beta} ,
\label{eqB.4}
\end{equation}
\begin{equation}
       \left\{\bf K\right\}_{l;ij} = \delta_{il}\delta_{jl}\tau_{0} ,
\label{eqB.5}
\end{equation}
\begin{equation}
   S_{\gamma\gamma'}=S_{ll'}=\delta_{\gamma\gamma'}c_P V_P^2 S(S+1)
   =\delta_{ll'}c_P V_P^2 S(S+1) .
\label{eqB.6}
\end{equation}
  The procedure of Section~3 then yields the correction to the
full self-energy $\Sigma[G,D]$, with the form
\begin{eqnarray}
&
   \delta\Sigma_{\mu\nu}({\rm i}p_s) =
   S_{\gamma\gamma'} &
\nonumber \\
&
   \times
   \left\{ \right.
   R^{\gamma}_{\mu\alpha'}({\rm i}p_s,{\rm i}p_s)
   G_{\alpha'\beta'}({\rm i}p_s)
   R^{\gamma'}_{\beta'\nu}({\rm i}p_s,{\rm i}p_s) +
   U_{\mu\mu';\delta\xi;\nu'\nu}({\rm i}p_s,{\rm i}p_s,{\rm i}p_s) &
\nonumber \\
&
   \times
   G_{\mu'\alpha}({\rm i}p_s)
   R^{\gamma}_{\alpha\alpha'}({\rm i}p_s,{\rm i}p_s)
   G_{\alpha'\delta}({\rm i}p_s)
   G_{\xi\beta}({\rm i}p_s)
   R^{\gamma'}_{\beta\beta'}({\rm i}p_s,{\rm i}p_s)
   G_{\beta'\nu'}({\rm i}p_s)
   \left. \right\} . &
\label{eqB.7}
\end{eqnarray}
In this expression $R^{\gamma}_{\alpha\beta}({\rm i}p_s,{\rm i}p_s)$,
is renormalized  ``paramagnetic  impurity''
vertex which obeys the equation
\begin{eqnarray}
  R^{\gamma}_{\alpha\alpha'}({\rm i}p_s,{\rm i}p_m) =
  K^{\gamma}_{\alpha\alpha'} +
  \Gamma_{\alpha\beta;\beta'\alpha'}({\rm i}p_s,{\rm i}p_m)
  G_{\beta\mu}({\rm i}p_s)G_{\mu'\beta'}({\rm i}p_m)
  K^{\gamma}_{\mu\mu'} ,
\label{eqB.8} \\
  R^{\gamma}_{\alpha\alpha'}({\rm i}p_s,{\rm i}p_m) =
  K^{\gamma}_{\alpha\alpha'} +
  U_{\alpha\beta;\beta'\alpha'}({\rm i}p_s,{\rm i}p_m)
  G_{\beta\mu}({\rm i}p_s)G_{\mu'\beta'}({\rm i}p_m)
  R^{\gamma}_{\mu\mu'}({\rm i}p_s,{\rm i}p_m) .
\label{eqB.9}
\end{eqnarray}

The expression~(\ref{eqB.7}) may be depicted in a form similar
to that for $\Sigma^{e-ph}[G,D]$ (see Fig.~5) with graphical
representation of the combination $c_P V_P^2 S(S+1)$ replacing phonon lines
and with the ``paramagnetic impurity'' three-leg vertices replacing EPI
vertices.

  Further manipulations repeat literally the analysis of Section~5
and, after  adopting  binary  alloy
model for non-magnetic disorder, we apply the qualitative  analysis
of Section~6 to the reduced equations obtained.

Introducing
\begin{equation}
   \gamma_P = c_P V_P^2 S(S+1),  \qquad
   \tilde{\gamma}_{P}=N(0)\gamma_P ,
\label{eqB.10}
\end{equation}
one can write the expression for the
correction to $\tilde{Z}({\rm i}p_s)$ as
\begin{equation}
   \delta\tilde{Z}({\rm i}p_s) =
    \pi\tilde{\gamma}_{P}
          \left\{
              f_{eee}(c,\tilde{V})+f_{oeo}(c,\tilde{V})+
              f_{eoo}(c,\tilde{V})+f_{ooe}(c,\tilde{V})
          \right\}
                  \frac{1}{|p_s|}
\label{eqB.11}
\end{equation}
and for the contribution to $\chi({\rm i}p_s)$
one has the following form
\begin{equation}
   \delta\chi({\rm i}p_s) =
   - {\rm i} \pi\tilde{\gamma}_{P}
             \left\{
             f_{ooo}(c,\tilde{V})+f_{eoe}(c,\tilde{V})+
             f_{eeo}(c,\tilde{V})+f_{oee}(c,\tilde{V})
             \right\}
\label{eqB.12}
\end{equation}
which  has  no energy  dependence (within the approximations
of Section~6, of course) and leads to an uninteresting
shift of chemical potential only.

  The contribution of paramagnetic  impurities  to  the  anomalous
part of the self-energy has the form
\begin{eqnarray}
& \displaystyle
   \delta\left(\tilde{Z}({\rm i}p_s)\Delta({\rm i}p_s)\right)=
&
\nonumber \\
& \displaystyle
        =-\pi\tilde{\gamma}_{P}
         \left\{ f_{eee}(c,\tilde{V}) - f_{eoo}(c,\tilde{V})-
                 f_{ooe}(c,\tilde{V}) -3f_{oeo}(c,\tilde{V})\right\}
                  \frac{\Delta({\rm i}p_s)}{|p_s|}
&
\label{eqB.13}
\end{eqnarray}
and the correction to the gap function becomes
\begin{equation}
   \delta\Delta({\rm i}p_s) =
         -2\gamma_{P}^{eff}
                   \frac{\Delta({\rm i}p_s)}{|p_s|} .
\label{eqB.14}
\end{equation}
In Eq.(\ref{eqB.14}) we have introduced
\begin{equation}
   \gamma^{eff}_{P}=\kappa(c,\tilde{V})\tilde{\gamma}_{P} ,
\label{eqB.15}
\end{equation}
where $\kappa(c,\tilde{V})$ is defined by Eq.(\ref{eq6.43}).
So we see that,  up  to  the
factor  $\kappa(c,\tilde{V})$, the contribution of the dilute paramagnetic
impurities to the equations for the self-energy has the same form
as in the case without alloy  disordering.

  Full equations on $\tilde{Z}({\rm i}p_s)$,
$\chi({\rm i}p_s)$ and $\Delta({\rm i}p_s)$
which  take  account
of the paramagnetic impurity contributions read
\begin{eqnarray}
& \displaystyle
    {\rm i}p_s[1-\tilde{Z}({\rm i}p_s)] =
     2{\rm i}
     \left\{ f_{eoo}(c,\tilde{V})+f_{ooe}(c,\tilde{V}) \right\}
                   \tilde{\lambda}(0) {\cal U}
                   {\rm Sign}({\rm i}p_s) &
\nonumber \\
& \displaystyle
   - {\rm i} \left\{ f_{eee}(c,\tilde{V})+f_{oeo}(c,\tilde{V}) \right\}
                   \pi T_{c} \sum_{{\rm i}p_m}
                   \tilde{\lambda}({\rm i}p_s-{\rm i}p_m)
                   {\rm Sign}({\rm i}p_m) &
\nonumber \\
& \displaystyle
   - {\rm i} \left\{ f_{eoo}(c,\tilde{V})+f_{ooe}(c,\tilde{V}) \right\}
                   \pi T_{c} \sum_{{\rm i}p_m}
                   \tilde{\lambda}({\rm i}p_s-{\rm i}p_m)
                   {\rm Sign}({\rm i}p_s) &
\nonumber \\
& \displaystyle
         -{\rm i}\pi\tilde{\gamma}_{P}
          \left\{
                 f_{eee}(c,\tilde{V}) + f_{eoo}(c,\tilde{V})+
                 f_{ooe}(c,\tilde{V}) + f_{oeo}(c,\tilde{V})
          \right\}
                 {\rm Sign}(p_s) , &
\label{eqB.16}
\end{eqnarray}
%%%%%%%%%%%%%%%%%
\begin{eqnarray}
&
   \chi({\rm i}p_s)] =
     2{\rm i}
     \left\{ f_{eoe}(c,\tilde{V})+f_{ooo}(c,\tilde{V}) \right\}
                   \tilde{\lambda}(0) {\cal U} &
\nonumber \\
& \displaystyle
   - {\rm i} \left\{ f_{ooo}(c,\tilde{V})+f_{eoe}(c,\tilde{V}) \right\}
                   \pi T_{c} \sum_{{\rm i}p_m}
                   \tilde{\lambda}({\rm i}p_s-{\rm i}p_m) &
\nonumber \\
& \displaystyle
   - {\rm i} \left\{ f_{eeo}(c,\tilde{V})+f_{oee}(c,\tilde{V}) \right\}
                   {\rm Sign}({\rm i}p_s)
                   \pi T_{c} \sum_{{\rm i}p_m}
                   \tilde{\lambda}({\rm i}p_s-{\rm i}p_m)
                   {\rm Sign}({\rm i}p_m) &
\nonumber \\
&
   - {\rm i} \pi\tilde{\gamma}_{P}
             \left\{
             f_{ooo}(c,\tilde{V})+f_{eoe}(c,\tilde{V})+
             f_{eeo}(c,\tilde{V})+f_{oee}(c,\tilde{V})
             \right\} , &
\label{eqB.17}
\end{eqnarray}
%%%%%%%%%%%%%%%%%%
\begin{eqnarray}
& \displaystyle
   \tilde{Z}({\rm i}p_s)\Delta({\rm i}p_s) =
    -2 \left\{ f_{eoo}(c,\tilde{V}) + f_{ooe}(c,\tilde{V})\right\}
    \tilde{\lambda}(0) {\cal U}
              \frac{\Delta({\rm i}p_s)}{|p_s|} &
\nonumber \\
& \displaystyle
       + \left\{ f_{eee}(c,\tilde{V}) - f_{eee}(c,\tilde{V})\right\}
         \pi T_{c} \sum_{{\rm i}p_m}
                   \tilde{\lambda}({\rm i}p_s - {\rm i}p_m)
                   \frac{\Delta({\rm i}p_m)}{|p_m|} &
\nonumber \\
& \displaystyle
  +\left\{ f_{eoo}(c,\tilde{V}) + f_{ooe}(c,\tilde{V})\right\}
   \frac{\Delta({\rm i}p_s)}{|p_s|}
   \pi T_{c} \sum_{{\rm i}p_m}
             \tilde{\lambda}({\rm i}p_s - {\rm i}p_m) &
\nonumber \\
& \displaystyle
   + 2 f_{oeo}(c,\tilde{V})
   \frac{\Delta({\rm i}p_s)}{|p_s|}
   \pi T_{c} \sum_{{\rm i}p_m}
             \tilde{\lambda}({\rm i}p_s - {\rm i}p_m)
             {\rm Sign}({\rm i}p_s)
             {\rm Sign}({\rm i}p_m) &
\nonumber \\
& \displaystyle
         -\pi\tilde{\gamma}_{P}
         \left\{ f_{eee}(c,\tilde{V}) - f_{eoo}(c,\tilde{V})-
                 f_{ooe}(c,\tilde{V}) -3f_{oeo}(c,\tilde{V})\right\}
                   \frac{\Delta({\rm i}p_s)}{|p_s|} . &
\label{eqB.18}
\end{eqnarray}
Solving the equation (\ref{eqB.16}) on $\tilde{Z}({\rm i}p_s)$
and inserting into (\ref{eqB.18}) one finds
\begin{eqnarray}
& \displaystyle
   \Delta({\rm i}p_s) =
         \pi T_{c} \sum_{{\rm i}p_m}
                   \lambda^{eff}({\rm i}p_s - {\rm i}p_m)
                   \frac{\Delta({\rm i}p_m)}{|p_m|} &
\nonumber \\
& \displaystyle
  -\frac{\Delta({\rm i}p_s)}{|p_s|}
   \pi T_{c} \sum_{{\rm i}p_m}
             \lambda^{eff}({\rm i}p_s - {\rm i}p_m)
             {\rm Sign}({\rm i}p_s)
             {\rm Sign}({\rm i}p_m)
   -2\gamma_{P}^{eff}\frac{\Delta({\rm i}p_s)}{|p_s|}, &
\label{eqB.19}
\end{eqnarray}
--- the equation on the gap function, from which results for
the critical concentration and for $T_{c}$--suppression follow by
standard consideration.\cite{c4}

\appendix{\ \ Generating Functions for Single-site Approximations}
\noindent
  In  this  Appendix  we  establish  relations  between   suitably
introduced generating function for arbitrary conserving single-site
approximation  for  disorder  self-energy  $\Sigma^{e}[G]$ and generating
functions for irreducible disorder four-leg vertex and  for  higher
vertices. The Ward's type identities for generating functions  will
be also derived.

  Generating functional for arbitrary single site approximation
has the following analytical form
\begin{equation}
   W^{e}[G] = \sum_{i} \sum_{{\rm i}p_s} \sum_{n=1}^{\infty}
                 q_{n-1}(c) \frac{\left( VG_{ii}({\rm i}p_s)\right)^{n}}{n}
\label{eqC.1}
\end{equation}
and outer sum runs over lattice sites.
For the corresponding self-energy one has the expression
\begin{equation}
   \Sigma^{e}[G] = V \sum_{n=0}^{\infty}
                   q_{n}(c) \left( VG_{ii}({\rm i}p_s)\right)^{n} ,
\label{eqC.2}
\end{equation}
where $q_n(c)$ is $n$-th order cumulant  renormalized  by  the  multiple
occupancy corrections within approximation of interest.  Particular
choice of these cumulants determines wholly the  approximation  for
the disorder self-energy. We introduce the generating function:
\begin{equation}
   P_{2}(z) = \sum_{n=0}^{\infty} q_{n}(c) z^{n}
\label{eqC.3}
\end{equation}
The expression for $\Sigma^{e}[G]$ can then be written in the form
\begin{equation}
   \Sigma^{e}[G] = V P_{2}(VG).
\label{eqC.4}
\end{equation}
Note also that generating functions we use here are connected
with the number of Green's functions in the corresponding contributions
to the self-energy $\Sigma^{e}[G]$.
%
%and we call the function~(\ref{eqC.4}),
%and analogous functions for vertex corrections which will be introduced
%latter on, generating $P$-functions to distinguish them from the generating
%functions introduced in Section~7.

  Diagrams for the irreducible disorder four-leg vertex are
generated by varying corresponding contributions to the self-energy
and the expression for $U_4[G_1,G_2]$ reads
\begin{eqnarray}
    U[G_{1},G_{2}] & = &
          V^{2}
          \sum_{n=2}^{\infty} q_{n-1}(c)
          \sum_{m=0}^{\infty}
          \sum_{l=0}^{\infty}
          \delta_{n,m+l+2}
          \left( VG_{1}\right)^{m}
          \left( VG_{2}\right)^{n}
\nonumber \\
  & = &
         V^{2} \sum_{m=0}^{\infty} \sum_{l=0}^{\infty}
         q_{m+l+1}(c)
         \left( VG_{1}\right)^{m}
         \left( VG_{2}\right)^{n},
\label{eqC.5}
\end{eqnarray}
where we have used the fact that for quenched disorder the self-energy
$\Sigma^{e}[G]$ does not contain fermionic loops and therefore the
disorder four-leg vertex may be considered as some
functional of two independent functional variables. Generally speaking,
we may think of disorder $2m$-leg vertex as a functional of $m$
independent functional variables by the same reason.

Analogously to (\ref{eqC.3}) we introduce the generating function
\begin{eqnarray}
   P_{4}(z_{1},z_{2}) & = &
          \sum_{n=2}^{\infty} q_{n-1}(c)
          \sum_{m=0}^{\infty}
          \sum_{l=0}^{\infty}
          \delta_{n,m+l+2} z^{m} z^{n}
\nonumber \\
    & = &
         \sum_{m=0}^{\infty} \sum_{l=0}^{\infty}
         q_{m+l+1}(c) z^{m} z^{n} ,
\label{eqC.6}
\end{eqnarray}
and hence
\begin{equation}
   U[G_{1},G_{2}] = V^{2} P_{4}(VG_{1},VG_{2})  .
\label{eqC.7}
\end{equation}

In general, for $2m$-leg disorder vertex within arbitrary
single-site approximation one can readly obtain
\begin{eqnarray}
   U[G_{1},\ldots,G_{m}] & = &
          V^{m}
          \sum_{n=m}^{\infty} q_{n-1}(c)
          \sum_{n_{1}=0}^{\infty} \cdots
          \sum_{n_{m}=0}^{\infty}
          \delta_{n,n_{1}+ \cdots +n_{m}+m}
          \prod_{j=1}^{m}\left( VG_{j}\right)^{n_{j}}
\nonumber \\
   & = &
          V^{m}
          \sum_{n_{1}=0}^{\infty} \cdots
          \sum_{n_{m}=0}^{\infty}
           q_{n_{1}+ \cdots +n_{m}+m-1}(c)
          \prod_{j=1}^{m}\left( VG_{j}\right)^{n_{j}} ,
\label{eqC.8}
\end{eqnarray}
so that
\begin{eqnarray}
   P_{2m}(z_{1},\ldots ,z_{m}) & = &
          \sum_{n=m}^{\infty} q_{n-1}(c)
          \sum_{n_{1}=0}^{\infty} \cdots
          \sum_{n_{m}=0}^{\infty}
          \delta_{n,n_{1}+ \cdots +n_{m}+m}
          \prod_{j=1}^{m} z_{j}^{n_{j}}
\nonumber \\
         & = &
          \sum_{n_{1}=0}^{\infty} \cdots
          \sum_{n_{m}=0}^{\infty}
           q_{n_{1}+ \cdots +n_{m}+m-1}(c)
          \prod_{j=1}^{m} z_{j}^{n_{j}} ,
\label{eqC.9}
\end{eqnarray}
and then
\begin{equation}
   U[G_{1},\ldots ,G_{m}] = V^{m}
         P_{2m}(VG_{1},\ldots ,VG_{m}) .
\label{eqC.10}
\end{equation}
  We now express cumulant of the $n$-th order in terms of Cauchi integral
\begin{equation}
   q_{n}(c) = \frac{1}{2\pi {\rm i}}
              \int_{C} {\rm d}z
                \frac{P_{2}(z)}{z^{n+1}} .
\label{eqC.11}
\end{equation}
The substitution of $q_n(c)$ in the form (\ref{eqC.11})
into (\ref{eqC.9}) leads
\begin{eqnarray}
   P_{2m}(z_{1},\ldots ,z_{m}) & = &
     \frac{1}{2\pi {\rm i}}
     \int_{C} {\rm d}z P_{2}(z)
           \sum_{n_{1}=0}^{\infty} \cdots
           \sum_{n_{m}=0}^{\infty}
           \frac{1}{z^{n_1+\cdots+n_m+m}}
           \prod_{j=1}^{m} z_{j}^{n_j}
\nonumber \\
    & = & \frac{1}{2\pi {\rm i}}
     \int_{C} {\rm d}z P_{2}(z)
           \frac{1}{z^{m}}
           \prod_{j=1}^{m}
           \left\{
                  \sum_{n_{j}=0}^{\infty}
                  \left(
                        \frac{z_{j}}{z}
                  \right)^{n_{j}}
           \right\}
\nonumber \\
     & = &
     \frac{1}{2\pi {\rm i}}
     \int_{C} {\rm d}z P_{2}(z)
     \prod_{j=1}^{m}
            \frac{1}{z-z_{j}}  ,
\label{eqC.12}
\end{eqnarray}
where the contour of integration  encircles  singularities  of  the
denominator only, and $P_{2m}(z_1,\ldots,z_m)$ possess the property
\begin{equation}
   P_{2m}(z_{1},\ldots ,z_{m}) =
               P_{2m}(z_{{\cal P}1},\ldots ,z_{{\cal P}m}) ,
\label{eqC.13}
\end{equation}
with  $\cal P$ being arbitrary permutation of indices $\{1\ldots m\}$.

The expression (\ref{eqC.12}) relates generating function $P_2(z)$
with the corresponding generating functions for $2m$-leg vertices.
In particular, we have
\begin{equation}
   P_{4}(z_{1},z_{2}) =
     \frac{1}{2\pi {\rm i}}
     \int_{C} {\rm d}z
     \frac{P_{2}(z)}{(z-z_{1})(z-z_{2})}
\label{eqC.14}
\end{equation}
and
\begin{equation}
   P_{6}(z_{1},z_{2},z_{3}) =
     \frac{1}{2\pi {\rm i}}
     \int_{C} {\rm d}z
     \frac{P_{2}(z)}{(z-z_{1})(z-z_{2})(z-z_{3})} .
\label{eqC.15}
\end{equation}
For $m\geq 2$ rewriting the product in (\ref{eqC.12}) as
\begin{eqnarray}
   \prod_{j=1}^{m} \frac{1}{z-z_{j}} & = &
         \frac{1}{z-z_{1}} \frac{1}{z-z_{2}}
         \prod_{j=3}^{m} \frac{1}{z-z_{j}}
\nonumber \\
         & = &
         \frac{1}{z_{2}-z_{1}}\left\{
               \frac{1}{z-z_{2}} - \frac{1}{z-z_{1}}
         \right\}
         \prod_{j=3}^{m} \frac{1}{z-z_{j}}
\label{eqC.16}
\end{eqnarray}
and substituting this expression into (\ref{eqC.12}) we obtain the analogue
of the Ward's identity which connects the generating functions for
vertices of different order
\begin{equation}
   P_{2m}(z_{1},\ldots ,z_{m}) =
     \frac{
          P_{2(m-1)}(z_{2},z_{3},\ldots ,z_{m}) -
          P_{2(m-1)}(z_{1},z_{3},\ldots ,z_{m})
          }{
          z_{2} - z_{1}
          } .
\label{eqC.17}
\end{equation}
This equation gives
\begin{equation}
   P_{4}(z_{1},z_{2}) =
     \frac{
          P_{2}(z_{2}) - P_{2}(z_{1})
          }{
          z_{2} - z_{1}
          }
\label{eqC.18}
\end{equation}
and
\begin{equation}
   P_{6}(z_{1},z_{2},z_{3}) =
     \frac{
          P_{4}(z_{2},z_{3}) - P_{4}(z_{1},z_{3})
          }{
          z_{2} - z_{1}
          }
\label{eqC.19}
\end{equation}
for $m=2$ and $m=3$ respectively.

  Yet another way to establish the relation (\ref{eqC.18}) is to  use  the
Ward's identity derived in Appendix~A. Indeed, noting  that  within
arbitrary single-site approximation irreducible four-leg vertex  is
a purely local quantity we rewrite (\ref{eqA.7}) as
\begin{equation}
   V \left\{
           P_{2}(VG_{2}) - P_{2}(VG_{1})
     \right\} =
     V P_{4}(VG_{2},VG_{1})
       \left\{
             VG_{2} - VG_{1}
       \right\} ,
\label{eqC.20}
\end{equation}
where the use was made of (\ref{eqC.4}) and (\ref{eqC.7}). Then substitutions
\begin{equation}
   VG_{1} \rightarrow z_{1}, \quad
   VG_{2} \rightarrow z_{2}
\label{eqC.21}
\end{equation}
immediately lead to the expression (\ref{eqC.18}).


\begin{thebibliography}{999}
\bibitem{c1}%
    P.W.~Anderson, {\bibit J. Phys. Chem. Solids} {\bibbf 11} (1959) 26.
\bibitem{c2}%
    P.W.~Anderson, in {\bibit Proc. 8th Conf. Low Temp.  Phys.} (University
    of Toronto Press, 1961).
\bibitem{c3}%
    A.A.~Abrikosov and L.P.~Gor'kov, {\bibit Zh. Eksp. Teor. Fiz.}
{\bibbf 35} (1958) 1558; {\bibit Sov. Phys.- JETP} {\bibbf 8} (1959) 1090.
\bibitem{c4}%
    P.B.~Allen and B.~Mitrovic, in {\bibit Solid  State  Physics} {\bibbf v.37}
    (Academic Press, 1982), p.2.
\bibitem{c5}%
    S.V.~Vonsovsky, Yu.A.~Izyumov and E.Z.~Kurmaev,
    {\bibit Superconductivity of Transition Metal Alloys and Compounds},
    (Springer - Verlag, 1982).
\bibitem{c6}%
    A.~Weinkauf and J.~Zittartz, {\bibit Solid State Commun.} {\bibbf 14}
    (1974) 365.
\bibitem{c7}%
    A.~Weinkauf and J.~Zittartz, {\bibit J.Low Temp. Phys.} {\bibbf 18}
(1975) 229.
\bibitem{c8}%
    F.~Takano, K.~Machida and F.~Shibata, {\bibit Prog. Theor. Phys.}
{\bibbf 49} (1973) 1077.
\bibitem{c9}%
    F.~Takano and K.~Fujiki, {\bibit Prog. Theor. Phys.} {\bibbf 49}
(1973) 1459.
\bibitem{c10}%
    K.~Tankei and F.~Takano, {\bibit Prog. Theor. Phys.} {\bibbf 51}
(1974) 988.
\bibitem{c11}%
    A.A.~Abrikosov and L.P.~Gor'kov, {\bibit Zh. Eksp. Teor. Fiz.} {\bibbf 39}
    (1960) 1781; {\bibit Sov. Phys.- JETP} {\bibbf 12} (1961) 1243.
\bibitem{c12}%
    A.R.~Miedema, {\bibit J. Phys. F} {\bibbf 3} (1973) 1803.
\bibitem{c13}%
    A.R.~Miedema and M.H.~Van~Maaren {\bibit Physica}
{\bibbf 69} (1973) 308.
\bibitem{c14}%
    A.R.~Miedema, {\bibit J. Phys. F} {\bibbf 4} (1974) 120.
\bibitem{c15}%
    P.W.~Anderson, K.A.~Muttalib and T.V.~Ramakrishnan, {\bibit Phys. Rev.}
{\bibbf B28} (1983) 117.
\bibitem{c16}%
    D.~Belitz, {\bibit Phys. Rev.} {\bibbf B35} (1987) 1636.
\bibitem{c17}%
    D.~Belitz, {\bibit Phys. Rev.} {\bibbf B35} (1987) 1651.
\bibitem{c18}%
    D.~Belitz, {\bibit Phys. Rev.} {\bibbf B36} (1987) 47.
\bibitem{c19}%
    L.N.~Bulaevskii, S.V.~Panyukov and M.V.~Sadovskii,
    {\bibit Zh. Eksp. Teor. Fiz.} {\bibbf 92} (1987) 672.
\bibitem{c20}%
    C.J.~Thompson and T.~Matsubara, {\bibit Prog. Theor. Phys.} {\bibbf 86}
    (1991) 1191.
\bibitem{c21}%
    D.~Belitz and T.R.~Kirkpatrick, {\bibit Rev. Mod. Phys.} {\bibbf 66}
(1994) 261.
\bibitem{c22}%
    A.A.~Abrikosov, {\bibit Physica} {\bibbf C244} (1995) 243.
\bibitem{c23}%
    J.~Hubbard, {\bibit Proc. Roy. Soc.} {\bibbf A281} (1964) 401.
\bibitem{c24}%
    M.~Cyrot, {\bibit Phys. Rev. Lett.} {\bibbf 25} (1970) 871.
\bibitem{c25}%
    J.~Hubbard, {\bibit Phys. Rev.} {\bibbf B20} (1979) 4584;
\bibitem{c26}%
    H.~Hasegawa, {\bibit J. Phys. Soc. Japan} {\bibbf 49} (1980) 178;
\bibitem{c27}%
    H.~Hasegawa, {\bibit J. Phys. Soc. Japan} {\bibbf 49} (1980) 963;
\bibitem{c28}%
    A.O.~Anokhin, V.Yu.~Irkhin and M.I.~Katsnelson,
    {\bibit Physica} {\bibbf C 179} (1991) 167.
\bibitem{c29}%
    M.~Rozenberg, X.Y.~Zhang  and G.~Kotliar, {\bibit Phys. Rev. Lett.}
{\bibbf 69} (1992) 1236.
\bibitem{c30}%
    A.~Georges and W.~Krauth, {\bibit Phys. Rev. Lett.} {\bibbf 69}
(1992) 1240.
\bibitem{c31}%
    X.Y.~Zhang, M.J.~Rozenberg and G.~Kotliar, {\bibit Phys. Rev. Lett.}
{\bibbf 70} (1993) 1666.
\bibitem{c32}%
    A.~Georges and W.~Krauth, {\bibit Phys. Rev.} {\bibbf B48} (1993) 7167.
\bibitem{c33}%
    M.J.~Rozenberg, G.~Kotliar and X.Y.~Zhang,  {\bibit Phys. Rev.}
{\bibbf B49} (1994) 10181.
\bibitem{c34}%
    P.W.~Anderson, {\bibit Phys. Rev.} {\bibbf 109} (1958) 1492.
\bibitem{c35}%
    G.~Baym and L.P.~Kadanoff, {\bibit Phys. Rev.} {\bibbf 124} (1961) 289.
\bibitem{c36}%
    G.~Baym, {\bibit Phys. Rev.} {\bibbf 127} (1962) 1391.
\bibitem{c37}%
    R.J.~Elliot, J.A.~Krumhansl and P.L.~Leath, {\bibit Rev. Mod. Phys.}
    {\bibbf 46} (1974) 465.
\bibitem{c38}%
    E.I.~Blount, {\bibit Phys. Rev.} {\bibbf 144} (1959) 418.
\bibitem{c39}%
    T.~Tsuneto, {\bibit Phys. Rev.} {\bibbf 121} (1961) 402.
\bibitem{c40}%
    A.~Schmid, {\bibit Z. Phys.} {\bibbf 259} (1979) 421.
\bibitem{c41}%
    B.~Keck and A.~Schmid, {\bibit Sol. St. Comm.} {\bibbf 17} (1975) 799.
\bibitem{c42}%
    B.~Keck and A.~Schmid, {\bibit J. Low. Temp. Phys.} {\bibbf 24} (1976) 611.
\bibitem{c43}%
    J.R.~Schrieffer, {\bibit Theory  of  Superconductivity} (Benjamin,
    New York, 1964).
\bibitem{c44}%
    A.B.~Migdal, {\bibit Zh. Eksp. Teor. Fiz.} {\bibbf 34} (1958) 1438;
    {\bibit Sov. Phys.- JETP} {\bibbf 7} (1958) 999.
\bibitem{c45}%
    Y.~Nambu, {\bibit Phys. Rev.} {\bibbf 117} (1960) 648.
\bibitem{Eli}%
    G.M.~Eliashberg, {\bibit Zh. Eksp. Teor. Fiz.} {\bibbf 38} (1960) 966;
    {\bibit Sov. Phys.- JETP} {\bibbf 11} (1960) 696.
\bibitem{c46}%
    J.M.~Luttinger and J.C.~Ward, {\bibit Phys.Rev.} {\bibbf 118} (1960) 1417.
\bibitem{c47}%
    H.~Lustfield, {\bibit J. Low Temp. Phys.} {\bibbf 12} (1973) 595.
\bibitem{c48}%
    W.E.~Pickett, {\bibit Phys. Rev.} {\bibbf B26} (1982) 1186.
\bibitem{c49}%
    N.N.~Bogolyubov (Jr.) and B.I.~Sadovnikov, in
    {\bibit Some problems of Statistical Mechanics (in Russian)},
    (Moscow, 1975).
\bibitem{c50}%
    P.B.~Allen, {\bibit Phys. Rev.} {\bibbf B13} (1976) 1416.
\bibitem{c51}%
    W.~Metzner and D.~Volhardt, {\bibit Phys. Rev. Lett.} {\bibbf 62}
(1989) 324.
\bibitem{c52}%
    E.~M\"uller-Hartmann, {\bibit Z. Phys. B} {\bibbf 74} (1989) 507.
\bibitem{c53}%
    H.~Schweitzer and G.~Czycholl, {\bibit Solid State Commun.} {\bibbf 69}
(1989) 171.
\bibitem{c54}%
    T.~Holstein, {\bibit Ann. Phys.} {\bibbf 8} (1959) 325.
\bibitem{c55}%
    J.K.~Freericks, M.~Jarrel and D.J.~Scalapino, {\bibit Phys. Rev.}
{\bibbf B48} (1993) 6302.
\bibitem{c56}%
    W.~Weber, {\bibit Z. Phys.} {\bibbf B70} (1988) 323.
\bibitem{c57}%
    P.~Soven, {\bibit Phys. Rev.} {\bibbf 156} (1967) 809.
\bibitem{c58}%
    F.~Yonezawa, {\bibit Prog. Theor. Phys.} {\bibbf 40} (1968) 734.
\bibitem{c59}%
    B.~Velick\'y, {\bibit Phys. Rev.} {\bibbf 184} (1969) 614.
\bibitem{c60}%
    P.L.~Leath, {\bibit Phys. Rev.} {\bibbf B2} (1970) 3078.
\bibitem{c61}%
    D.~Vollhardt and P.~Wolfle, {\bibit Phys. Rev.} {\bibbf B22} (1980) 4666.
\end{thebibliography}
\end{document}